\documentclass[11pt,letterpaper,english]{article}
\usepackage[latin9]{inputenc}
\usepackage{array}
\usepackage{pifont}
\usepackage{mathrsfs}
\usepackage{url}
\usepackage{multirow}
\usepackage{amsmath}
\usepackage{amssymb}
\usepackage{graphicx}
\usepackage{esint}

\makeatletter

\pdfpageheight\paperheight
\pdfpagewidth\paperwidth

\providecommand{\tabularnewline}{\\}

\usepackage{jheppub}
\usepackage{babel}
\usepackage{subscript}

\pdfoutput=1

\allowdisplaybreaks

\usepackage{datetime}

\usepackage[usenames,dvipsnames,svgnames,table]{xcolor}
\hypersetup{colorlinks=true, citecolor=red}

\usepackage{titlesec}
\titleformat{\subsubsection}{\normalsize\sf}{\thesubsubsection}{1em}{}

\makeatother

\usepackage{babel}

\begin{document}

\title{A Holographic Model for Pseudogap in BCS-BEC Crossover (I): Pairing Fluctuations, Double-Trace Deformation and Dynamics of Bulk Bosonic Fluid}

\author[]{Oliver DeWolfe, Oscar Henriksson and Chaolun Wu}

\affiliation[]{
Department of Physics, 390 UCB \\ University of Colorado, Boulder, CO 80309, USA \medskip\\ 
Center for the Theory of Quantum Matter \\ University of Colorado, Boulder, CO 80309, USA \medskip
}

\emailAdd{oliver.dewolfe@colorado.edu, oscar.henriksson@colorado.edu,\\chaolun.wu@colorado.edu}

\abstract{
We build a holographic model for the pairing fluctuation pseudogap phase in fermionic high temperature superconductivity/superfluidity based on the BCS-BEC crossover scenario. The pseudogap originates from incoherent Cooper pairing and has been observed in recent cold atom experiments. The strength of Cooper pairing and hence the BCS-BEC crossover is controlled by an effective 4-Fermi interaction and we argue that the double-trace deformation for charged scalar operator is a close analog in large N field theories. We employ the double-trace deformed Abelian Higgs model of holographic superconductors and propose that the incoherent fluctuations of the charged scalar in the bulk is the holographic dual of the fluctuating Cooper pairs. Using a Madelung transformation and the velocity-potential formalism, we develop a quantum fluid dynamics as an effective theory for these bulk fluctuations. The new fluid dynamics takes care of the boundary conditions required by AdS/CFT and encodes the vacuum polarization effect in curved spacetime. The pseudogap in conductivity can be related to the plasma oscillation of this bulk fluid.
}

\keywords{AdS/CFT Correspondence, Holography for Condensed Matter Physics, Pairing Fluctuation Pseudogap, BCS-BEC Crossover}

\maketitle


\section{Introduction}

Since its discovery less than two decades ago, the AdS/CFT correspondence
\cite{Maldacena:1997re,Gubser:1998bc,Witten:1998qj}, or holography,
has shed new light not only on the fields of gravity and high energy
theories, but also on other areas of physics that are highly driven
by experiments, such as nuclear physics, condensed matter and cold
atoms. One tremendous success it enjoys over the last decade is the
building of holographic superconductor/superfluid models starting
in \cite{Hartnoll:2008vx,Hartnoll:2008kx,Horowitz:2008bn,Herzog:2008he}.
It bridges the physics of various phases and phase transitions in
strongly interacting field theories at finite densities, such as those
studied in the context of high temperature superconductivity, to the
physics of the dynamics and instabilities of charged black holes in
asymptotic AdS spacetime \cite{Gubser:2008px}. This is part of the
fruitful program now called AdS/CMT \cite{Hartnoll:2009sz,Zaanen:2015oix}.
Since then, great efforts have been devoted to build various holographic
models with different types of black hole instabilities and to identify
them with possible interesting phases in the dual field theories. 

One of the driving forces behind AdS/CMT is to develop effective field
theory descriptions for the phenomena of high temperature superconductivity
and superfluidity. In the simplest setup, the pure charged AdS black
hole geometry is dual to the gapless normal phase at high temperature
\cite{Lee:2008xf,Liu:2009dm,Cubrovic:2009ye,Faulkner:2009wj}, while
at low temperature, the black hole develops charged hair and is
dual to the gapped superconducting phase \cite{Herzog:2009xv,Horowitz:2010gk,Cai:2015cya}.
However, the actual phase diagrams of high $T_{\mathrm{c}}$ materials
measured in laboratories are more complicated. For the family
of cuprate materials, there exists a so-called ``pseudogap'' phase
\cite{Alloul:1988pg,Batlogg:1994pg,Loeser:1996pg,Ding:1996pg} that
is still mysterious and has defied a consensus among theorists for
a long time \cite{Keimer:2014review,Hashimoto:2014review,Kordyuk:2015review}.
This is a region in the phase diagram located in between the superconducting
phase and normal phase in the underdoped regime, where a gap exists
but no coherent superconductivity develops. Recent advances in experimental
techniques have shown stronger evidence in favor of the competing
order scenario \cite{Fradkin:2014review,Sachdev:2015review}: there
exists more than one symmetry breaking pattern in this region and
the other orders are responsible for the pseudogap and are competing
with the superconducting order. This scenario can be easily incorporated
into holographic model building. The competing orders can be achieved
in holography by adding more matter fields in the bulk. These matters
transform under the same symmetry groups as their dual competing orders,
and can trigger black hole instabilities toward formation of various
types of hair, in similar ways as the superconducting order does.
Such a strategy has been successfully implemented in \cite{Kiritsis:2015hoa}
(and see early references therein) and a phase diagram similar to
that of cuprate is produced.

The aforementioned story is a familiar one to holographic model builders;
however, it is not the whole story of the pseudogap in high temperature
superconductivity and superfluidity. In this paper we want to turn
our attention to another pseudogap phenomenon that is similar yet
distinct from the cuprate one. Recall that fermionic high temperature superfluidity
has also been realized and observed in ultra-cold atomic systems since
2004 \cite{Regal:2004zza}. For the superfluidity transition temperature
$T_{\mathrm{c}}$, in terms of the normalized ratio $T_{\mathrm{c}}/T_{\mathrm{F}}$
where $T_{\mathrm{F}}$ is the Fermi temperature of the system, the
cold atoms in the unitary regime can reach a ratio of $0.15$ to $0.2$,
the highest of any fermionic superfluid! Later, a pseudogap phase
was also detected in such systems \cite{Gaebler:2010pg,Feld:2011pg}.
These cold atom systems are realized in both three and two spatial
dimensions, without an underlying optical lattice. Up to the effect
of the harmonic trap, these systems can be roughly viewed as translational
invariant and isotropic. The dominant symmetry of the order parameter
is $s$-wave, not $d$-wave. All these features make the cold atom
systems different from cuprate materials. As the competing order scenario
for the pseudogap in cuprates relies on the existence of a two-dimensional
lattice and $d$-wave symmetry, it lacks foundation in cold atom systems.
Thus the explanation for the pseudogap in cold atoms will be very
different from the cuprate counterpart. As the paring mechanism is
very well understood in fermionic atom systems, the explanation for the
pseudogap is more transparent and can be largely attributed to incoherent
Cooper pairing with short coherent length and large fluctuations of
the superconducting order parameter. However, this poses challenges
to holographic model building. As there is no competing order, introducing
additional matter fields in the bulk and allowing the black hole to
develop different types of hair does not capture the essence of physics
in the dual field theory. This fact essentially confines us to the
minimal holographic superconductor models, such as the Abelian Higgs
model of \cite{Hartnoll:2008vx,Hartnoll:2008kx,Herzog:2008he}. In
this paper, we propose that for the Abelian Higgs model, the incoherent
fluctuations of the charged scalar field in the bulk is dual to the
pseudogap phase. This is the holographic realization of the pairing
fluctuation pseudogap in the BCS-BEC crossover scenario \cite{Randeria:2013review,Chen:2014review,Levin:2012review,Levin:2005short,Zwerger:2012book}.
We will develop an effective fluid dynamic description for these bosonic
bulk fluctuations.

In fact, the pairing fluctuation pseudogap in the BCS-BEC crossover is
an example of a family of phenomena that could exist in many quantum
field theories, where fluctuations play a crucial role. It offers
a broad and generic paradigm and can fit into many field theories
of different microscopic details. As holography is viewed to be a
generic framework for studying strongly interacting quantum field
theories, it is already interesting enough to ask the question on
a purely theoretical ground how holography can incorporate this
paradigm into it, regardless of its applications on experimental phenomenology.
This is another motivation of ours to initiate this project.

Another issue involving the holographic study of high temperature superconductivity
is the identification of the second axis of the phase diagram. Unlike
conventional fermionic superconductivity and superfluidity, whose
phase diagrams are usually one-dimensional and labeled by the normalized
temperature $T/T_{\mathrm{F}}$ (whereas in holography $T_{\mathrm{F}}$
is usually replaced by the chemical potential or appropriate power
of the charge density), high temperature superconductivity has two-dimensional
phase diagrams. The second axis is an external tunable knob in the
experiments: doping for cuprates and scattering length for cold atoms.
There is no consensus in holography how these shall be realized
in the bulk theory. In this paper, we propose to use a double-trace
deformation \cite{Witten:2001ua} as a universal knob for modeling
these tunable parameters in holography. This is not a completely new
idea, as it has already been employed in the early days of holographic
superconductors \cite{Faulkner:2010gj}; but an explicit identification
with the second axis in high $T_{\mathrm{c}}$ phase diagram was rarely
made in the literature. The justification comes from the fact that,
although doping and scattering length are quite different at the microscopic
level, at low energy due to the renormalization group (RG) flow, they
generate the same IR effect: they induce an effective 4-Fermi type
interaction between the elementary fermions, and the tunable parameters
enter as the dimensionful 4-Fermi coupling. The simplest non-Abelian
generalization of such 4-Fermi interaction is the double-trace deformation,
where the single-trace operator is made of fermion bilinears and their
supersymmetric partners. To make the argument stronger, in this paper,
we will show that using the variational principle and the trick of the Hubbard-Stratonovich
transformation, the 4-Fermi interaction in condensed matter field
theories and the double-trace deformation in high energy field theories
have similar structures in the generating functionals, which are mainly
characterized by pairing symmetry and a dimensionful coupling parameter.
Such structures pass naturally into holography. Now there are two
distinct coupling strengths in our field theory: the 't Hooft coupling
of the undeformed theory and the double-trace coupling. As we are
working with AdS/CFT, we are always in the large 't Hooft coupling
limit, by which we claim our field theory is always in the strong
coupling regime. Meanwhile we can always tune the double-trace coupling
from week to strong, which mimics the effect of ``doping'' the large
$N$ field theory. This is thus a large $N$ setup analogous to
the theory of the BCS-BEC crossover, from which a pairing fluctuation
pseudogap phase will emerge.

This paper is organized as follows. In the next section, we will
give an introduction to the experimental observations of the pseudogap
in cold atom systems and the pairing fluctuation theory in the BCS-BEC
crossover scenario, and discuss what they imply for holographic
model building. In Section 3, we set up the Abelian Higgs model of
holographic superconductivity with a double-trace deformation, review the
basics in a slightly different perspective and discuss how this model
will be extended without adding a new bulk field to generate the pairing
pseudogap phase. Sections 4 and 5 consist of two steps that
transform the conceptual ideas developed in Section 3 into a practically
calculable model based on fluid dynamics. Section 6 focuses on the
bulk dynamics of the hydrostatic configuration which we propose corresponds to
the ground state of the pseudogap. Section 7 discusses bulk dynamics
of charges and how the pseudogap in AC conductivity can be related
to the oscillations of these charges. Section 8 consists of a summary
and comments.

\emph{Notation}: $d$ is the spacetime dimension of the field theory,
hence the bulk has $d+1$ dimensions. $M,N$ denote the bulk spacetime
indices. $\mu,\nu$ denote the boundary spacetime indices. $I,J$
denote bulk spatial indices while $i,j$ boundary spatial indices.
We also use the notation $\vec{v}$ to denote the boundary spatial
vector with components $v^{i}$. We choose the general ansatz for
the metric to be
\begin{equation}
ds^{2}=g_{tt}(z)dt^{2}+g_{zz}(z)dz^{2}+g_{\perp}(z)d\vec{x}^{2},
\end{equation}
and near the boundary, the asymptotic AdS metric takes the form
\begin{equation}
ds_{\mathrm{AdS}}^{2}=\frac{R^{2}}{z^{2}}\left(-dt^{2}+dz^{2}+d\vec{x}^{2}\right),
\end{equation}
where $z\in\left[\epsilon,z_{\mathrm{h}}\right]$ is the radial coordinate,
$R$ the AdS radius, $\epsilon\rightarrow0$ the location of the boundary
and $z_{\mathrm{h}}$ the location of the horizon. The stress tensor
and charge current operators are defined as 
\[
T^{MN}=\frac{2}{\sqrt{-g}}\frac{\delta S}{\delta g_{MN}},\qquad J^{M}=\frac{1}{\sqrt{-g}}\frac{\delta S}{\delta A_{M}}.
\]

\bigskip{}


\section{BCS-BEC Crossover and Incoherent Cooper Pairing}

In this section we give a pedagogical introduction to the experimental
facts and theoretical ideas of the BCS-BEC crossover scenario, focusing
on physics closely related to the pairing fluctuation pseudogap. For
readers interested in more details, we recommend the reviews \cite{Randeria:2013review,Chen:2014review,Levin:2012review,Levin:2005short},
the book \cite{Zwerger:2012book} and references therein.

\subsection{Experimental Evidences for Pairing Fluctuation Pseudogap}

\begin{figure}[t]
\begin{centering}
\includegraphics[width=15cm]{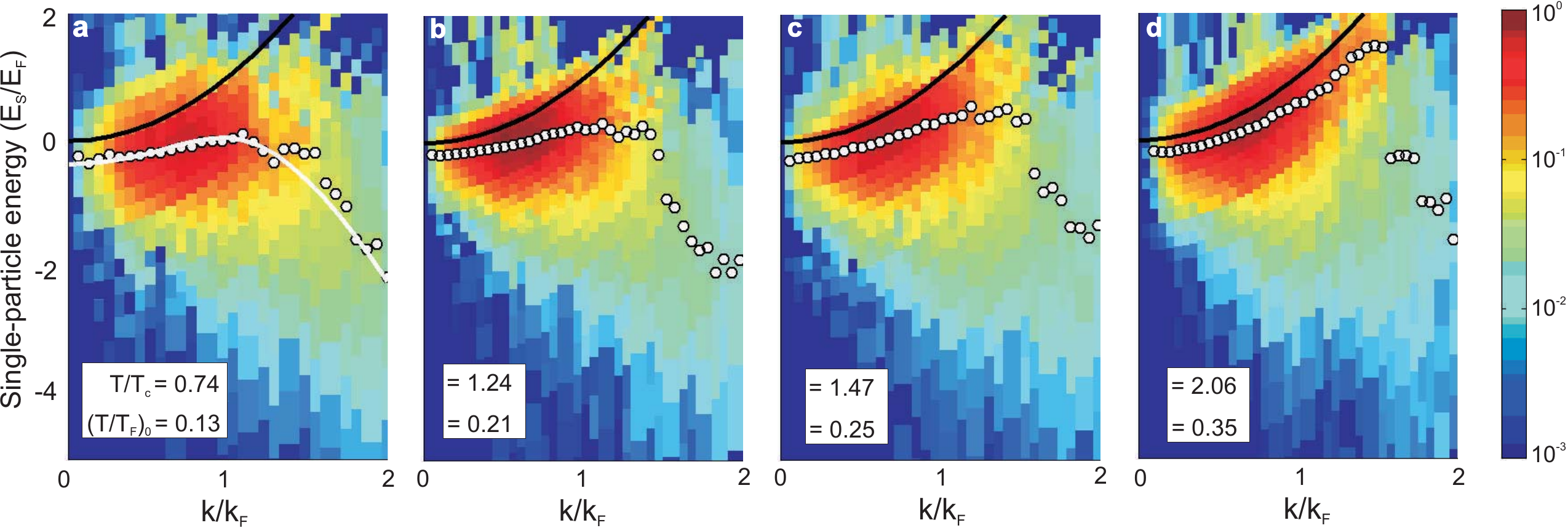}
\par\end{centering}

\caption{Photoemission spectra throughout the pseudogap regime. Spectra are
shown for Fermi gases of $^{40}\mathrm{K}$ near unitarity at four
different temperatures: the first one below $T_{\mathrm{c}}$, the
onset temperature of superfluidity, while the rest above $T_{\mathrm{c}}$.
White dots indicate Gaussian fits of the dispersion relation. The
black curve is the standard quadratic dispersion relation for non-relativistic
free particles. The white curve is a fit to a BCS-like dispersion,
indicating the existence of a gap. It is manifest through the two
central plots that the gap persists into temperatures well above $T_{\mathrm{c}}$.
The figure is reproduced from \cite{Gaebler:2010pg}. \label{Fig:JinExperiment}}

\medskip{}
\end{figure}

In experiments, a pseudogap is defined as a gradual depletion of
the density of state of the elementary fermions near the Fermi surface
at a temperature above $T_{\mathrm{c}}$, the onset temperature of
superconductivity or superfluidity. It can be directly measured by
scattering of the elementary fermions out of the system using techniques
such as angle-resolved photoemission spectroscopy (ARPES) in condensed
matter \cite{Damascelli:2003review,Damascelli:2002review,Campuzano:2002review}
and momentum resolved radio frequency (RF) spectroscopy in cold atoms
\cite{Stewart:2008rf}. In \cite{Gaebler:2010pg}, a gas of fermionic
$^{40}\mathrm{K}$ atoms is cooled to a fraction of its Fermi temperature
in a three-dimensional trap and tuned close to the unitary regime
where the interactions between the atoms are near the strongest. Then
using the technique of RF spectroscopy, the single-particle spectral
function of the fermionic atoms is measured both below and above
$T_{\mathrm{c}}$, and the dispersion relation is retrieved from these
measurements. The results are shown in Figure \ref{Fig:JinExperiment},
which is directly reproduced from \cite{Gaebler:2010pg}. The first
plot is measured below $T_{\mathrm{c}}$, and the rest above $T_{\mathrm{c}}$.
The white dots are fits of the dispersion relation. The black curves
are the standard quadratic dispersion relation for non-relativistic
free particles, while the white curve is a fit to BCS-type dispersion
relation with a non-vanishing energy gap. From the two central plots,
it is obvious that the dispersion relation follows the BCS trend very
well into temperatures well above $T_{\mathrm{c}}$, indicating the
existence of a pseudogap phase above $T_{\mathrm{c}}$. 
Later, the same phenomenon was also observed in two-dimensional Fermi
gases \cite{Feld:2011pg}. 

Unlike in the cuprate case where the underlying lattice structure,
the $d$-wave symmetry and the still mysterious mechanism for Cooper
pairing can give rise to many possibilities for competing orders,
the cold atom systems are much cleaner and simpler. The interactions
between elementary fermions are well understood and the strengths
are highly tunable in experiments. The $s$-wave symmetry and the
absence of the lattice make it much easier to attribute the observed pseudogap
to the incoherent fluctuations of Cooper pairs. This phenomenon has
been predicted, long before the experiment of \cite{Gaebler:2010pg},
by the BCS-BEC crossover scenario, which is originally proposed to
explain the pseudogap phenomenon in cuprate materials (for reviews
on this topic, see \cite{Chen:2000thesis,Levin:2005long,Levin:2010review,Strinati:2010review}).
In the following, we will give a brief introduction to the theory
of BCS-BEC crossover and how it deals with the pseudogap.

\subsection{The BCS-BEC Crossover Scenario}

Conventional superconductivity and superfluidity in systems of
fermions and bosons are described by the Bardeen\textendash{}Cooper\textendash{}Schrieffer
(BCS) theory \cite{BCS:1957} and the Bose-Einstein Condensation (BEC)
theory respectively. The BCS-BEC crossover scenario views these two
distinct paradigms as two opposite limits of a unified paradigm that
continuously interpolates between them. The central concept of the
BCS paradigm of fermionic superconductivity is the Cooper pairing
of fermions via attractive interactions. However, the pairing mechanism
in the original BCS theory is really a special case that is far from the most
general thing that can happen to a pair of fermions. The attractive
interaction is so weak that the fermions are only loosely bound. This
results in large pair size characterized by a divergent coherence
length in position space. In momentum space, the pairing happens near
the Fermi surface between momenta of opposite direction. Thus the
center of mass momentum of the Cooper pair, i.e. the momentum of this
composite boson, is always zero. From the BEC point of view, this is a
boson at its ground state, i.e. a condensate. In BCS superconductivity,
this boson can only be excited by breaking into two fermions (the
Bogoliubov quasi-particles) rather than jumping into an excited bosonic
state with non-vanishing momentum, because the attractive interaction
between the constituent fermions is so weak that any effort to shake
the boson a little bit simply breaks it. The key constituents in the BCS
paradigm are the unpaired fermions and paired bosons in the ground state.
On the other hand, in the BEC paradigm, we start with a system of bosons.
However, a second thought immediately tells us that this is not always
true. For example, $^{4}\mathrm{He}$ is made of six fermions of electrons,
protons and neutrons at the subatomic level and the latter two can be
further decomposed into more elementary fermions. Thus the fact that
we can start with well defined bosons in the BEC paradigm is really
a low energy effective picture, because the attractive interactions
that bind the elementary fermions are so strong that the binding energy
is way higher than the energy scale at which we probe the system to
study superfluidity. This strong interaction results in tightly bound
pairs in position space with small coherence length of order of the
boson size. In momentum space, the boson, i.e. pair of fermions, can
be excited to states with very large momenta without being broken
into fermions. We can say the key constituents in the BEC paradigm are
the paired bosons in the ground state and excited states, without unpaired
fermions. Every feature discussed here in the two paradigms is opposite
to the other. However, they are both rooted in the same ground: pairing
of two fermions into a boson, and the differences are only quantitative,
not qualitative. 

\begin{figure}[t]
\begin{centering}
\includegraphics[width=9cm]{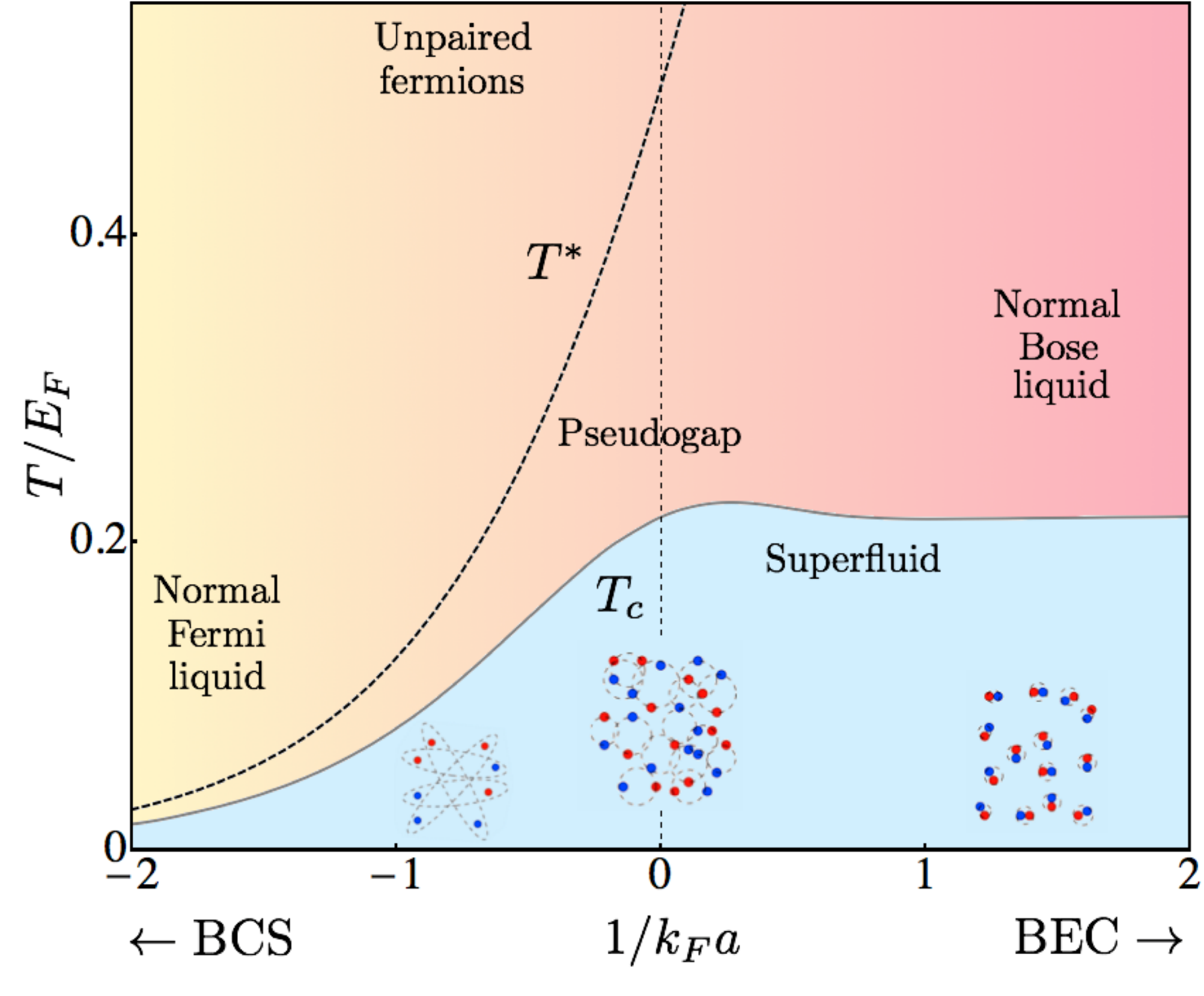}
\par\end{centering}

\caption{A qualitative illustration of the phase diagram of the BCS-BEC crossover
scenario based on theoretical studies. The horizontal axis is the
strength of the coupling, in this note denoted by $\lambda$, with
the origin at the unitarity critical value $\lambda=\lambda_{\mathrm{c}}$.
$T^{*}$ is the pseudogap onset temperature, while $T_{\mathrm{c}}$
is the critical temperature for the onset of superfluidity. The figure
is reproduced from \cite{Randeria:2013review}. \label{Fig:PhaseDiagram}}

\medskip{}
\end{figure}

We can summarize the BCS-BEC crossover scenario in the following.
It describes a system of elementary fermions with tunable attractive
interaction. The fermions pair into bosons. The energy scale associated
with pairing is the binding energy, which corresponds to an onset
temperature $T^{*}$. Below this temperature, the pairs start to form
and the binding energy manifests itself as an energy gap in the system
which can be directly detected in experiments by scattering the fermions
off the system. Once the paired bosons are formed, they can occupy both
the ground state and excited states, labeled by different momenta. As
the temperature keeps lowering, the paired bosons tend to populate more
lower energy states. Eventually at some critical temperature $T_{\mathrm{c}}$,
the Bose-Einstein condensation of the boson pairs takes place and
the ground state is macroscopically occupied: a coherence starts to
form. The pairing temperature $T^{*}$ shall never be lower than the
condensation temperature $T_{\mathrm{c}}$: this is simply the statement
that the bosons have to form first before they condense. One limiting
case is $T^{*}=T_{\mathrm{c}}$, i.e. the pairs condense as soon as
they form --- this is the BCS limit. The other limiting case is $T^{*}\gg T_{\mathrm{c}}$,
where the pairs have already formed even at room temperature which
make it looks like we start with bosons --- this is the BEC limit.
In between these two limit, there is a large regime where the two
temperatures are comparable but not equal. This is the regime of unconventional
superconductivity and superfluidity for which the BCS-BEC crossover
scenario is proposed. Figure \ref{Fig:PhaseDiagram} is a qualitative
phase diagram based on theoretical studies. 

\subsection{Pairing in Pseudogap Phase and its Holographic Dual}

As can be seen from Figure \ref{Fig:PhaseDiagram}, there are three
distinct phases in the BCS-BEC crossover scenario: the normal phase
at $T>T^{*}$, the pseudogap phase at $T_{\mathrm{c}}<T<T^{*}$ and
the superconducting/superfluid phase at $T<T_{\mathrm{c}}$. The elementary
fermions can exist in three different states: the unpaired fermionic
state, the excited and the ground states of the paired bosonic states.
A summary of the three phases, as well as their holographic duals
to be discussed later, can be found in Table \ref{Tab:BCSBEC_ThreePhases}.
From the fermionic superconductivity's point of view, the most exotic
phase in this scenario is the phase taking place at $T_{\mathrm{c}}<T<T^{*}$,
the so-called pairing fluctuation pseudogap phase, which is absent
in conventional superconductors. There are two equivalent ways to
view the excited pair states in this phase. From an unpaired fermion's
perspective, they can be viewed as preformed Cooper pairs that serve
as precursors to the superconductivity: they are meta-stable pairs
that have not condensed. This is ``pairing without condensation''.
From the superconducting condensate's point of view, the condensate
is a huge coherent pairing state whose phase at different positions
is well synchronized. When the excited pair states are populated,
it corresponds to exciting Goldstone bosons of this condensate to
randomize the phase and destroy its coherence at larger scales. This
phase decoherence restores the $U(1)$ symmetry that otherwise would
be broken by the condensate. Now the non-vanishing expectation value
$\langle cc\rangle$ (where $c$ is the field operator for the elementary
fermions) of the fermion pairs goes back to zero at large scales.
This is ``incoherent Cooper pairing''. In field theories, these
two perspectives are almost instantaneously equivalent. However, in
holography, the second perspective of phase decoherence of the condensate
has a more straightforward bulk realization which is the direction
that we will pursue in this paper. The first viewpoint based on fermion
pairing is more obscure in holography because the elementary fermions
in field theory usually do not have an explicit bulk counterpart.
It is mathematically viable to study pairing in the bulk, which has
in fact been done \cite{Faulkner:2009am,Hartman:2010fk,Bagrov:2014mqa,Liu:2014mva,Gubankova:2014iha}.
But it is not immediately obvious what the physical connections between
the pairing in the bulk and that in the field theories are and how
this captures the essence of the BCS-BEC crossover picture presented
above. Thus we will choose a different path based on the philosophy
that we are going to explain now.
\begin{table}[t]
\begin{centering}
\begin{tabular}{|c|c|c|c|c|c|}
\hline 
\multirow{2}{*}{Phase} & \multirow{2}{*}{Temperature} & \multirow{2}{*}{Gap} & \multirow{2}{*}{Broken $U(1)$} & \multicolumn{2}{c|}{Charge Configuration}\tabularnewline
\cline{5-6} 
 &  &  &  & Field Theory & Holography\tabularnewline
\hline 
\hline 
Normal & $T^{*}<T$ & $\times$ & $\times$ & \ding{192} & \ding{182}\tabularnewline
\hline 
Pseudogap & $T_{\mathrm{c}}<T<T^{*}$ & $\surd$ & $\times$ & \ding{193} ( \ding{192} ) & \ding{183} ( \ding{182} )\tabularnewline
\hline 
SC/SF & $T<T_{\mathrm{c}}$ & $\surd$ & $\surd$ & \ding{194} ( \ding{193}, \ding{192} ) & \ding{184} ( \ding{183}, \ding{182} )\tabularnewline
\hline 
\end{tabular}
\par\end{centering}

\medskip{}

where the numbers in the table represent the following

\medskip{}

\begin{centering}
\begin{tabular}{|c|l|}
\hline 
\ding{192} & \emph{Unpaired} fermions: Bogoliubov quasi-particles\tabularnewline
\hline 
\ding{193} & Incoherent Cooper pairs: paired fermions in \emph{excited} states\tabularnewline
\hline 
\ding{194} & Coherent Cooper pairs: paired fermions in \emph{ground} state\tabularnewline
\hline 
\ding{182} & Charges confined \emph{behind} the black hole horizon\tabularnewline
\hline 
\ding{183} & Charges outside the horizon carried by excited scalar quanta: bosonic
\emph{normal} fluid\tabularnewline
\hline 
\ding{184} & Charges outside the horizon carried by condensate of the scalar: bosonic
\emph{super}fluid\tabularnewline
\hline 
\end{tabular}
\par\end{centering}

\begin{centering}
\caption{Key features of the three phases in pairing fluctuation theory of
the BCS-BEC crossover scenario and their charge configurations in
the black hole geometry of the holographic dual, where we assume the
Abelian Higgs model of holographic superconductor. SC and SF stand
for superconducting and superfluid respectively. The ``( )'' indicates
possible coexisting configurations of the charges. \label{Tab:BCSBEC_ThreePhases}}

\par\end{centering}

\medskip{}
\end{table}

A key feature of the physical picture that we have just described
is that there is \emph{no} competing order or hidden symmetry breaking
in the BCS-BEC crossover scenario. The pseudogap parameter and the
superconducting gap parameter share the same microscopic origin and
the same symmetry. The only obvious difference is that the latter
is complex and the former is real because its phase is washed out
by phase decoherence. This will be an important guiding principle
in our holographic model building. Trying to generate more complicated
classical background configurations by introducing additional bulk
fields other than the original one that produces the superconducting
condensate is equivalent to modeling competing orders in the field
theories (see for example \cite{Kiritsis:2015hoa} and references
therein). For the study of BCS-BEC crossover, however, we choose a different
track. To capture the essence of the pairing fluctuation pseudogap phase
in holographic models, we will stick to the minimal holographic superconductor
model and see how this new phase can be generated from the same old
model by attempting to upgrade the bulk dynamics to the quantum level and including
fluctuations for the condensate field. A clue supporting this strategy
is that, since in the field theory, the condensed pairs and the incoherent
pairs are indeed the same type of pairs just in different quantum
states, the charges dual to them in the holographic bulk shall be
carried by the same bulk field, only in different configurations ---
one coherent and one incoherent. In the Abelian Higgs model of holographic
superconductors, superconductivity is realized by pumping charges out
of AdS-Reissner-Nordström black hole%
\footnote{Here we assume the gapless normal phase at non-vanishing temperature
is dual to a non-extremal AdS-Reissner-Nordström black hole. At low
temperature, there are alternative scenarios such as the holographic
electron star model \cite{Hartnoll:2010gu,Hartnoll:2010ik}. For more
on the alternatives, see the Introduction section of \cite{Sachdev:2011ze}
or Chapters in \cite{Zaanen:2015oix}. These alternative scenarios
for the normal phase will not affect the holographic realization of
the pseudogap phase to be discussed in the rest of this note based
on incoherent bosonic fluid.%
} to form a coherent condensate of the charged scalar outside the horizon.
Then by analogy, the pseudogap will be realized by pumping the
same type of charges, i.e.\ quanta of the charged scalar, out of the
black hole to form an incoherent entity outside the horizon. When
the bulk theory is viewed as a quantum field theory, the superconducting
hair is the Bose-Einstein condensate of the charged scalar, i.e.\ a
macroscopic number of quanta of the ground state. The incoherent entity
is just the collection of quanta of the excited states. They can be
viewed as a depletion of the coherent ground state quanta as well.
This is the bulk scalar analog of the two-fluid (superfluid versus
normal fluid) picture of $^{4}\mathrm{He}$ superfluidity. Thus in
a coarse-grained picture, the bulk configuration that is responsible
for the pseudogap is a normal fluid outside the black hole horizon,
which is made of the same charged scalar that develops the superconducting
hair. This is also shown in Table \ref{Tab:BCSBEC_ThreePhases}. Thus
the first step toward a holographic model for pairing fluctuation
pseudogap is to formulate the dynamics for this normal fluid. This
is the main purpose of this paper.

\subsection{From 4-Fermi Interaction to Double-Trace Deformation}

Before directly jumping into holographic model building, it is instructive
to have a look at field theoretical approaches to the pairing fluctuation
pseudogap. Here in alignment with the condensed matter literature,
we will adopt the notation that $c$ and $c^{\dagger}$ denote fermionic
operators and $b$ and $b^{\dagger}$ bosonic operators.

The field theoretical approaches in condensed matter and cold atom
physics usually start with the Hamiltonian $H=H_{0}+H_{\mathrm{int}}$.
Here $H_{0}$ is the Hamiltonian for free elementary fermions (electrons
in condensed matter physics and fermionic atoms in cold atom physics):
we will not specify its specific form since we will eventually pass
to a dual holographic description. For us, we can view $H_{0}$ as
denoting a general class of field theories (especially CFTs) which
have the standard holographic dual description. $H_{\mathrm{int}}$
takes the following single-channel form in momentum space
\begin{equation}
H_{\mathrm{int}}=\iiint\frac{d^{d-1}\vec{k}}{(2\pi)^{d-1}}\frac{d^{d-1}\vec{k}^{\prime}}{(2\pi)^{d-1}}\frac{d^{d-1}\vec{q}}{(2\pi)^{d-1}}V(\vec{k},\vec{k}^{\prime})c_{\uparrow}^{\dagger}\left(\vec{k}+\frac{\vec{q}}{2}\right)c_{\downarrow}^{\dagger}\left(-\vec{k}+\frac{\vec{q}}{2}\right)c_{\downarrow}\left(-\vec{k}^{\prime}+\frac{\vec{q}}{2}\right)c_{\uparrow}\left(\vec{k}^{\prime}+\frac{\vec{q}}{2}\right).
\end{equation}
Here $c_{\sigma}$ and $c_{\sigma}^{\dagger}$ are the field operators
of the elementary fermions in the system with spin index $\sigma=\uparrow,\downarrow$.
This type of 4-Fermi interaction shall be viewed as an IR effective
operator that deforms the original field theory given by $H_{0}$.
It is generated from some more fundamental interactions between the
fermions in the microscopic theory by interacting out the UV degrees
of freedom. For example, for conventional BCS type superconductors,
the fundamental interaction between fermions is the phonon, and by
integrating it out, we end up with effective interactions between
fermions of the above form. Thus the form of the interaction potential
$V(\vec{k},\vec{k}^{\prime})$ is related to microscopic physics,
such as the momentum cutoff (spatial range) of the interaction, which
will serve as a UV cutoff of this low energy effective description.%
\footnote{A similar example is the 4-Fermi interaction for $\beta$-decay, which
is a low energy effective description of the microscopic weak interaction.
Its interaction strength (analog of our $V(\vec{k},\vec{k}^{\prime})$
here) is set by the W boson mass.%
} In practice, it is usually assumed that the interaction potential
is of a separable form 
\begin{equation}
V(\vec{k},\vec{k}^{\prime})=\lambda\varphi(\vec{k})\varphi(\vec{k}^{\prime})^{*},\label{eq:InteractionSeparability}
\end{equation}
where $\varphi(\vec{k})$ is normalized to be \emph{dimensionless},
and its Fourier transform is denoted as $\varphi(\vec{x})$. Define
the following operator 
\begin{align}
b^{\dagger}(\vec{x}) & \equiv\int d^{d-1}\vec{r}\varphi(\vec{r})c_{\uparrow}^{\dagger}\left(\vec{x}+\frac{1}{2}\vec{r}\right)c_{\downarrow}^{\dagger}\left(\vec{x}-\frac{1}{2}\vec{r}\right).\label{eq:PairCreationOperator_Position}
\end{align}
The physical meaning of $b^{\dagger}(\vec{x})$ is to create a pair
of elementary fermions whose center-of-mass is located at $\vec{x}$.
$\varphi(\vec{r})$ is the relative wave-function of this pair in
its center-of-mass frame. In the context of fermionic superconductivity
and superfluidity, $b^{\dagger}$ is the creation operator of Cooper
pairs and the form of $\varphi(\vec{r})$ is determined by the symmetry
of pairing, i.e. s-, p- or d-wave. We will assume s-wave symmetry
so $\varphi(\vec{r})=\delta^{d-1}(\vec{r})$. Now the interaction
Hamiltonian in position space can be simply written as 
\begin{equation}
H_{\mathrm{int}}=\lambda\int d^{d-1}\vec{x}b^{\dagger}(\vec{x})b(\vec{x}).\label{eq:DoubleTrace_Hamiltonian}
\end{equation}
The interaction Hamiltonian looks just like a chemical potential term
for Cooper pairs, with the interaction strength $\lambda$ playing
the role of chemical potential. 

From high energy theory's point of view, we can view the operator
$b$ as a charged scalar single-trace operator of low conformal dimension
(possibly equal or close to that of elementary fermion bilinears)
in the undeformed field theory specified by $H_{0}$, and is dual
to a charged bulk scalar in the holographic theory. The interaction
$H_{\mathrm{int}}$ is then a double-trace deformation, similar to
that studied in the context of AdS/CFT correspondence in \cite{Aharony:2001pa,Witten:2001ua,Berkooz:2002ug,Mueck:2002gm,Sever:2002fk}.
It is not hard to recognize the structural similarity between double-trace
deformations made of fermion bilinear single-trace operators and the
4-Fermi interactions widely used in models of condensed matter and
cold atom theories. \cite{Vecchi:2010jz} has studied a few explicit
examples of such double-trace deformed high energy models. In fact,
the scalar double-trace deformation has already been used as a knob to
study holographic superconductor models in the large $N$ limit \cite{Faulkner:2010gj}.
It it true that such a connection cannot be established rigorously,
since the microscopic field theories studied in high energy physics
and in condensed matter and atomic physics usually have quite different
field contents and symmetries. It is hard to precisely identify counterparts
of (\ref{eq:PairCreationOperator_Position}) and (\ref{eq:DoubleTrace_Hamiltonian})
in theories such as $\mathcal{N}=4$ supersymmetric Yang-Mills (SYM)
theory. Nonetheless, the scalar double-trace deformation in non-Abelian
gauge theories is the structure that most closely resembles the structure
exhibited by (\ref{eq:DoubleTrace_Hamiltonian}) in the sense that
it is a bilinear of charged scalar observable operators which develops
an expectation value in the broken gauge symmetry phase. They can
both be viewed as IR effective operators that are generated by integrating
out UV degrees of freedom, such as the force mediator in the mechanism
for Cooper pairing. A more convincing evidence is what we are going
to show later in equation (\ref{eq:GeneratingFunctional_HSTransformed}),
that the general relations between the effective action of the deformed
theory and that of the undeformed theory are exactly the same, regardless
of the underlying structures at the microscopic level. Thus in the
study of holographic models of superconductivity and superfluidity,
the double-trace deformation is a good candidate for modeling the
extra knob in the experiments, such as the doping in cuprate superconductivity
and the magnetic field tuned scattering length in cold atom experiments.%
\footnote{A major difference between the literature of double-trace deformations
in high energy physics and that of 4-Fermi interactions in condensed
matter and atomic physics is that the former are usually studied in
vacuum and the latter always at finite density and temperature.%
}

If we assume the pair operator $b$ has energy dimension $d-1$, i.e.
that of an elementary fermion bilinear, which is the one typically
employed in condensed matter, then the coupling constant $\lambda$
has energy dimension $2-d$. Since we are interested in $d=3$ or
$4$, the Hamiltonian operator (\ref{eq:DoubleTrace_Hamiltonian})
will be slightly irrelevant, not marginal. The coupling constant $\lambda$
in (\ref{eq:DoubleTrace_Hamiltonian}) is the bare coupling. It is
renormalized in quantum field theory. An explicit calculation of its
renormalization requires knowledge of details of the interaction,
i.e. the specific form of $\varphi(\vec{k})$, as well as knowledge
of the Hamiltonian $H_{0}$. It is a case by case study that has been
carried out many times in different contexts. In the context of superconductivity
and superfluidity, a brief outline can be found in many of the aforementioned
reviews. Detailed calculations can be found, for example, in \cite{Randeria:1990pg,Kokkelmans:2002zz,Gurarie:2006}.
In conformal field theories it has been studied in \cite{Dymarsky:2005uh,Pomoni:2008de,Vecchi:2010dd,Aharony:2015afa}.
Here we will not refer to any specific context but give a rather general
discussion that is just enough for us in later sections. The key idea
is that as one tunes $\lambda$ from 0 to $-\infty$, the bare attractive
interaction between two fermions changes from tiny to large, and at
some critical value $\lambda_{\mathrm{c}}<0$ the first bound state
between the fermions just forms. This is the familiar story from scattering
theory in quantum mechanics and the bound state corresponds to a divergent
scattering length. In quantum field theory, $\lambda_{\mathrm{c}}$
is a pole in the fermionic 4-point function $\langle c^{\dagger}c^{\dagger}cc\rangle$,
or equivalently, the bosonic 2-point function $\langle b^{\dagger}b\rangle$:
\begin{equation}
\langle b^{\dagger}b\rangle\Big|_{\lambda\rightarrow\lambda_{\mathrm{c}}}\sim O\left(\frac{1}{\lambda-\lambda_{\mathrm{c}}}\right).
\end{equation}
In BCS-BEC crossover, the $\lambda\sim0$ regime corresponds to the
BCS limit, where the interaction is weak. As we tune $\lambda$ to
approach $\lambda_{\mathrm{c}}$, we enter the unitary regime where
the renormalized interaction is strongest. Later we will see fluctuations
are also strongest in this regime. As we keep tuning $\lambda$ to
pass the unitary regime, the bound state between fermions are tighter
and tighter and they dimerize. As $\lambda\rightarrow-\infty$ we
are entering the other asymptotic regime opposite to the BCS limit
--- the BEC limit. Although the strength of the bare coupling $|\lambda|$
is even greater than $|\lambda_{\mathrm{c}}|$ here, the renormalized
(residual) interaction between dimers becomes weaker. Thus this is
also a weak coupling regime.

The coupling $\lambda$ is dimensionful and its scale is set by an
UV energy scale which is related to the range of the interaction potential
$V$, or equivalently a momentum cutoff in $\varphi(\vec{k})$. In
AdS/CFT, this UV cutoff is related to the location of the boundary.
For us, the boundary is located at $z=\epsilon$, where $\epsilon$
is a small length scale. Thus $\epsilon^{-1}$ is proportional to
the UV cutoff energy scale, or we can say $\epsilon$ itself is related
to the small range of the contact potential (assuming the interaction
potential $V$ is s-wave), which is in fact usually the smallest length
scale in cold atom problems. We can define a dimensionless coupling
$\hat{\lambda}$ by dividing $\lambda$ with appropriate power of
this UV cutoff. The critical value of the coupling $\lambda_{\mathrm{c}}$
is also set by this UV cutoff.

\subsection{Transformation of the Effective Action}

The main modern approaches to the pairing fluctuation problem of BCS-BEC
crossover in condensed matter and atomic physic are diagrammatic approaches.
We refer readers interested in the diagrammatic approaches to the
reviews \cite{Chen:2000thesis,Levin:2005long,Levin:2010review,Strinati:2010review}
and references therein. These are mostly orthogonal to the approach
we are interested in. Here we will briefly outline the path integral
treatment of the problem, which can easily lead us to holography. 

Let $S_{0}\left[c\right]$ denote the action of the undeformed field
theory specified by $H_{0}$ in the above. The elementary quantum
fields are $c_{\sigma}$ and $c_{\sigma}^{\dagger}$ (which we will
just write as $c$ for simplicity), among others which we do not write
explicitly, and their path integrals are collectively denoted as $\int\mathcal{D}c$.
The generating functional $\mathcal{Z}_{0}$ and effective action
$\Gamma_{0}$ for this undeformed theory are
\begin{equation}
\mathcal{Z}_{0}\left[J_{b}\right]=e^{i\Gamma_{0}\left[J_{b}\right]}=\int\mathcal{D}c\exp\left\{ iS_{0}\left[c\right]+i\int d^{d}x\left(J_{b}^{\dagger}b+J_{b}b^{\dagger}\right)\right\} ,\label{eq:GeneratingFunctional_Undeformed}
\end{equation}
where $b$ shall be viewed as a composite operator defined by (\ref{eq:PairCreationOperator_Position}),
and $J_{b}$ is the source coupled to it. Of course there can be other
sources coupled to other operators. We will not write them explicitly.
Now we add the interaction term $H_{\mathrm{int}}$ given by (\ref{eq:DoubleTrace_Hamiltonian})
with coupling parameter $\lambda$ to deform the original field theory.
We will call this field theory specified by the full Hamiltonian $H=H_{0}+H_{\mathrm{int}}$
the deformed field theory. This deformation corresponds to adding
the following interaction action 
\begin{equation}
S_{\mathrm{int}}\left[c\right]=-\lambda\int d^{d}xb^{\dagger}(x)b(x)
\end{equation}
to the action $S_{0}\left[c\right]$. The deformed generating functional
$\mathcal{Z}_{\lambda}$ and effective action $\Gamma_{\lambda}$
is
\begin{equation}
\mathcal{Z}_{\lambda}\left[J_{b}\right]=e^{i\Gamma_{\lambda}\left[J_{b}\right]}=\int\mathcal{D}c\exp\left\{ iS_{0}\left[c\right]+iS_{\mathrm{int}}\left[c\right]+i\int d^{d}x\left(J_{b}^{\dagger}b+J_{b}b^{\dagger}\right)\right\} .
\end{equation}
To manipulate this path integral, we employ the standard trick of
Hubbard-Stratonovich transformation
\[
\exp\left\{ -i\lambda\int d^{d}xb^{\dagger}b\right\} =\int\mathcal{D}\upsilon\exp\left\{ i\int d^{d}x\left[\frac{1}{\lambda}\upsilon^{\dagger}\upsilon+\upsilon^{\dagger}b+b^{\dagger}\upsilon\right]\right\} ,
\]
where $\upsilon$ and $\upsilon^{\dagger}$ are the Hubbard-Stratonovich
auxiliary fields. A $1/\lambda$ coefficient in front of the path
integral has been dropped since it will not have any physical consequence.
Then the generating functional becomes
\[
\mathcal{Z}_{\lambda}\left[J_{b}\right]=\int\mathcal{D}\upsilon\int\mathcal{D}c\exp\left\{ iS_{0}\left[c\right]+i\int d^{d}x\left[\left(J_{b}+\upsilon\right)b^{\dagger}+\mathrm{c.c.}\right]+i\int d^{d+1}x\frac{1}{\lambda}\upsilon^{\dagger}(x)\upsilon(x)\right\} .
\]
The path integral over $c$ can now be formally performed by using
the definition of the undeformed effective action (\ref{eq:GeneratingFunctional_Undeformed}),
which yields 
\begin{equation}
e^{i\Gamma_{\lambda}\left[J_{b}\right]}=\int\mathcal{D}\upsilon\exp\left\{ i\Gamma_{0}\left[J_{b}+\upsilon\right]+i\int d^{d}x\frac{1}{\lambda}\upsilon^{\dagger}(x)\upsilon(x)\right\} .\label{eq:GeneratingFunctional_HSTransformed}
\end{equation}
This last equation establishes a formal relation between the effective
actions of the undeformed and deformed theories. It also serves as
the starting point of our holographic model building. It is valid
for field theories considered in both condensed matter and high energy
theories whose interactions have a similar structure as what we have
just discussed. Thus this formula is the bridge that allows us to
travel back and forth between the realm of non-relativistic field
theories in condensed matter and cold atom physics and that of conformal
field theories and holography in high energy physics. Although the
path integral over the auxiliary field $\upsilon$ can not be done
exactly, this formula is the starting point of many theoretical studies
using different approximations to extract physical information from
it. It has been studied in \cite{Gubser:2002vv,Pomoni:2008de} for
CFTs in the large $N$ limit. For BCS-BEC crossover, the way to proceed
with the path integral (\ref{eq:GeneratingFunctional_HSTransformed})
is to first study its saddle point. This was first applied in \cite{SadeMelo:1993zz,Engelbrecht:1997zz}
and further developed by other researchers. The result is a BCS-BEC
crossover of the superconducting phase as one tunes the parameter
$\lambda$. However, the saddle point approximation cannot yield
the pairing fluctuation pseudogap phase, because it completely ignores
the fluctuations. To study the pseudogap phase, one has to look at
the fluctuations of $\upsilon$ around its saddle point. This is a
much harder task. A typical approximation to simplify the task is
to truncate the fluctuation at quadratic order. This Gaussian approximation
corresponds to a one-loop expansion of the generating functional.
For a brief summary of this approach in the BCS-BEC crossover, see \cite{Tempere_OnlineNotes}.

\subsection{Saddle Points and Gaussian Fluctuations}

To proceed from (\ref{eq:GeneratingFunctional_HSTransformed}), we
write $\upsilon=\bar{\upsilon}+\delta\upsilon$, where $\bar{\upsilon}$
is the saddle point value and $\delta\upsilon$ is the fluctuation
around the saddle point. The connected $n$-point correlation functions
of operator $b$ are given by functional derivatives of the effective
action $\Gamma_{\lambda}\left[J_{b}\right]$ with respect to $J_{b}$.
Since we only want to illustrate the general structures of the path
integral, for simplicity, we will ignore details of the ordering of
operators and assume 2-point functions $\langle bb\rangle$ and $\langle b^{\dagger}b^{\dagger}\rangle$
are vanishing or negligible. 

The saddle points of (\ref{eq:GeneratingFunctional_HSTransformed})
are given by the condition 
\begin{align}
\frac{\delta\Gamma_{0}\left[X\right]}{\delta X^{\dagger}}\Big|_{X=J_{b}+\bar{\upsilon}}+\frac{1}{\lambda}\bar{\upsilon} & =0.\label{eq:SaddlePoint_Condition}
\end{align}
This saddle point condition is a self-consistent equation for $\bar{\upsilon}$
since it appears in both terms. On the other hand, we can directly
take the functional derivative of (\ref{eq:GeneratingFunctional_HSTransformed})
and then use this saddle point condition to simplify it, and we obtain
\begin{equation}
\bar{\upsilon}(x)=-\lambda\langle b(x)\rangle_{\lambda,J_{b}}^{\mathrm{sadd}},\label{eq:SaddlePoint_1PointFunction}
\end{equation}
where the subscripts emphasize that $\langle b\rangle$ here is both
a function of coupling $\lambda$ and source $J_{b}$, i.e. the non-equilibrium
one-point function of the deformed theory. The superscript ``sadd''
stands for ``saddle point''. (\ref{eq:SaddlePoint_1PointFunction})
tells us that at saddle points, the value of $\upsilon$ is just the
expectation value of the Cooper pair operator $\langle b\rangle$.
This shows that, when $J_{b}=0$, there are only two distinct phases
at saddle point level: the normal phase where both $\langle b\rangle$
and $\bar{\upsilon}$ vanish, and the superconducting phase where
both of them are non-vanishing. In the former case, the deformed effective
action (\ref{eq:GeneratingFunctional_HSTransformed}) is the same
as the undeformed one (\ref{eq:GeneratingFunctional_Undeformed}),
which usually describes a ``trivial'' gapless phase such as a free
theory, a Fermi liquid, a metal or a CFT. In the latter case we have
a superconducting phase with a broken $U(1)$ symmetry, which describes
the evolution of the condensate from BCS limit to BEC limit as one
tunes the coupling $\lambda$. Here we have either a broken $U(1)$
symmetry (i.e. $\langle b\rangle\neq0$) or trivial phase (i.e. $\upsilon=0$),
but there is no room for the pairing fluctuation pseudogap phase,
which corresponds to unbroken $U(1)$ symmetry (i.e. $\langle b\rangle=0$)
and non-trivial gapped phase (i.e. $\upsilon\neq0$). As the pseudogap
is related to fluctuations of the condensate and the saddle point
approximation is a perfect mean field theory which suppresses all
fluctuations, we have to go beyond the saddle points. Later we will
see precisely the same thing happens in the holographic dual theory
as well.

We now look at the Gaussian fluctuations around the saddle points.
The effective action (\ref{eq:GeneratingFunctional_HSTransformed})
up to quadratic orders in $\delta\upsilon$ can be written as $\Gamma_{\lambda}\left[J_{b}\right]=\Gamma_{\lambda}^{\mathrm{sadd}}\left[J_{b}\right]+\Gamma_{\lambda}^{\mathrm{flct}}\left[J_{b}\right]$,
where $\Gamma_{\lambda}^{\mathrm{sadd}}\left[J_{b}\right]$ is the
saddle point value and $\Gamma_{\lambda}^{\mathrm{flct}}\left[J_{b}\right]$
the contribution from Gaussian fluctuations that we are going to investigate
now. To proceed, first we combine the saddle point condition (\ref{eq:SaddlePoint_Condition})
and the one-point function (\ref{eq:SaddlePoint_1PointFunction})
as
\[
\frac{\delta\Gamma_{\lambda}^{\mathrm{sadd}}\left[J_{b}\right]}{\delta J_{b}^{\dagger}}=\frac{\delta\Gamma_{0}\left[X\right]}{\delta X^{\dagger}}\Big|_{X=J_{b}+\bar{\upsilon}}.
\]
Taking one more functional derivative of it and using (\ref{eq:SaddlePoint_1PointFunction}),
it can then be written as
\begin{equation}
\frac{\delta^{2}\Gamma_{0}\left[X\right]}{\delta X^{\dagger}\delta X}\Big|_{X=J_{b}+\bar{\upsilon}}=-\frac{\mathcal{G}_{bb^{\dagger}}^{\mathrm{sadd}}\left[J_{b};\lambda\right]}{1+\lambda\mathcal{G}_{bb^{\dagger}}^{\mathrm{sadd}}\left[J_{b};\lambda\right]},
\end{equation}
where we have used the definition of the two point function 
\[
\mathcal{G}_{bb^{\dagger}}^{\mathrm{sadd}}\left[J_{b};\lambda\right]=-\frac{\delta^{2}\Gamma_{\lambda}^{\mathrm{sadd}}\left[J_{b}\right]}{\delta J_{b}^{\dagger}\delta J_{b}}.
\]
Using the above relation, Gaussian fluctuation part of the effective
action can be expressed in term of the saddle point 2-point function
as 
\begin{equation}
e^{i\Gamma_{\lambda}^{\mathrm{flct}}\left[J_{b}\right]}=\int\mathcal{D}\delta\upsilon\exp\left\{ i\iint d^{d}xd^{d}x^{\prime}\frac{\delta\upsilon^{\dagger}\delta\upsilon}{\lambda\left(1+\lambda\mathcal{G}_{bb^{\dagger}}^{\mathrm{sadd}}\left[J_{b};\lambda\right]\right)}\right\} .\label{eq:EffectiveAction_Gaussian}
\end{equation}
Notice here we use $\iint d^{d}xd^{d}x^{\prime}$ to emphasize that
the integral is actually non-local since $\delta\upsilon$ and $\delta\upsilon^{\dagger}$
are not at the same point and the two-point function $\mathcal{G}_{bb^{\dagger}}^{\mathrm{sadd}}\left[J_{b};\lambda\right]$
is also a non-local function depending on two different locations.
The integral can be localized in momentum space, and it can also be
written formally as a functional determinant: but these are mathematical
details that are not relevant here. A quantitative evaluation of (\ref{eq:EffectiveAction_Gaussian})
is hard and is not what we will pursue here. However, from its structure
we can see when the fluctuations are important. Recall that the renormalization
of coupling $\lambda$ yields three different regimes of distinct
characters, we will discuss what happens to the Gaussian fluctuations
in these three regimes respectively.
\begin{itemize}
\item BCS limit: $\lambda\rightarrow0$. This is the weak coupling limit
and the deformation (\ref{eq:DoubleTrace_Hamiltonian}) can be treated
perturbatively. It can be shown that $\mathcal{G}_{bb^{\dagger}}^{\mathrm{sadd}}\left[J_{b};\lambda\right]$
does not depend on $\lambda$ in a too singular way, then the denominator
in (\ref{eq:EffectiveAction_Gaussian}) vanishes as $\lambda\rightarrow0$.
Now the exponential factor is highly oscillatory and its major contribution
to the path integral of $\delta\upsilon$ comes from the region where
$\delta\upsilon=0$. This means the Gaussian fluctuation is highly
suppressed and the saddle point approximation is a pretty good one.
This is in fact what we expect for the BCS limit since it is a perfect
mean field theory.
\item BEC limit: $\lambda\rightarrow-\infty$. In this regime, $\mathcal{G}_{bb^{\dagger}}^{\mathrm{sadd}}\left[J_{b};\lambda\right]$
can be calculated after a little algebra under certain simplification
(for example, in \cite{Gubser:2002vv}). The denominator in (\ref{eq:EffectiveAction_Gaussian})
is finite for large $|\lambda|$. This means the fluctuations are
not suppressed in the path integral and the saddle point results may
get considerable corrections.
\item Unitarity: $\lambda\sim\lambda_{\mathrm{c}}$. This is the regime
where scattering length diverges and the pair 2-point function approaches
its pole 
\begin{equation}
\mathcal{G}_{bb^{\dagger}}^{\mathrm{sadd}}\left[J_{b};\lambda\right]\Big|_{\lambda\rightarrow\lambda_{\mathrm{c}}}\sim O\left(\frac{1}{\lambda-\lambda_{\mathrm{c}}}\right).
\end{equation}
Now the denominator of (\ref{eq:EffectiveAction_Gaussian}) diverges,
and the path integral does not suppress the fluctuations at all. Hence
the fluctuations reach maximum and calculations obtained from the saddle
point approximation may not be reliable at all. This regime is experimentally
the most interesting one and theoretically the hardest.
\end{itemize}
In any case, when the fluctuation $\delta\upsilon$ is non-trivial, it
will nullify the proportionality relation between $\upsilon$ and
$\langle b\rangle$. Recall (\ref{eq:SaddlePoint_1PointFunction})
is only a special case for $\upsilon=\bar{\upsilon}$ and $\langle b\rangle$
when $\delta\upsilon$ is completely absent. In cases when $\delta\upsilon$
is non-trivial, $\delta\upsilon$ or $\upsilon$ does not even have
a unique fixed value in the effective action. In the path-integral
sense, it is really a superposition of infinitely many configurations
with different values of $\upsilon$. Each individual configuration
with a specific non-vanishing value of $\upsilon$ breaks the $U(1)$
symmetry, but the superposition of all these configurations restores
the $U(1)$ symmetry. This gives rise to a new phase where $\langle b\rangle=0$
but $\upsilon$ is non-trivial in the effective action (\ref{eq:GeneratingFunctional_HSTransformed}),
and the effective action will not equal to the undeformed gapless
effective action (\ref{eq:GeneratingFunctional_Undeformed}). It can
describe a phase of unbroken $U(1)$ symmetry with a pseudogap parameter
generated by the superposition of non-vanishing $\upsilon$ configurations
(in some sense, a non-trivial ``average'' $\upsilon$). This suggests
how the pseudogap phase arises from the path integral formalism.

\bigskip{}


\section{Double-Trace Deformed Holography: Going beyond Saddle Points}

\subsection{Holography as a Hubbard-Stratonovich Transformation}

Now we go back to the path integral formula of the effective action
(\ref{eq:GeneratingFunctional_HSTransformed}), and seek to proceed
in a completely different direction than has previously been explored
in the BCS-BEC crossover literature: the holography. Along the line
of what we have been doing so far, the whole holographic structure
can be viewed as a second and fancier Hubbard-Stratonovich transformation
for the path integral in (\ref{eq:GeneratingFunctional_HSTransformed}),
whose purpose is to help to integrate out the first Hubbard-Stratonovich
auxiliary field $\upsilon$ exactly! The spirit of the Hubbard-Stratonovich
transformation is to linearize a non-linear interaction term, and thus
to facilitate path integrals over the original quantum fields. Recall
that the reason why we introduce the original Hubbard-Stratonovich
transformation with $\upsilon$ is to decouple the double-trace deformation
(\ref{eq:DoubleTrace_Hamiltonian}) from being quadratic in $b$ to
being linear in $b$, and then we know how to formally perform the
path integral for $c$ with linear $b$ using the formula (\ref{eq:GeneratingFunctional_Undeformed}).
We end up with (\ref{eq:GeneratingFunctional_HSTransformed}). From
the mathematical point of view, this is simply a change of integration
variables. The gain is $c$ has been integrated out exactly, and the
price we pay is to introduce another path integral over $\upsilon$,
which we do not know how to carry out rigorously because $\Gamma_{0}\left[J_{b}+\upsilon\right]$
depends on $\upsilon$ in a very complicated way. Recall the coefficients
of $\Gamma_{0}$'s Taylor expansion at each order are the corresponding
$n$-point functions, thus $\Gamma_{0}\left[J_{b}+\upsilon\right]$
contains all non-negative powers of $\upsilon$ in general. We only
know how to perform the path integral over $\upsilon$ in (\ref{eq:GeneratingFunctional_HSTransformed})
rigorously if we can write $\Gamma_{0}\left[J_{b}+\upsilon\right]$
in a way that is at most quadratic in $\upsilon$. In this sense,
we need to introduce a second Hubbard-Stratonovich transformation,
and the one that does the magic is holography!

We now write down the bulk action as a holographic Hubbard-Stratonovich
transformation for the effective action $\Gamma_{0}\left[J_{b}\right]$
defined in (\ref{eq:GeneratingFunctional_Undeformed}), which is in
fact the Abelian Higgs model of a holographic superconductor. According
to the standard AdS/CFT dictionary, a charged scalar operator $b$
of charge $q_{\phi}$ and conformal dimension $\Delta_{+}$ is dual
to a charged scalar $\phi$ in the holographic bulk. The effective
action can be written as a path integral of $\phi$ in the bulk manifold
$\mathcal{M}$
\begin{equation}
\exp\left\{ i\Gamma_{0}\left[J_{b}\right]\right\} =\int\mathcal{D}\phi\exp\left\{ iS_{\mathrm{bulk}}\left[\phi\right]+iS_{\mathrm{ct}}\left[\phi;\epsilon\right]+iS_{\mathrm{sc}}\left[\phi,J_{b};\epsilon\right]\right\} ,\label{eq:HolographicEffectiveAction_Undeformed}
\end{equation}
where we will denote the radial coordinate by $z$ and $z=\epsilon$
is the location of the boundary. Here we only write down the bulk
scalar part explicitly. Notice that the way we write down the above
equation means that we treat the dynamics of $\phi$ in the bulk as
a full quantum field theory in curved spacetime. Of course the bulk
dynamics involves other fields, particularly the bulk metric $g_{MN}$
and a Maxwell gauge field $A_{M}$ under which $\phi$ is charged.
For brevity we will not write down their actions and path integrals
explicitly because they do not participate in what will be discussed
in the rest of this paper. The actions appearing in the above bulk
path integral have the following form
\begin{align}
S_{\mathrm{bulk}}\left[\phi\right] & =-\frac{1}{2\kappa_{\phi}^{2}}\int_{\mathcal{M}}d^{d+1}x\sqrt{-g}\left\{ g^{MN}\left(D_{M}\phi\right)^{\dagger}\left(D_{N}\phi\right)+m_{\phi}^{2}\phi^{\dagger}\phi\right\} ,\label{eq:BulkAction_Scalar}\\
S_{\mathrm{ct}}\left[\phi;\epsilon\right] & =-\frac{\Delta_{\mathrm{ct}}}{2\kappa_{\phi}^{2}R}\int_{\partial\mathcal{M}}d^{d}x\sqrt{-\gamma}\phi^{\dagger}\phi,\label{eq:BoundaryAction_CounterTerm_Undeformed}\\
S_{\mathrm{sc}}\left[\phi,J_{b};\epsilon\right] & =-\frac{1}{2\kappa_{\phi}^{2}R}\int_{\partial\mathcal{M}}d^{d}x\sqrt{-\gamma}\left\{ \epsilon^{\Delta_{-}}\Delta_{\mathrm{sc}}\left(J_{b}\phi^{\dagger}+J_{b}^{\dagger}\phi\right)-\frac{\epsilon^{2\Delta_{-}}\Delta_{\mathrm{sc}}^{2}}{\Delta_{-}-\Delta_{\mathrm{ct}}}J_{b}^{\dagger}J_{b}\right\} ,\label{eq:BoundaryAction_Source_Undeformed}
\end{align}
where $R$ is the AdS radius, $D_{M}=\nabla_{M}-iq_{\phi}A_{M}$ is
the gauge covariant derivative, $\nabla_{M}$ is the general relativistic
covariant derivative in curved spacetime, $\partial\mathcal{M}$ is
the boundary of $\mathcal{M}$ and $\gamma$ is the determinant of
the induced metric at the boundary $\partial\mathcal{M}$. $\Delta_{\mathrm{ct}}$
and $\Delta_{\mathrm{sc}}$ are dimensionless constants.%
\footnote{The subscripts ``ct'' and ``sc'' in $\Delta_{\mathrm{ct}}$ and
$\Delta_{\mathrm{sc}}$ stand for ``counter term'' and ``source''.%
} Here we will not consider self-interactions of $\phi$. For later
convenience, we define $\nu$ as 
\begin{equation}
\nu\equiv\sqrt{\left(\frac{d}{2}\right)^{2}+m_{\phi}^{2}R^{2}},\qquad\Delta_{\pm}\equiv\frac{d}{2}\pm\nu.
\end{equation}
What we really want to emphasize in this note is the term given in
(\ref{eq:BoundaryAction_Source_Undeformed}), the boundary action
$S_{\mathrm{sc}}$ which is both linear and quadratic in the source
$J_{b}$. It is this term that does the magic of the second Hubbard-Stratonovich
transformation that we advertised earlier. The first term in (\ref{eq:BoundaryAction_Source_Undeformed})
which is linear in $J_{b}$ was used in \cite{Vecchi:2010dd}.
Under the variational principle, this term yields the standard boundary
condition which equates the non-normalizable mode of the bulk solution
of $\phi$ to the source $J_{b}$, but it does not yield a finite
result for the effective action $\Gamma_{0}\left[J_{b}\right]$. To
cure the latter problem, we add the second term quadratic in $J_{b}$
in (\ref{eq:BoundaryAction_Source_Undeformed}). 

Now we can easily integrate out $\upsilon$ rigorously in (\ref{eq:GeneratingFunctional_HSTransformed}).
To do so, plug (\ref{eq:HolographicEffectiveAction_Undeformed}) into
(\ref{eq:GeneratingFunctional_HSTransformed}). This will shift $J_{b}$
in (\ref{eq:BoundaryAction_Source_Undeformed}) to $J_{b}+\upsilon$.
Now $\upsilon$ appears only linearly and quadratically in either
the boundary action $S_{\mathrm{sc}}\left[J_{b}+\upsilon\right]$
or $\lambda^{-1}\upsilon^{\dagger}\upsilon$ term in (\ref{eq:GeneratingFunctional_HSTransformed}),
and can be integrated out exactly. We end up with
\begin{equation}
\exp\left\{ i\Gamma_{\lambda}\left[J_{b}\right]\right\} =\int\mathcal{D}\phi\exp\left\{ iS_{\mathrm{bulk}}\left[\phi\right]+iS_{\mathrm{ct}}^{\lambda}\left[\phi;\epsilon\right]+iS_{\mathrm{sc}}^{\lambda}\left[\phi,J_{b};\epsilon\right]\right\} ,\label{eq:HolographicEffectiveAction_Deformed}
\end{equation}
where $S_{\mathrm{bulk}}$ remains the same as in (\ref{eq:BulkAction_Scalar})
and the boundary terms now read
\begin{align}
S_{\mathrm{ct}}^{\lambda}\left[\phi;\epsilon\right] & =-\frac{\Delta_{\mathrm{ct}}\left(\lambda,\epsilon\right)}{2\kappa_{\phi}^{2}R}\int_{\partial\mathcal{M}}d^{d}x\sqrt{-\gamma}\phi^{\dagger}\phi,\label{eq:BoundaryAction_CounterTerm_Deformed}\\
S_{\mathrm{sc}}^{\lambda}\left[\phi,J_{b};\epsilon\right] & =-\frac{1}{2\kappa_{\phi}^{2}R}\int_{\partial\mathcal{M}}d^{d}x\sqrt{-\gamma}\left\{ \epsilon^{\Delta_{-}}\Delta_{\mathrm{sc}}(\lambda,\epsilon)\left(J_{b}\phi^{\dagger}+J_{b}^{\dagger}\phi\right)-\frac{\epsilon^{2\Delta_{-}}\Delta_{\mathrm{sc}}^{2}(\lambda,\epsilon)}{\Delta_{-}-\Delta_{\mathrm{ct}}(\lambda,\epsilon)}J_{b}^{\dagger}J_{b}\right\} .\label{eq:BoundaryAction_Source_Deformed}
\end{align}
Here the new coefficients $\Delta_{\mathrm{ct}}\left(\lambda,\epsilon\right)$
and $\Delta_{\mathrm{sc}}\left(\lambda,\epsilon\right)$ are 
\begin{equation}
\Delta_{\mathrm{ct}}(\lambda,\epsilon)=\frac{\Delta_{\mathrm{ct}}-\Delta_{-}\hat{\lambda}}{1-\hat{\lambda}},\qquad\Delta_{\mathrm{sc}}(\lambda,\epsilon)=\frac{\Delta_{\mathrm{sc}}}{1-\hat{\lambda}},
\end{equation}
where
\begin{equation}
\hat{\lambda}\equiv\frac{\Delta_{\mathrm{sc}}^{2}}{\Delta_{\mathrm{ct}}-\Delta_{-}}\frac{R^{d-1}}{2\kappa_{\phi}^{2}}\frac{\lambda}{\epsilon^{2\nu}}.
\end{equation}
Recall that earlier we have said that $\epsilon^{-1}$ is related
to the UV momentum cutoff of the interaction potential $V$ in the
field theory. We see introducing the double-trace deformation only
changes the coefficients of the boundary terms from $\Delta_{\mathrm{ct}}$
and $\Delta_{\mathrm{sc}}$ to $\Delta_{\mathrm{ct}}(\lambda,\epsilon)$
and $\Delta_{\mathrm{sc}}(\lambda,\epsilon)$, while no other holographic
structure is changed. The above expressions are the starting point
of the holographic construction for pseudogap phase. They not only
let us recover some well known results at classical level such as
the mixed boundary condition first introduced in \cite{Witten:2001ua},
but also allow us to derive new results such as the boundary condition
and bulk dynamics beyond saddle point in a systematic manner.

\subsection{Bulk Dynamics at the Saddle Points}

Varying the bulk action (\ref{eq:BulkAction_Scalar}) yields the bulk
equation of motion (EOM) for $\phi$, the Klein-Gordon equation, together
with a boundary term. Combining this boundary term with the variations
of (\ref{eq:BoundaryAction_CounterTerm_Undeformed}) and (\ref{eq:BoundaryAction_Source_Undeformed}),
and setting the coefficient of the variation at the boundary to vanish,
we obtain a general expression for the boundary condition
\begin{equation}
\left[-z\frac{\partial}{\partial z}+\Delta_{\mathrm{ct}}(\lambda,\epsilon)\right]\bar{\phi}\Big|_{z=\epsilon}+\Delta_{\mathrm{sc}}(\lambda,\epsilon)\epsilon^{\Delta_{-}}J_{b}=0.\label{eq:BoundaryCondition_Deformed_general}
\end{equation}
Here we use the notation $\bar{\phi}$ to denote the bulk solution
of $\phi$ that satisfies its classical EOM, i.e. the saddle point
value of $\phi$, in the same sense of how we use $\bar{\upsilon}$
to denote saddle point value for $\upsilon$ in the previous section.
From now on, we will mostly work in momentum space where $k^{\mu}$
denote the momentum in the time and transverse spatial directions
and $k^{2}=k^{\mu}k_{\mu}$. The two independent solutions of $\phi$
near the asymptotic AdS boundary are
\begin{align}
 & \bar{\phi}(z,k)\Big|_{z\rightarrow\epsilon}\nonumber \\
= & \phi_{-}(k,\epsilon)\Gamma(1-\nu)\left(\frac{\sqrt{k^{2}}}{2}\right)^{\nu}z^{\frac{d}{2}}I_{-\nu}\left(\sqrt{k^{2}}z\right)+\phi_{+}(k)\Gamma(1+\nu)\left(\frac{\sqrt{k^{2}}}{2}\right)^{-\nu}z^{\frac{d}{2}}I_{\nu}\left(\sqrt{k^{2}}z\right)\label{eq:AsymptoticSolution_NonInteger}\\
\simeq & \phi_{-}(k,\epsilon)z^{\Delta_{-}}\left[1+O(\sqrt{k^{2}}z)\right]+\phi_{+}(k)z^{\Delta_{+}}\left[1+O(\sqrt{k^{2}}z)\right],\nonumber 
\end{align}
where $I_{\pm\nu}$ are the modified Bessel functions. Plugging this equation
into (\ref{eq:BoundaryCondition_Deformed_general}), using $z\partial_{z}I_{\nu}(z)=\nu I_{\nu}(z)+zI_{\nu+1}(z)$
where the second term $zI_{\nu+1}(z)$ can be ignored for small $z$,
and 
\[
I_{\nu}(\xi)=\frac{1}{\Gamma(1+\nu)}\left(\frac{\xi}{2}\right)^{\nu}\qquad\left(\big|\xi\big|\ll1\right),
\]
we have 
\begin{equation}
\phi_{-}(k,\epsilon)=-\frac{\Delta_{\mathrm{sc}}(\lambda,\epsilon)}{\Delta_{\mathrm{ct}}(\lambda,\epsilon)-\Delta_{-}}J_{b}(k)-\frac{\Delta_{\mathrm{ct}}(\lambda,\epsilon)-\Delta_{+}}{\Delta_{\mathrm{ct}}(\lambda,\epsilon)-\Delta_{-}}\epsilon^{2\nu}\phi_{+}(k).\label{eq:BoundaryCondition_Deformed_NonInteger}
\end{equation}
Notice that to arrive at the above relation, we have only assumed
$\sqrt{k^{2}}\epsilon\ll1$, but not any relation between $\lambda$
and $\epsilon$. Using the boundary condition (\ref{eq:BoundaryCondition_Deformed_general}),
the on-shell action is 
\[
\bar{\Gamma}_{\lambda}\left[J_{b}\right]=-\int_{\partial\mathcal{M}}d^{d}x\left\{ \frac{\Delta_{\mathrm{sc}}(\lambda,\epsilon)R^{d-1}}{4\kappa_{\phi}^{2}\epsilon^{\Delta_{+}}}\left(J_{b}\bar{\phi}^{\dagger}+J_{b}^{\dagger}\bar{\phi}\right)+\frac{\alpha\left(\epsilon\right)\sqrt{-\gamma}}{1-\lambda\alpha\left(\epsilon\right)\sqrt{-\gamma}}J_{b}^{\dagger}J_{b}\right\} .
\]
Plug in (\ref{eq:AsymptoticSolution_NonInteger}) and (\ref{eq:BoundaryCondition_Deformed_NonInteger}),
the on-shell effective action for the deformed theory in position
space is 
\begin{equation}
\bar{\Gamma}_{\lambda}\left[J_{b}\right]=\frac{\nu\Delta_{\mathrm{sc}}(\lambda,\epsilon)}{\Delta_{-}-\Delta_{\mathrm{ct}}(\lambda,\epsilon)}\cdot\frac{R^{d-1}}{2\kappa_{\phi}^{2}}\int d^{d}x\left[J_{b}^{\dagger}(x)\phi_{+}(x)+J_{b}(x)\phi_{+}^{\dagger}(x)\right].\label{eq:EffectiveAction_Deformed_OnShell}
\end{equation}
By taking functional derivative with respect to $J_{b}$, the expectation
value of the scalar operator $b$ is 
\begin{equation}
\langle b\rangle_{\lambda,J_{b}}^{\mathrm{sadd}}=\frac{\nu\Delta_{\mathrm{sc}}(\lambda,\epsilon)}{\Delta_{-}-\Delta_{\mathrm{ct}}(\lambda,\epsilon)}\cdot\frac{R^{d-1}}{2\kappa_{\phi}^{2}}\phi_{+}.\label{eq:OnePointFunction_Deformed_OnShell}
\end{equation}
At this moment we want to pause to make some comments on subtleties
hidden in the above calculation. For the expression of $\bar{\Gamma}_{0}\left[J_{b}\right]$
given in (\ref{eq:EffectiveAction_Deformed_OnShell}), if $\nu<1$,
the term written explicitly there is the only non-vanishing term in
the limit $\epsilon\rightarrow0$. However, if $\nu\geqslant1$, there
will be additional finite or divergent terms coming from $O\left(k^{2}\epsilon^{2-2\nu}\right)$.
Such terms are actually quadratic in $J_{b}$ and come in positive
integer powers of $k^{2}$. Thus they contribute some additional terms
to 2-point functions which are analytic in $k^{2}$, i.e. contact
terms. Contact terms in momentum space, whether finite or divergent
in $\epsilon$, arise naturally from the Fourier transform of position
space correlation functions which involve negative powers of distance.
They usually do not contain any interesting physical information,
thus we can simply ignore them. They can be removed by adding additional
boundary terms to (\ref{eq:BoundaryAction_Source_Undeformed}). For
example, to cancel the $O\left(k^{2}\epsilon^{2-2\nu}\right)$ term,
we can add a term like $|\partial J_{b}|^{2}$ term to (\ref{eq:BoundaryAction_Source_Undeformed})
with an appropriate power of $\epsilon$. Equivalently, we can choose
to extend the coefficient of the $J_{b}^{\dagger}J_{b}$ term in (\ref{eq:BoundaryAction_Source_Undeformed})
from constant to a function of $k^{2}$ in momentum space. However,
in the following we will choose not to remove the contact terms and
will keep the form of (\ref{eq:BoundaryAction_Source_Undeformed})
as it is. A second subtlety is that when $\nu$ is an integer, the
modified Bessel function in one of the two independent solutions in
(\ref{eq:AsymptoticSolution_NonInteger}) will be replaced by $K_{\nu}$,
and now we will have $\log\left(k^{2}\epsilon^{2}\right)$ terms appear
in the calculation. Although this make the intermediate steps more
complicated, after careful treatment, we find the final results in
the limit $\epsilon\rightarrow0$ are unchanged. In any case, what
is never changed is the general structure of the boundary action (\ref{eq:BoundaryAction_Source_Undeformed})
that it depends only linearly and quadratically in $J_{b}$. It is
this general feature that allows us to rigorously carry out the path
integral over $\upsilon$ in (\ref{eq:GeneratingFunctional_HSTransformed}). 

(\ref{eq:BoundaryCondition_Deformed_NonInteger}) is the double-trace
deformed boundary condition. Together with (\ref{eq:OnePointFunction_Deformed_OnShell}),
they can be written as  
\begin{align*}
\phi_{-} & =\frac{\Delta_{\mathrm{sc}}(\lambda,\epsilon)}{\Delta_{-}-\Delta_{\mathrm{ct}}(\lambda,\epsilon)}J_{b}+\epsilon^{2\nu}\frac{\Delta_{\mathrm{ct}}(\lambda,\epsilon)-\Delta_{+}}{\nu\Delta_{\mathrm{sc}}(\lambda,\epsilon)}\cdot\frac{2\kappa_{\phi}^{2}}{R^{d-1}}\langle b\rangle_{\lambda,J_{b}}^{\mathrm{sadd}},\\
\phi_{+} & =\frac{\Delta_{-}-\Delta_{\mathrm{ct}}(\lambda,\epsilon)}{\nu\Delta_{\mathrm{sc}}(\lambda,\epsilon)}\cdot\frac{2\kappa_{\phi}^{2}}{R^{d-1}}\langle b\rangle_{\lambda,J_{b}}^{\mathrm{sadd}}.
\end{align*}
It is conventional to set $\Delta_{\mathrm{ct}}=\Delta_{+}$ \cite{Vecchi:2010dd},
then the above equations become simply
\begin{align}
\phi_{-} & =-\frac{\Delta_{\mathrm{sc}}}{2\nu}J_{b}+\frac{\Delta_{\mathrm{sc}}}{\nu}\lambda\langle b\rangle_{\lambda,J_{b}}^{\mathrm{sadd}},\\
\phi_{+} & =-\frac{2}{\Delta_{\mathrm{sc}}}\cdot\frac{2\kappa_{\phi}^{2}}{R^{d-1}}\langle b\rangle_{\lambda,J_{b}}^{\mathrm{sadd}}.
\end{align}
These reproduce the familiar mixed boundary conditions first presented
in \cite{Witten:2001ua}. Through our derivation above using the variational
principle, it is very clear that these are only the saddle point results.
It will not hold beyond the saddle points. In this sense, these are
exactly the analog of the saddle point result (\ref{eq:SaddlePoint_1PointFunction})
that we derived earlier in the $\upsilon$-field representation of
the effective action $\Gamma_{\lambda}\left[J_{b}\right]$. In fact,
using (\ref{eq:SaddlePoint_1PointFunction}) we can identify
\begin{equation}
\phi_{-}=-\frac{\Delta_{\mathrm{sc}}}{\nu}\bar{\upsilon}\qquad\left(J_{b}=0\right),
\end{equation}
which relates our bulk field $\phi$ viewed as a second Hubbard-Stratonovich
field to the first Hubbard-Stratonovich field $\upsilon$. For 
hunting for the pseudogap phase, the current holographic result suffers
the same problem as we discussed below (\ref{eq:SaddlePoint_1PointFunction}):
it produces only two distinct phases: the gapless normal phase with
unbroken $U(1)$ symmetry and the superconducting phase with a broken
$U(1)$ symmetry. After setting $J_{b}=0$, both $\phi_{-}$ and $\phi_{+}$,
and thus the classical solution of the bulk field $\bar{\phi}$, are
proportional to $\langle b\rangle$. When $\langle b\rangle\neq0$,
the $U(1)$ symmetry is broken and we must have a non-trivial $\bar{\phi}$
in the bulk: this is the superconducting phase studied in \cite{Faulkner:2010gj}.
If we do not want to break the symmetry, we have $\langle b\rangle=0$,
which means $\bar{\phi}=0$ in the bulk: this is the gapless strongly
interacting (non-)Fermi liquid phase dual to the AdS-Reissner-Nordström
background \cite{Lee:2008xf,Liu:2009dm,Cubrovic:2009ye,Faulkner:2009wj,Zaanen:2015oix}.%
\footnote{For a more thoughtful discussion on holographic realizations of normal
Fermi-liquid type phases, see the Introduction section of \cite{Sachdev:2011ze}.
For a more comprehensive review, see \cite{Zaanen:2015oix}.%
} We are now in the same dilemma as that expressed below (\ref{eq:SaddlePoint_1PointFunction}):
holography at the bulk saddle points does not capture the physics
of pairing fluctuation pseudogap either. The solution to this problem
is similar: we have to include the effect of fluctuations for the
bulk scalar to achieve the pseudogap phase.

\subsection{Comments on Treatments beyond the Saddle Points}

For the pseudogap phase to be realized in holography, what we expect
is that the bulk scalar $\phi$ shall behave non-trivially in the
bulk, similarly to how it behaves as charged hair in the classical holographic
superconductor models. This will allow $\phi$ to carry a finite amount
of charges and energy-stress outside the black hole horizon. This
charged matter of $\phi$ will leave its imprint as a pseudogap in
the field theory correlation functions of stress tensor and charge
current, since these correlators are calculated from the perturbations
of the metric and gauge field in the bulk. This is similar to the
story that has been well studied in holographic superconductor
models. Meanwhile, we do not want this non-trivial profile of $\phi$
to contribute to $\langle b\rangle$, but this is forbidden at the
saddle point level by the mixed boundary condition we have just derived
because of the coherence of the classical dynamics. The tie between
the non-trivial $\phi$ and non-vanishing $\langle b\rangle$ can
only be broken and washed out by incoherent quantum fluctuations.
Thus $\phi$ shall be in a superposition of incoherent states, as
opposed to a coherent state of condensate. Macroscopically, it behaves
like a normal fluid, as summarized in Table \ref{Tab:BCSBEC_ThreePhases}.
In the following we will show how this happens via phase decoherence
effect. But before doing so, we want to pause for a moment to make
some comments on the consistency and legitimacy of our treatments
beyond the saddle point level in the bulk. 

Strictly speaking, when we are considering effects due to bulk fluctuations,
we are going away from the classical level into the quantum regime
in the bulk. For top-down AdS/CFT, we are moving away from $N=\infty$
limit \cite{Aharony:1999ti}. Treating the bulk dynamics as a full
quantum field theory (including the gravity!) is far beyond the scope of this paper. More importantly, we do not
believe much of this full quantum treatment is crucial for capturing
the essence of the physics of the pairing fluctuation pseudogap in
the BCS-BEC crossover scenario. In the common field theoretical treatments
of this subject such as those reviewed in \cite{Chen:2000thesis,Levin:2010review},
only the fluctuations in the channel of the Cooper pair operator $b$
are considered. Fluctuations can take place in many other channels
as well, for example, via the operators of stress tensor and charge
current, but none of them is considered in the field theory because
their contributions are negligible. We do not rule out the possibility
that for some other phenomena they may contribute significantly, but
the existing studies show that they do not matter much for BCS-BEC
crossover. We will inherit this fact in our holographic model building.
Each channel of fluctuations of a certain operator in the field theory
is dual to the excitations of the corresponding bulk field. As only
the fluctuations of Cooper pair operator $b$ is important for BCS-BEC
crossover, in the holographic model, only the fluctuations of the
bulk field $\phi$ need to be taken into account. Thus we will treat
all bulk fields other than $\phi$ always as classical fields that
satisfy their classical EOMs in the bulk, and their fluctuations will
be neglected throughout. Only the dynamics of $\phi$ goes beyond
saddle points.

Our strategy of singling out $\phi$'s fluctuation is only justified
\emph{a posteriori}. The logic is completely bottom-up and may only
work for the phenomenon of pairing fluctuation pseudogap that we want
to study. In the standard top-down narratives of AdS/CFT correspondence,
such as the duality between $\mathcal{N}=4$ SYM theory with gauge
group $SU(N)$ and type IIB superstring theory in $AdS_{5}\times S^{5}$
background \cite{D'Hoker:2002aw}, our strategy is clearly not a consistent
treatment of the fluctuations. The highly symmetric structures of
the $\mathcal{N}=4$ SYM conspire that in its holographic dual, all
the bulk fields have the same coupling constant:
\[
\mathcal{L}_{5\mathrm{D}}=\frac{N^{2}}{8\pi^{2}R^{3}}\left[\mathcal{R}-2\Lambda-\frac{1}{2}\left(\partial\phi\right)^{2}-\frac{R^{2}}{8}\mathrm{tr}F^{2}+\ldots\right].
\]
If the $N=\infty$ limit is uplifted, all the bulk fields will enter
the quantum regime simultaneously, and it is not consistent to include
only some of them in a calculation while to ignore the others. There
are special cases where all quantum fluctuations can be calculated
(for example in \cite{Alho:2015zua}), but such cases are seldom relevant
to us. In general, only when there is a hierarchy of bulk coupling
constants can one treat some of the fields as quantum while others
still fully classical. Actually the fact that all bulk couplings are
equal in the above example is really an artifact due to the high symmetries
of this particular field theory. Although this may be common in other
known top-down duality, this can hardly be a generic case. For holographic
duals of more realistic field theories (if they exist and can be derived),
it is very possible that the bulk couplings will have a hierarchy
that allows a consistent quantum treatment for only part of the bulk
fields in certain regimes of interest.

Now imagine if we can schematically integrate out the other bulk fields
before discussing the dynamics of $\phi$, what will the resulting
effective action for $\phi$ look like? Recall that we start with
a local quadratic action for $\phi$ in (\ref{eq:BulkAction_Scalar}).
In the absence of the hierarchy, as in the $\mathcal{N}=4$ SYM case,
we will end up with a highly non-local effective action for $\phi$.
This is the case that we assume will not happen to us in the bottom-up
model. What is likely to happen is that there is a hierarchy of couplings,
which results in the fact that the effective action for $\phi$ is
gapped and can be expanded as a Taylor series. At the quadratic order,
this only causes a renormalization of the kinetic and mass couplings.
At higher order, it induces effective self-interactions for $\phi$
(such as a $|\phi|^{4}$ term) via loop effects. At the phenomenological
level, we can reproduce this effect simply by adding non-linear interactions
to (\ref{eq:BulkAction_Scalar}) while still keeping other fields
classical, and treating the existing parameters as the renormalized
ones. Of course now the coefficient of the $|\phi|^{4}$ term will
have to be set by hand rather than computed from first principle.
A non-vanishing $|\phi|^{4}$ vertex in the bulk will generate non-vanishing
4-point functions for the Cooper pair operators $b$ and $b^{\dagger}$
in the field theory. In BCS-BEC crossover scenario, this represents
a non-vanishing residual interaction between Cooper pairs. Similarly,
higher powers of $|\phi|^{2}$ induce higher $n$-point functions
of $b$ and $b^{\dagger}$. If the former case of a non-local action
of $\phi$ takes place, it implies in the field theory, all higher
$n$-point functions of the Cooper pairs are not negligible and they
add up non-perturbatively in the effective action. Physically, this
means the residual interaction, and the residue of the residual interaction
etc, of the Cooper pairs are all strong, which will trigger a chain
reaction of dimerizations of Cooper pairs. This is an instability
of the system and implies the effective degrees of freedom is no longer
the Cooper pairs, but some other operators with large charges and
high dimensions. This case goes beyond the BCS-BEC crossover scenario,
thus will not be considered here. This is an argument we provide from
the phenomenological perspective for only treating $\phi$ as a quantum
field.

A second fact which helps to single out the quantum fluctuations of
$\phi$ is that we identify the external knob for the crossover with
the double-trace deformation of the scalar operator $b$. Turning
on this double-trace deformation puts the field $\phi$ in a unique
position compared to other bulk fields. Upon correctly normalizing
the double-trace coupling $\lambda$, the quantum fluctuation of $\phi$
can be enhanced and elevated out from all other quantum fluctuations.
This agrees with the BCS-BEC crossover scenario. In the BCS limit
which corresponds to turning off the double-trace deformation, we
have a perfect mean field theory with highly suppressed fluctuations.
This implies in the holographic dual, we should expect that $\phi$
will sit back at its saddle points when the double-trace deformation
is off. Only a non-vanishing double-trace deformation will kick $\phi$
out of its saddle points. What is really important here is the difference
between the presence and absence of the double-trace deformation.
This is similar to the logic of \cite{Gubser:2002zh}, where only
the scalar one-loop correction is computed and that yields only the
difference of the free energy between $\lambda\neq0$ and $\lambda=0$.

\subsection{Phase Decoherence of Bulk Scalar Fluctuations}

Now even for the field $\phi$ away from its saddle points, we will
not treat it fully quantum mechanically. For example, integrating
out $\phi$ will also induce bulk vertices for other fields, or make
their action non-local. We will not be interested in describing such
effects. The single quantum effect of $\phi$ most relevant to us
is its phase decoherence. At scales that are macroscopically small
but microscopically large, the phase of the quantum states of $\phi$
are random. This can be viewed as a depletion of the coherent condensate
by exciting Goldstone bosons. The excitations of Goldstone bosons
wash out the phase coherence partially or fully at distances much
longer than the typical wavelengths of the excitations, resulting
in a reduction of coherent length. This is dual to the incoherent
Cooper pairing in the field theory. The effect of phase decoherence due to thermal fluctuations has been demonstrated in \cite{Anninos:2010sq} for the Abelian Higgs holographic superconductor model in 3+1 dimensional bulk in the probe limit by a direct calculation at the microscopic quantum field theory level. Our approach to it will be more phenomenological at the low energy effective field theory level.

To be more specific, let us write down the mode expansion of $\phi$
in the bulk and second quantize it
\begin{equation}
\hat{\phi}(x)=\hat{\phi}_{0}(x)+\sum_{j>0}\hat{\phi}_{j}(x).
\end{equation}
Here we add ``$\hat{\phantom{s}}$'' to all second quantized operators.
``$0$'' labels the ground state and positive ``$j$'' labels
excited states. We do not need to know the specific form of each mode
for our purpose. Recall that $\phi$ is a bosonic field. A bosonic
field can condense on its ground state. Once this happens, there will
be a macroscopic population in the ground state and it forms a coherent
many-body wave function. Typically in the study of BEC, we can macroscopically
treat the ground state wavefunction as a classical field satisfying
certain classical EOM, such as the Gross-Pitaevskii equation \cite{Pitaevskii:Book}.
Thus we can replace $\langle\hat{\phi}_{0}(x)\rangle$ by a classical
field $\bar{\phi}(x)$. This is exactly what we have done at the saddle
point. From the quantum point of view, the bulk Klein-Gordon equation
for $\phi(x)$ derived from the saddle point of (\ref{eq:HolographicEffectiveAction_Deformed})
is the Gross-Pitaevskii equation for the Bose-Einstein condensate
of the quantum field $\hat{\phi}_{0}(x)$. Now let us define the excitation
part as $\hat{\phi}_{\mathrm{e}}(x)\equiv\sum_{j>0}\hat{\phi}_{j}(x)$.
$\hat{\phi}_{\mathrm{e}}(x)$ is a superposition of all excited modes.
In the AdS black hole background, the ground state is only inhomogeneous
along the radial direction. On the contrary, at every position along
the radial direction, the excited modes can be viewed as a collections
of Fourier modes of all frequencies and wave-vectors in transverse
directions. We now define the notation $\langle\ldots\rangle$ as
an ensemble average over large enough spatial volume in the transverse
directions or over long enough time, compared to any possible local
correlation length scales. Since we are considering systems in an
infinitely large volume for the field theory, this average can always
be done. A key idea is that the phases for different $j$ modes have
no correlations, i.e. the fluctuations are \emph{incoherent}. This
means, locally, at a specific spacetime location $x$, $\hat{\phi}_{\mathrm{e}}(x)$
may have a well defined amplitude and phase, which we can denote as
\begin{equation}
\hat{\phi}_{\mathrm{e}}(x)=\hat{\psi}(x)e^{i\hat{\theta}(x)},
\end{equation}
but once the large ensemble average is done, the uncorrelated random
phases will add up to zero
\begin{equation}
\big\langle e^{i\hat{\theta}(x)}\big\rangle=0.
\end{equation}
This is phase decoherence. Since the density of fluctuations are always
non-negative, we have 
\[
\big\langle\hat{\psi}(x)\big\rangle\geqslant0.
\]
It is reasonable to treat $\langle\hat{\psi}(x)\rangle$ as homogeneous
in time and transverse spatial directions and only varying along the
radial direction, i.e. $\langle\hat{\psi}(x)\rangle=\langle\hat{\psi}(z)\rangle$.
Now we can write
\begin{equation}
\big\langle\hat{\phi}_{\mathrm{e}}(x)\big\rangle=0,\qquad\big\langle\hat{\phi}_{\mathrm{e}}^{\dagger}(x)\hat{\phi}_{\mathrm{e}}(x)\big\rangle\geqslant0.\label{eq:PhaseDecoherence}
\end{equation}
For a fluid, the gradient of phase is related to its velocity, we
thus have 
\[
\big\langle\nabla_{M}\hat{\phi}_{\mathrm{e}}^{\dagger}(x)\nabla^{M}\hat{\phi}_{\mathrm{e}}(x)\big\rangle=\big\langle\nabla_{M}\hat{\theta}(x)\nabla^{M}\hat{\theta}(x)\big\rangle\neq0.
\]
Whether $\langle\nabla_{M}\hat{\theta}(x)\rangle$ is vanishing or
not depends on the macroscopic motion of the fluid, but $\langle\nabla_{M}\hat{\phi}_{\mathrm{e}}(x)\rangle\simeq\nabla_{M}\langle\hat{\phi}_{\mathrm{e}}(x)\rangle=0$
because the phase factor averages to zero.

Although the phase decoherence cannot be used to evaluate the path
integral (\ref{eq:HolographicEffectiveAction_Deformed}), it can determine
which terms in (\ref{eq:HolographicEffectiveAction_Deformed})
will vanish after performing the path integral and thus helps
us to simplify and decouple the effective action into two separate
parts. To see this, let us write $\phi=\bar{\phi}+\phi_{\mathrm{e}}$
in (\ref{eq:HolographicEffectiveAction_Deformed}) and change the
path integral variable from $\phi$ to $\phi_{\mathrm{e}}$. $\bar{\phi}$
is just the condensate field. $\phi_{\mathrm{e}}$ is the new functional
variable in the path integral corresponding to the second quantized
operator $\hat{\phi}_{\mathrm{e}}$ discussed above, and it eventually
becomes $\big\langle\hat{\phi}_{\mathrm{e}}(x)\big\rangle$ after
the path integral is done, thus we can think $\phi_{\mathrm{e}}$
in the path integral integrand satisfies the relation (\ref{eq:PhaseDecoherence})
as well. Now let us see how the actions in (\ref{eq:HolographicEffectiveAction_Deformed})
split under phase decoherence. For both the bulk action (\ref{eq:BulkAction_Scalar})
and boundary action (\ref{eq:BoundaryAction_CounterTerm_Deformed}),
every term is quadratic in $\phi^{\dagger}\phi$. Under (\ref{eq:PhaseDecoherence}),
they all split into sums of $\bar{\phi}^{\dagger}\bar{\phi}$ and
$\phi_{\mathrm{e}}^{\dagger}\phi_{\mathrm{e}}$ terms, because the
cross terms like $\bar{\phi}^{\dagger}\phi_{\mathrm{e}}$ vanish by
(\ref{eq:PhaseDecoherence}). Thus both actions split into sum of
two copies of themselves, one with $\phi$ replaced by $\bar{\phi}$
and the other by $\phi_{\mathrm{e}}$. The most interesting fact is
what happens to the source term (\ref{eq:BoundaryAction_Source_Deformed}):
its first term is linear in $\phi$, thus according to (\ref{eq:PhaseDecoherence}),
the $\phi_{\mathrm{e}}$ part will be washed out and only $\bar{\phi}$
term survives after phase decoherence! Thus the source term (\ref{eq:BoundaryAction_Source_Deformed})
does not split, but just turns $\phi$ in it into $\bar{\phi}$. Now
collecting all $\bar{\phi}$ terms in (\ref{eq:HolographicEffectiveAction_Deformed}),
including the $J_{b}^{\dagger}J_{b}$ term in (\ref{eq:BoundaryAction_Source_Deformed}),
they reproduce precisely the saddle point result, i.e. the on-shell
action $\bar{\Gamma}_{\lambda}\left[J_{b}\right]$ given in (\ref{eq:EffectiveAction_Deformed_OnShell}).
The rest are terms that contain only $\phi_{\mathrm{e}}$ but do not
depend on the source $J_{b}$! Now we can write (\ref{eq:HolographicEffectiveAction_Deformed})
as
\begin{align}
\Gamma_{\lambda}\left[J_{b}\right] & =\bar{\Gamma}_{\lambda}\left[J_{b}\right]+\Gamma_{\lambda}^{\mathrm{flct}},\\
\exp\left\{ i\Gamma_{\lambda}^{\mathrm{flct}}\right\}  & =\int\mathcal{D}\phi\exp\left\{ iS_{\mathrm{bulk}}\left[\phi\right]+iS_{\mathrm{ct}}^{\lambda}\left[\phi;\epsilon\right]\right\} ,\label{eq:HolographicEffectiveAction_Decoherence}
\end{align}
where $S_{\mathrm{bulk}}\left[\phi\right]$, $S_{\mathrm{ct}}^{\lambda}\left[\phi;\epsilon\right]$
and $\bar{\Gamma}_{\lambda}\left[J_{b}\right]$ are given by (\ref{eq:BulkAction_Scalar}),
(\ref{eq:BoundaryAction_CounterTerm_Deformed}) and (\ref{eq:EffectiveAction_Deformed_OnShell})
respectively. Notice we should have written $\phi_{\mathrm{e}}$ as
the path integral variable in the second equation, but since it is
a dummy variable, we are free to rename it $\phi$.

The above two equations are the first key result in our holographic
model building. First, the fact that $\Gamma_{\lambda}^{\mathrm{flct}}$
does not depend on the source $J_{b}$ is precisely what we need for
constructing the pseudogap phase: the fluctuations can develop non-trivial
profiles in the bulk to gap the correlation functions, but will not
give a non-vanishing expectation value for $b$ (if the condensate
part is vanishing), because they do not couple to $J_{b}$. The fluctuations
will never break the $U(1)$ symmetry. $\Gamma_{\lambda}^{\mathrm{flct}}$
contributes only to the zero-point function, i.e. the free energy.
In this sense, it is the vacuum polarization, and its value (whether
finite or infinite) will not enter any physical observable defined
via correlation functions; it is just an overall factor that always
get canceled in the calculation of connected correlation functions.
Secondly, there is a match of gap parameters in the duality. In the
superconducting phase, the bulk profile of $\bar{\phi}$ is the holographic
dual of the superconducting gap parameter in the field theory, and
both of them are complex, with amplitude and phase parts. On the contrary,
in the pseudogap phase, the phase of the fluctuations $\phi_{\mathrm{e}}$
is washed out by phase decoherence and does not have a well-defined
value, but its amplitude is still non-vanishing and well defined (this
is what we called $\langle\hat{\psi}(x)\rangle$ earlier). It is this
amplitude that is dual to the pseudogap parameter in the pairing fluctuation
theory of BCS-BEC crossover \cite{Chen:2000thesis}, both of which
are real. Lastly, it shall be noted that although the fluctuations
do not directly couple to the source nor contribute to $\langle b\rangle$,
nor does it directly interact with the condensate $\bar{\phi}$ in
the bulk, it will have effects on two-point functions of $b$ and
other correlation functions though backreactions to the metric and
gauge field.

\subsection{From Incoherent Fluctuations to Quantum Fluid}

The remaining question is how to perform the path integral of $\phi$
in $\Gamma_{\lambda}^{\mathrm{flct}}$ given by (\ref{eq:HolographicEffectiveAction_Decoherence}).
We will answer this question in the rest of this paper. Directly performing
the path integral of $\phi$ using the method of quantum field theory
in curved spacetime \cite{Wald:Book,BirrellDavies:Book} is possible
in a given background with high symmetry, but is still too laborious.
This method can not include the backreactions in a self-consistent
manner either. We want to build a mathematical framework that allows
us to quantitatively study the dynamics of quantum fluctuations in
a relatively economical way, hopefully close to the level of simplicity
of classical holographic superconductors. The strategy is to develop
an effective field theory description for $\Gamma_{\lambda}^{\mathrm{flct}}$
in term of fluid dynamics. We will replace the action on the right
hand side of (\ref{eq:HolographicEffectiveAction_Decoherence}) by
a perfect fluid type action whose field variables are the coarse-grained
thermal fluid variables such as the temperature, the chemical potential
and the velocity, and the saddle point value of this fluid action
yields the value of $\Gamma_{\lambda}^{\mathrm{flct}}$ and renormalized
correlation functions. Because of the randomness of the fluctuations,
this fluid shall be treated as an incoherent normal fluid at finite
temperature with non-vanishing entropy density, as opposed to the coherent
fluid that describes the superfluid condensate (for reviews of the
latter, see for example \cite{Schunck:2003kk,Matos:2014Book,Chavanis:2015zua}),
although part of the mathematical structures appear to be similar.

The idea of using fluid as an effective description for incoherent
fluctuations are not new in the context of holography. This idea has
been applied to study bulk fermions in \cite{deBoer:2009wk,Arsiwalla:2010bt,Hartnoll:2010gu,Hartnoll:2010ik}.
The bulk fluids in these studies are purely \emph{classical} and fermionic,
in the sense that 
\begin{enumerate}
\item The fluid dynamics is of the standard perfect fluid form in curved
spacetime, i.e. the stress tensor takes the form $T_{N}^{M}=\mathrm{diag}\left(\varepsilon,p,\ldots,p\right)$
in the rest frame where $\varepsilon$ and $p$ are the energy density
and pressure;
\item the fluids locally satisfy the standard fermionic equation of state
(EOS) as derived from Fermi-Dirac statistics in \emph{flat} spacetime. 
\end{enumerate}
However, this classical formalism can not be directly applied to our
case by just replacing the fermionic constituents with the corresponding
bosonic ones. The purely classical fluid dynamics must be now upgraded
to include certain quantum effects in curved spacetime to make it
work in the holographic context that we are interested in, for the
following reasons.
\begin{itemize}
\item \emph{Negative mass square}. Our boson fluid is charged. There is
a local chemical potential, typically non-vanishing in the bulk, to
control its charge density. According to Bose statistics, this chemical
potential shall lie between the particle-hole (i.e. anti-particle)
mass gap. In flat spacetime, this mass gap is just the mass of the
charged scalar, whose mass square is always positive. However, in holography,
the scalar mass squared can be slightly negative as long as is above
the Breitenlohner-Freedman (BF) bound \cite{Breitenlohner:1982bm}.
In fact, in holography, we are mostly interested in such negative
mass-squared scalars which can develop a superconducting instability
more easily. Another reason for negative mass-squared is that the dual
operator in the field theories is the Cooper pair operator, which
usually does not have a very high scaling dimension. Clearly, for
the bulk bosonic fluid, this negative mass square can not be treated
as a local particle-hole mass gap literally. The way out is that the
negative mass squared gets renormalized and shifted to a non-negative
value due to the vacuum polarization effect induced by spacetime curvature.
This is the first hint for a quantum fluid.
\item \emph{Boundary condition}. The parameter that controls the strength
of fluctuations in the field theory is the double-trace deformation,
which serves as the external tunable knob in the phase diagram of
BCS-BEC crossover. It enters into the dual holographic dynamics only
through boundary conditions. This is both true at the classical level
and beyond, because the double-trace coupling $\lambda$ only appears
in boundary actions (\ref{eq:BoundaryAction_CounterTerm_Deformed})
and (\ref{eq:BoundaryAction_Source_Deformed}), thus does not directly
modify bulk dynamics. To get physically sensible results, our fluid
must be sensitive to the boundary condition of the scalar field. For
classical fluid dynamics, it is not clear how this can be implemented
in a manifest and unique way. This is the second hint for a quantum
fluid (or at least a modification of classical fluid dynamics).
\item \emph{Anisotropy of stress tensor}. For a classical perfect fluid, the
transverse spatial components and the radial component of the stress
tensor are equal: $T_{i}^{i}=T_{z}^{z}$ (no sum in $i$). This is
built-in in the constituent relation of the stress tensor. On the
contrary, the renormalized stress tensors in curved spacetime such
as a black hole background usually do not have this isotropy. This
can be seen by direct calculations, for example in \cite{Page:1982fm,Howard:1985yg}.
Of course there are non-equilibrium hydrodynamics formalisms such
as that of \cite{Israel:1979wp} and its descendents, but the idea
behind all of these are a perturbative gradient expansion. However,
in our case, neither is the fluid in non-equilibrium states nor is
the anisotropy small enough and suitable for a perturbative expansion.
This non-perturbative anisotropy is of equilibrium by nature and has
a different root in vacuum polarization, which is a purely quantum
effect. This is the third hint for a quantum fluid.
\item \emph{Stability near horizon}. In the absence of a black hole, even
in highly curved spacetime, classical fluid dynamics can still yield
a star-like stable solution. It may have large deviation from the
correct physical solution due to ignorance of quantum effects, but
at least we have a solution. However, things change in the presence
of a black hole. Near the horizon, a classical bosonic fluid can not
enjoy any hydrostatic configuration unless its stress tensor diverges.
This is even true for extremally charge fluids. This is because the
Bose statistics requires the chemical potential of a bosonic fluid
to lie between its particle-hole mass gap, which constrains the electric
force to be always smaller than the gravitational force.%
\footnote{Even for fermionic fluids, if the chemical potential lies in this
range, then hydrostatic configurations can not be achieved outside
a black hole either \cite{Hartnoll:2010ik}. %
} This is a mathematical statement of saying the very intuitive fact
that black holes tend to suck classical thermal bosonic matter in
and nothing without enough angular momentum can avoid this fate. Meanwhile,
at the quantum level, an outgoing flux of Hawking radiation can balance the
ingoing flux which yields a hydrostatic configuration --- the Hartle-Hawking
state \cite{Hartle:1976tp,Israel:1976ur,Jacobson:1994fp}. This is
the kind of configuration we are looking for. This only happens at the
quantum level and is maintained by particle creations in the presence
of black hole. This is the fourth hint for a quantum fluid.
\end{itemize}
The microscopic origin of these quantum effects can be traced to the
``normal ordering'' of the field operators in curved spacetime if
we directly perform the path integral of $\phi$ in (\ref{eq:HolographicEffectiveAction_Deformed}),
as discussed in \cite{Wald:Book,BirrellDavies:Book} and references
therein. In the quantum fluid dynamics that we are going to develop
in the rest of this paper, the first three points above are taken
care of simultaneously by introducing a pair of radial profile functions
whose product is nothing but the renormalized vacuum polarization
$\langle\phi^{\dagger}\phi\rangle$. The dynamics of these radial
profiles and the consequences on the formalism of fluid dynamics are
the focus of this note. The last point above is related to the quantum
corrections to the EOS. It has a different mathematical treatment
in our quantum fluid dynamics which is relatively independent of the
radial profile part, thus we will leave it for a detailed discussion
in the follow-up paper \cite{Wu:2016_Note3}.

Thus we will assume the fluctuations given by (\ref{eq:HolographicEffectiveAction_Deformed})
can be described by a perfect quantum fluid in hydrostatic thermal
equilibrium. To be specific, the thermal aspect of this statement
includes the following key points:
\begin{itemize}
\item Perfect: only the zeroth order terms in hydrodynamic expansion will
be considered. The dynamics has a Lagrangian formalism. There is no
dissipation.
\item Thermal: the fluid has a non-vanishing entropy density, or equivalently
there is some kind of notion of local temperature conjugate to the
local entropy density. The entropy density shall be viewed as a well
defined local observable, but the local temperature we will introduce
is more a mathematical construction rather than a physical observable
that can be read off from a thermometer carried by a certain observer,
except perhaps in the asymptotic region. The presence of entropy and
temperature is a major difference of our incoherent fluid dynamics
from that of superfluid dynamics.
\item Equilibrium: the fluid is locally at equilibrium. No entropy is produced
anywhere, i.e. the divergence of the entropy current vanishes everywhere.
Meanwhile, the black hole at the quantum level can be viewed as a
thermal reservoir at Hawking temperature, and the fluid is in thermal
equilibrium with the black hole.
\item Hydrostatic: this is defined in a static geometry which admits a time-like
Killing vector in the region that we are interested in (i.e. outside
the horizon). The fluid velocity is normalized and parallel to the
future-directed time-like Killing vector everywhere, and all physical
observables are translationally invariant along the Killing time direction.
For AdS-Reissner-Nordström black hole, the Killing time is just the
Poincar?chwarzschild time $t$, and the above statement means all
physical observables are independent of $t$. From the perspective
of quantum field theory in curved spacetime, this is a Hartle-Hawking
type state \cite{Hartle:1976tp,Israel:1976ur,Jacobson:1994fp}.
\end{itemize}
The quantum aspect of the fluid is mainly related to the renormalization
due to vacuum polarization (i.e. Casimir effect) in curved spacetime
\cite{BirrellDavies:Book}. Key points include:
\begin{itemize}
\item The mass square of the scalar field is renormalized and shifted in
curved spacetime from a negative value to a non-negative one, and
it is the latter that appears in the fluid's EOS as a mass gap in
Bose statistics.
\item The stress tensor and charge current are also renormalized in curved
spacetime. The renormalized ones are conserved and regular everywhere
outside and \emph{at} the black hole horizon \cite{BirrellDavies:Book,Wald:Book}.
\item The local EOS acquires a coherent quantum correction part due to vacuum
polarization.
\end{itemize}
In this note, we will develop a Lagrangian formalism that encode all
the above features. Most of these features will be manifest in the
development of our formalism, except for the last two points regarding
the regularity at the horizon and quantum correction to the EOS. We
will discuss these two points in more details in the follow-up paper
\cite{Wu:2016_Note3}.

\bigskip{}


\section{Madelung Transformation}

It is hard to derive fluid dynamics in a mathematically rigorous way
from microscopic theories. However, there are certain procedures that
can guide us and offer enough physical insights along the way. Since
our quantum fluid dynamics is more complicated than standard
perfect fluid dynamics, it is easier and more secure to work at the
action level rather than at the EOM level. Fortunately, for a perfect
fluid without dissipation, an action principle is possible. For us,
there will be two major steps to go from the microscopic theory specified
by (\ref{eq:HolographicEffectiveAction_Decoherence}) to the fluid
dynamics. The first step is a Madelung transformation, which will transform
the microscopic bulk action (\ref{eq:BulkAction_Scalar}) into an
on-shell fluid form. The second step is to rewrite the on-shell fluid
action in an off-shell form using the velocity potential formalism
of fluid dynamics. We will carry out the first step in this section
and the second step in the next section.

\subsection{The Radial Profile and Mass Renormalization}

We introduce the Madelung transformation \cite{Madelung:1926,Madelung:1927}
\begin{equation}
\phi(x)=\psi(x)\tilde{\phi}(x),\qquad\tilde{\phi}(x)=e^{i\vartheta(x)},\label{eq:MadelungTransformation}
\end{equation}
where $\psi(x)$ and $\vartheta(x)$ are both real functions. Now
we have separated the complex scalar field $\phi(x)$ into its amplitude
part $\psi(x)$ and a uni-modular part $\tilde{\phi}(x)$.%
\footnote{Throughout this note, we will add ``$\tilde{\phantom{s}}$'' to
all quantities that are directly related to or derived from the uni-modular
part $\tilde{\phi}(x)$.%
} We require $\psi(x)$ to satisfy the following Klein-Gordon equation
in the bulk:
\begin{equation}
\left(\nabla^{2}-m_{\psi}^{2}\right)\psi=0.\label{eq:BulkEOM_FluidProfile}
\end{equation}
where $\nabla^{2}\equiv g^{MN}\nabla_{M}\nabla_{N}$. Given a radial
profile $\psi(x)$, the above equation can be viewed as a definition
for the effective mass $m_{\psi}(x)$, which is a local function,
not a constant. In fact, in the literature, there is another name
for the mass square $m_{\psi}^{2}$: it is called the quantum potential,
or Bohm potential, defined up to a proportionality constant as%
\footnote{In the literature the term $\sqrt{|\phi|^{2}}$ is usually written
as $\sqrt{n}$ where $n$ has the meaning of number density in non-relativistic
cases.%
} 
\begin{equation}
U_{\mathrm{Q}}\equiv-\frac{\nabla^{2}\sqrt{|\phi|^{2}}}{\sqrt{|\phi|^{2}}}=-\frac{\nabla^{2}\psi}{\psi}=-m_{\psi}^{2}.
\end{equation}
We can see this quantum potential is nothing but the mass square $m_{\psi}^{2}$
according to the radial profile equation (\ref{eq:BulkEOM_FluidProfile}).
This quantum potential always appears after Madelung transformation
(\ref{eq:MadelungTransformation}) of a scalar equation (Schrödinger/Gross-Pitaevskii
equation or Klein-Gordon equation) as a consequence of the Heisenberg
uncertainty principle \cite{Matos:2014Book,Chavanis:2015zua}. It
first appeared in \cite{Madelung:1927} and was later named the ``quantum
potential'' in \cite{Bohm:1951xw}, because this term always comes
with a coefficient of $\hbar$, which vanishes in the classical limit
$\hbar\rightarrow0$, and thus has no classical counterpart. Later
we will see the gradient of this quantum potential appearing as a
``quantum force'' alongside with the electric force and buoyant
force (pressure gradient) in the conservation equation for stress
tensor: this is why it is called a potential. In the following, we
will seldom use the symbol $U_{\mathrm{Q}}$ or the terminology ``quantum
potential'', but rather view it as a mass square of the radial profile
field $\psi$ as in the Klein-Gordon equation (\ref{eq:BulkEOM_FluidProfile}),
because the latter view will be more helpful when we discuss the EOS
of the fluid. Under the Madelung transformation (\ref{eq:MadelungTransformation}),
we have $D_{M}\phi=\left(\partial_{M}\log\psi+i\xi_{M}\right)\phi$,
where we define the so-called Taub current as 
\begin{equation}
\xi_{M}\equiv\partial_{M}\vartheta-q_{\phi}A_{M}.\label{eq:TaubCurrent}
\end{equation}
Notice for the uni-modular part, $D_{M}\tilde{\phi}=i\xi_{M}\tilde{\phi}$
and $g^{MN}\left(D_{M}\tilde{\phi}\right)^{\dagger}\left(D_{N}\tilde{\phi}\right)=\xi^{2}$.

For the bulk action (\ref{eq:BulkAction_Scalar}), using the Madelung
transformation (\ref{eq:MadelungTransformation}), integrating by
parts the kinetic term of $\psi$ and using $\psi$'s EOM (\ref{eq:BulkEOM_FluidProfile}),
the action can be written as 
\begin{align}
S_{\mathrm{bulk}}\left[\phi\right]= & -\frac{1}{2\kappa_{\phi}^{2}}\int_{\mathcal{M}}d^{d+1}x\sqrt{-g}\psi^{2}\left\{ g^{MN}\left(D_{M}\tilde{\phi}\right)^{\dagger}\left(D_{N}\tilde{\phi}\right)+\left(m_{\phi}^{2}-m_{\psi}^{2}\right)\tilde{\phi}^{\dagger}\tilde{\phi}\right\} \nonumber \\
 & +\frac{1}{2\kappa_{\phi}^{2}}\int_{\partial\mathcal{M}}d^{d}x\sqrt{-g}g^{zM}\psi\partial_{M}\psi.
\end{align}
From this action, we see that the bulk action of the uni-modular field
$\tilde{\phi}$ is very similar to that of the original field $\phi$,
with two differences: (i) there is an overall factor of $\psi^{2}$;
(ii) the effective mass square $\tilde{m}^{2}$ is shifted as 
\begin{equation}
\tilde{m}^{2}=m_{\phi}^{2}-m_{\psi}^{2}.\label{eq:EffectiveMass}
\end{equation}
Typically, in a fluid description, macroscopic quantities such as
the entropy density and the pressure of the fluid, are related at
the microscopic level to the incoherent phase fluctuations of the
quantum field, i.e. the field $\tilde{\phi}$ here. We will think of
the incoherent fluid as an effective description of the $\tilde{\phi}$
part, while the amplitude $\psi$ is a radial profile that satisfies
classical dynamics given by (\ref{eq:BulkEOM_FluidProfile}). We will
view the mass $\tilde{m}$ as an effective mass gap between particles
and antiparticles in this fluid. From this perspective, such a mass
squared shall always be positive
\begin{equation}
\tilde{m}^{2}\geqslant0.
\end{equation}
This is a physical requirement we will impose on the fluid dynamics.
On the contrary, in holography the original mass square of the scalar
$m_{\phi}^{2}$ is usually chosen to be negative. If one views this
$m_{\phi}^{2}$ directly as the mass square related to the particle-antiparticle
gap in the fluid, it will be meaningless. The way to reconcile this
contradiction is through the mass shift (\ref{eq:EffectiveMass}):
due to the radial profile $\psi$ and its EOM (\ref{eq:BulkEOM_FluidProfile})
which are both non-trivial in curved spacetime, $m_{\psi}^{2}$ will
be non-trivial and it shifts the negative value of $m_{\phi}^{2}$
to the non-negative value of $\tilde{m}^{2}$.%
\footnote{The idea of introducing a radial profile function was employed in
\cite{Breitenlohner:1982bm} to derive the Breitenlohner-Freedman
bound. %
} The origin of all these can be traced back to the curvature of spacetime.
Thus the mass shift (\ref{eq:EffectiveMass}) due to radial profile
function $\psi$ is a description of the mass renormalization effect
in curved spacetime. Furthermore, we notice (\ref{eq:EffectiveMass})
can also be written as 
\begin{equation}
U_{\mathrm{Q}}=\tilde{m}^{2}-m_{\phi}^{2},
\end{equation}
i.e. up to a constant zero-point energy $-m_{\phi}^{2}$, $\tilde{m}^{2}$
is just the quantum potential $U_{\mathrm{Q}}$. This agrees with
what we have discussed earlier that the quantum potential has no classical
counterpart, because it is related to the mass renormalization which
is a pure quantum field theory effect.

To incorporate EOM (\ref{eq:BulkEOM_FluidProfile}) into the transformed
action, we introduce a Lagrange multiplier $\chi$ to put the transformed
action off-shell, and rewrite it as following 
\begin{align}
S_{\mathrm{bulk}}\left[\phi\right]= & -\frac{1}{2\kappa_{\phi}^{2}}\int_{\mathcal{M}}d^{d+1}x\sqrt{-g}\psi^{2}\left(g^{MN}\xi_{M}\xi_{N}+\tilde{m}^{2}\right)\nonumber \\
 & +\frac{1}{2\kappa_{\phi}^{2}}\int_{\mathcal{M}}d^{d+1}x\sqrt{-g}\chi\left[\nabla^{2}-\left(m_{\phi}^{2}-\tilde{m}^{2}\right)\right]\psi\label{eq:BulkAction_Transformed_OffShell}\\
 & +\frac{1}{2\kappa_{\phi}^{2}}\int_{\partial\mathcal{M}}d^{d}x\sqrt{-g}g^{zM}\psi\partial_{M}\psi.\nonumber 
\end{align}

\subsection{Charge Current and Stress Tensor}

The bulk charge current is
\[
J_{M}^{\mathrm{flct}}=\frac{q_{\phi}}{\kappa_{\phi}^{2}}\psi^{2}\xi_{M}.
\]
From now on we will add ``flct'' to the current and stress tensor to
remind us that this is only the fluctuation part. There is also a
condensate part which we will omit most of the time. This is justified
for the linear scalar field we are considering because the condensate
and the fluctuation parts decouple at the effective action level.
The above relation is similar to eq. (3.3) in \cite{Carter:1994rv}.
We define the normalized mechanical velocity $u_{M}$ of the normal
fluid as
\begin{equation}
u_{M}\equiv\frac{\xi_{M}}{\mu_{\mathrm{h}}},\qquad\mu_{\mathrm{h}}^{2}\equiv-\xi^{2}=-\big\langle\left(\partial\vartheta-q_{\phi}A\right)^{2}\big\rangle,
\end{equation}
which satisfies the usual normalization condition for velocity $u^{2}=-1$.
Notice for normal fluid $\xi^{2}<0$. $\mu_{\mathrm{h}}$ has the
dimension of energy. Its physical meaning is enthalpy per charge,
which will be clear later. In the rest of this section, we will view
$\mu_{\mathrm{h}}$ and $u_{M}$ as independent (of metric) and physical
variables, instead of $\xi_{M}$. The correctly normalized charge
density $\rho$ can then be read off using $\rho=-u^{M}J_{M}^{\mathrm{flct}}$
as 
\begin{equation}
\rho=q_{\phi}\frac{\psi^{2}\mu_{\mathrm{h}}}{\kappa_{\phi}^{2}},
\end{equation}
this is in fact the same as eq. (2.9) in \cite{Carter:1994rv}. Now
the charge current is just 
\begin{equation}
J_{\mathrm{flct}}^{M}=\rho u^{M}.
\end{equation}
We now define the rescaled charge density (rescaled by the radial
profile $\psi$) associated with the phase fluctuations as $\tilde{\rho}$
\begin{equation}
\rho\equiv\frac{\kappa_{\mathrm{f}}^{2}}{\kappa_{\phi}^{2}}\psi^{2}\tilde{\rho},\qquad\tilde{\rho}=\frac{q_{\phi}\mu_{\mathrm{h}}}{\kappa_{\mathrm{f}}^{2}},
\end{equation}
and the corresponding rescaled charge current $\tilde{J}_{\mathrm{flct}}^{M}$
is 
\begin{equation}
J_{\mathrm{flct}}^{M}=\frac{\kappa_{\mathrm{f}}^{2}}{\kappa_{\phi}^{2}}\psi^{2}\tilde{J}_{\mathrm{flct}}^{M},\qquad\tilde{J}_{\mathrm{flct}}^{M}=\tilde{\rho}u^{M}.
\end{equation}
Here $\kappa_{\mathrm{f}}$ has dimension $[\mathrm{length}]^{\frac{d-1}{2}}$
such that the combination $\kappa_{\mathrm{f}}\psi/\kappa_{\phi}$
is dimensionless. In AdS/CFT, we can choose $\kappa_{\mathrm{f}}$
to be set by the length scale of the AdS radius $R$. But the choice does
not really matter, because $\kappa_{\mathrm{f}}$, $\kappa_{\phi}$
and $R$ will completely drop off all the EOMs in their dimensionless
version. Now the expression for the ``$\tilde{\phantom{s}}$'' part
takes the standard perfect fluid form.

Using the transformed action (\ref{eq:BulkAction_Transformed_OffShell}),
we can derive the stress tensor of the fluctuation by taking functional
derivatives with respect to the metric. The stress tensor of the fluctuation
can be split into two parts --- the part resulting from the incoherent
phase and that from the amplitude: 
\begin{equation}
T_{\mathrm{flct}}^{MN}=T_{\mathrm{phase}}^{MN}+T_{\mathrm{amp}}^{MN}.
\end{equation}
The first line of the action (\ref{eq:BulkAction_Transformed_OffShell})
gives the phase part: 
\begin{align*}
T_{\mathrm{phase}}^{MN} & =\frac{1}{\kappa_{\phi}^{2}}\psi^{2}\mu_{\mathrm{h}}^{2}u^{M}u^{N}+\frac{1}{2\kappa_{\phi}^{2}}g^{MN}\psi^{2}\left(\mu_{\mathrm{h}}^{2}-\tilde{m}^{2}\right).
\end{align*}
We can now identify the rescaled incoherent phase part of the stress
tensor $\tilde{T}_{\mathrm{phase}}^{MN}$ as 
\begin{align}
T_{\mathrm{phase}}^{MN} & =\frac{\kappa_{\mathrm{f}}^{2}}{\kappa_{\phi}^{2}}\psi^{2}\tilde{T}_{\mathrm{phase}}^{MN},\\
\tilde{T}_{\mathrm{phase}}^{MN} & =\tilde{\varepsilon}_{\mathrm{phase}}u^{M}u^{N}+\tilde{p}\left(g^{MN}+u^{M}u^{N}\right),
\end{align}
where the energy density $\tilde{\varepsilon}_{\mathrm{phase}}$ and
pressure $\tilde{p}$ are%
\footnote{Here we add a subscript ``$_{\mathrm{phase}}$'' to the energy density
because it is different from the energy density $\tilde{\varepsilon}$
that will appear later. $\tilde{\varepsilon}_{\mathrm{phase}}$ is
the total perfect fluid energy density defined as $u_{M}u_{N}\langle\tilde{T}_{\mathrm{phase}}^{MN}\rangle$.
Later we will define $\tilde{\varepsilon}$ through the thermodynamic
relation $\tilde{\varepsilon}+\tilde{p}=\tilde{T}\tilde{s}+\tilde{\mu}\tilde{\rho}$,
which only accounts for part of $\tilde{\varepsilon}_{\mathrm{phase}}$.%
} 
\begin{align}
\tilde{\varepsilon}_{\mathrm{phase}} & =\frac{1}{2\kappa_{\mathrm{f}}^{2}}\left(\mu_{\mathrm{h}}^{2}+\tilde{m}^{2}\right),\\
\tilde{p} & =\frac{1}{2\kappa_{\mathrm{f}}^{2}}\left(\mu_{\mathrm{h}}^{2}-\tilde{m}^{2}\right).
\end{align}
Notice the above relations implies the following constraint
\begin{equation}
-1\leqslant\frac{\tilde{p}}{\tilde{\varepsilon}_{\mathrm{phase}}}\leqslant1,
\end{equation}
and the lower bound corresponds to $\tilde{\varepsilon}_{\mathrm{phase}}\simeq-\tilde{p}$
when $\big|\mu_{\mathrm{h}}\big|\ll\tilde{m}$. The second line of
(\ref{eq:BulkAction_Transformed_OffShell}) gives the amplitude part
of the stress tensor
\begin{equation}
T_{\mathrm{amp}}^{MN}=\frac{1}{2\kappa_{\phi}^{2}}\left\{ \left(\partial^{M}\chi\right)\left(\partial^{N}\psi\right)+\left(\partial^{N}\chi\right)\left(\partial^{M}\psi\right)-g^{MN}\left[g^{PQ}\left(\partial_{P}\chi\right)\left(\partial_{Q}\psi\right)+m_{\psi}^{2}\chi\psi\right]\right\} ,
\end{equation}
where we have defined the short-hand notation $\partial^{M}\equiv g^{MN}\partial_{N}$.

\subsection{On-Shell and Partially Off-Shell Bulk Actions}

Now using the expression for $\tilde{p}$ and $u^{2}=-1$, the on-shell
bulk action can be written as
\begin{equation}
S_{\mathrm{bulk}}^{\mathrm{fluid}}\left[\tilde{p},\psi\right]=\frac{\kappa_{\mathrm{f}}^{2}}{\kappa_{\phi}^{2}}\int_{\mathcal{M}}d^{d+1}x\sqrt{-g}\psi^{2}\tilde{p}+\frac{1}{2\kappa_{\phi}^{2}}\int_{\partial\mathcal{M}}d^{d}x\sqrt{-g}g^{zM}\psi\partial_{M}\psi.\label{eq:FluidAction_OnShell_1}
\end{equation}
This is an on-shell action for the fluid. The difference from the usual
fluid action in flat spacetime is the appearance of radial profile
function $\psi^{2}$ in the bulk action. From now on, we will view
(or \emph{propose}) this action, together with EOM (\ref{eq:BulkEOM_FluidProfile}),
as our \emph{first principle} for the fluid dynamics. It is this
action and EOM that we will pass on to further calculations. Notice
that the boundary term shall not be forgotten; it will play a crucial
role in getting boundary conditions for the fluid later. To incorporate
the EOM (\ref{eq:BulkEOM_FluidProfile}) into the action by a Lagrange
multiplier $\chi$, we obtain the partially off-shell fluid action
\begin{align}
S_{\mathrm{bulk}}^{\mathrm{fluid}}\left[\tilde{p},\psi,\chi\right]= & \int_{\mathcal{M}}d^{d+1}x\sqrt{-g}\left\{ \frac{\kappa_{\mathrm{f}}^{2}}{\kappa_{\phi}^{2}}\psi^{2}\tilde{p}+\frac{1}{2\kappa_{\phi}^{2}}\chi\left(\nabla^{2}-m_{\psi}^{2}\right)\psi\right\} \nonumber \\
 & +\frac{1}{2\kappa_{\phi}^{2}}\int_{\partial\mathcal{M}}d^{d}x\sqrt{-g}g^{zM}\psi\partial_{M}\psi,\label{eq:FluidAction_OffShell_1}
\end{align}
which is the fluid version of (\ref{eq:BulkAction_Transformed_OffShell}).
From now on, $m_{\psi}$ shall be understood as a short hand notation
for the relation (\ref{eq:EffectiveMass}). The variations of the
action with respect to $\psi$ and $\chi$ are 
\begin{align}
 & \delta_{\psi,\chi}S_{\mathrm{bulk}}^{\mathrm{fluid}}\left[\tilde{p},\psi,\chi\right]\nonumber \\
= & \frac{1}{2\kappa_{\phi}^{2}}\int_{\mathcal{M}}d^{d+1}x\sqrt{-g}\left(\nabla^{2}\psi-m_{\psi}^{2}\psi\right)\delta\chi\nonumber \\
 & +\frac{1}{2\kappa_{\phi}^{2}}\int_{\mathcal{M}}d^{d+1}x\sqrt{-g}\left\{ 4\kappa_{\mathrm{f}}^{2}\psi\tilde{p}+\left(\nabla^{2}-m_{\psi}^{2}\right)\chi\right\} \delta\psi\label{eq:FluidAction_OffShellVariation}\\
 & +\frac{1}{2\kappa_{\phi}^{2}}\int_{\partial\mathcal{M}}d^{d}x\sqrt{-g}g^{zM}\delta\psi\partial_{M}\left(\psi+\chi\right)+\frac{1}{2\kappa_{\phi}^{2}}\int_{\partial\mathcal{M}}d^{d}x\sqrt{-g}g^{zM}\left(\psi-\chi\right)\partial_{M}\delta\psi.\nonumber 
\end{align}
In the above expression, the first two lines give the bulk EOMs for
$\psi$ and $\chi$ respectively; the last line will yield two independent
boundary conditions for $\psi$ and $\chi$ when the other boundary
term for $\psi$ given in (\ref{eq:BoundaryAction_CounterTerm_Deformed})
is included.

\bigskip{}


\section{Velocity-Potential Representation of Quantum Fluid}

The action we have obtained in (\ref{eq:FluidAction_OffShell_1})
is still not the final version that can be used in actual calculation,
because for the pressure $\tilde{p}$, we have not identified what
its field variables are. In grand canonical ensemble, we shall view
$\tilde{p}$ as a function of temperature $\tilde{T}$, chemical potential
$\tilde{\mu}$ and effective mass $\tilde{m}$. The functional relation
between them is the EOS: $\tilde{p}=\tilde{p}\left[\tilde{T},\tilde{\mu},\tilde{m}\right]$.
Even so, at this moment we still do not obtain the correct EOMs by
varying $\tilde{T}$, $\tilde{\mu}$ and $\tilde{m}$ because the
action for the $\tilde{p}$ part is still on-shell. To arrive at the
full off-shell action which can correctly produce a set of EOMs that
are physically meaningful, we will use the so-called velocity-potential
representation of a perfect fluid \cite{Schutz:1970my}. A useful review
on this topic is given in \cite{Brown:1992kc}. Readers not familiar
with this formalism can refer to Appendix (A) where we present a brief
pedagogical review on how to use this formalism to derive the classical
perfect fluid dynamics. Appendix (B) discusses how our quantum fluid
dynamics can reduce to the classical one.

\subsection{Full Off-Shell Action of Quantum Fluid}

We now introduce velocity-potentials for the off-shell action (\ref{eq:FluidAction_OffShell_1}),
which is our quantum extension of the example given in (\ref{eq:FluidAction_OffShell_TimeLike}).
The full off-shell action is 
\begin{align}
 & S_{\mathrm{bulk}}^{\mathrm{fluid}}\left[\psi,\chi,\theta_{\mathrm{s}},\theta,\theta_{\mathrm{m}},u^{M},\tilde{T},\tilde{\mu},\tilde{m},\tilde{s},\tilde{\rho},\tilde{\varsigma},\tilde{\eta}\right]\nonumber \\
= & \frac{\kappa_{\mathrm{f}}^{2}}{\kappa_{\phi}^{2}}\int_{\mathcal{M}}d^{d+1}x\sqrt{-g}\psi^{2}\Big\{\tilde{p}\left[\tilde{T},\tilde{\mu},\tilde{m}\right]+\frac{1}{2}\tilde{\eta}\left(u^{M}u_{M}+1\right)\label{eq:FluidAction_OffShell_2}\\
 & \qquad-\tilde{s}\left(\tilde{T}-u^{M}\partial_{M}\theta_{\mathrm{s}}\right)-\tilde{\rho}\left[\tilde{\mu}+u^{M}\left(\partial_{M}\theta-A_{M}\right)\right]-\tilde{\varsigma}\left(\tilde{m}^{2}-m_{\phi}^{2}\right)\left(\tilde{m}-u^{M}\partial_{M}\theta_{\mathrm{m}}\right)\Big\}\nonumber \\
 & +\frac{1}{2\kappa_{\phi}^{2}}\int_{\mathcal{M}}d^{d+1}x\sqrt{-g}\,\chi\left[\nabla^{2}-\left(m_{\phi}^{2}-\tilde{m}^{2}\right)\right]\psi+\frac{1}{2\kappa_{\phi}^{2}}\int_{\partial\mathcal{M}}d^{d}x\sqrt{-g}g^{zM}\psi\partial_{M}\psi.\nonumber 
\end{align}
Here the functional form of $\tilde{p}\left[\tilde{T},\tilde{\mu},\tilde{m}\right]$
will be given by the EOS. We will discuss its form for our quantum
fluid in the follow-up paper \cite{Wu:2016_Note3}. In this note,
it will be kept general. $u^{M}$ is the fluid velocity, $\theta_{\mathrm{s}}$,
$\theta$ and $\theta_{\mathrm{m}}$ are the velocity-potentials and
$\tilde{s}$, $\tilde{\rho}$ and $\tilde{\varsigma}$ are corresponding
Lagrange multipliers. In the literature, $\theta_{\mathrm{s}}$ is
called the ``thermasy'' while $\theta$ is the Clebsch potential. The
effective mass $\tilde{m}$ is again from (\ref{eq:EffectiveMass}),
and from now on we shall view $\tilde{m}$ rather than $m_{\psi}$
as an elementary field variable of the fluid. We require $\tilde{m}\geqslant0$.
This action is the effective fluid description for (\ref{eq:HolographicEffectiveAction_Decoherence}).
Meanwhile, we shall not forget the boundary action (\ref{eq:BoundaryAction_CounterTerm_Deformed}),
which under the Madelung transformation (\ref{eq:MadelungTransformation})
takes the following form 
\begin{equation}
S_{\mathrm{ct}}^{\mathrm{fluid}}\left[\psi,\epsilon\right]=-\frac{\Delta_{\mathrm{ct}}(\lambda,\epsilon)}{2\kappa_{\phi}^{2}R}\int_{\partial\mathcal{M}}d^{d}x\sqrt{-\gamma}\psi^{2}.
\end{equation}
The above two equations form the complete off-shell action for the
quantum fluid in the bulk. In the rest of this section, we will show
what dynamics they produce.

\subsection{Bulk Equations of Motion}

First, variations of $\tilde{\eta}$, $\tilde{s}$, $\tilde{\rho}$,
$\tilde{T}$, $\tilde{\mu}$, $\theta_{\mathrm{s}}$, $\theta$ and
$\theta_{\mathrm{m}}$ yield the following equations 
\begin{align}
u^{2} & =-1,\\
\tilde{T} & =u^{M}\partial_{M}\theta_{\mathrm{s}},\label{eq:BulkFluidEOM_Thermasy}\\
\tilde{\mu} & =u^{M}\left(-\partial_{M}\theta+A_{M}\right),\label{eq:BulkFluidEOM_Clebsch}\\
\tilde{s} & =\frac{\partial\tilde{p}}{\partial\tilde{T}},\\
\tilde{\rho} & =\frac{\partial\tilde{p}}{\partial\tilde{\mu}},\\
0 & =\nabla_{M}\left(\psi^{2}\tilde{s}u^{M}\right),\\
0 & =\nabla_{M}\left(\psi^{2}\tilde{\rho}u^{M}\right),\\
0 & =\nabla_{M}\left[\psi^{2}\left(\tilde{m}^{2}-m_{\phi}^{2}\right)\tilde{\varsigma}u^{M}\right].
\end{align}
Varying $\tilde{\varsigma}$, we have two solutions
\begin{equation}
\tilde{m}=u^{M}\partial_{M}\theta_{\mathrm{m}},\qquad\textrm{or}\qquad\tilde{m}=m_{\phi}\;\left(\textrm{if }m_{\phi}\geqslant0\right).
\end{equation}
We will call the solution on the left the \emph{quantum branch} and
the right the \emph{classical branch}. The former is the one we are
mainly interested in. The variation with respect to $\tilde{m}$ gives
\begin{align}
\tilde{\varsigma} & =\frac{1}{\left(\tilde{m}^{2}-m_{\phi}^{2}\right)+2\tilde{m}\left(\tilde{m}-u^{M}\partial_{M}\theta_{\mathrm{m}}\right)}\left(\frac{\partial\tilde{p}}{\partial\tilde{m}}+\frac{\chi}{\psi}\frac{\tilde{m}}{\kappa_{\mathrm{f}}^{2}}\right)\label{eq:BulkFluidEOM_ConformalDeviation}\\
 & =\begin{cases}
\frac{1}{\left(\tilde{m}^{2}-m_{\phi}^{2}\right)}\left(\frac{\partial\tilde{p}}{\partial\tilde{m}}+\frac{\chi}{\psi}\frac{\tilde{m}}{\kappa_{\mathrm{f}}^{2}}\right) & \qquad\left(\tilde{m}=u^{M}\partial_{M}\theta_{\mathrm{m}}\right)\\
\frac{1}{2\tilde{m}\left(\tilde{m}-u^{M}\partial_{M}\theta_{\mathrm{m}}\right)}\left(\frac{\partial\tilde{p}}{\partial\tilde{m}}+\frac{\chi}{\psi}\frac{\tilde{m}}{\kappa_{\mathrm{f}}^{2}}\right) & \qquad\left(\tilde{m}=m_{\phi}\right)
\end{cases}.\nonumber 
\end{align}

Variation of $u^{M}$ gives 
\[
\tilde{\eta}u_{M}+\tilde{s}\partial_{M}\theta_{\mathrm{s}}-\tilde{\rho}\left(\partial_{M}\theta-A_{M}\right)+\left(\tilde{m}^{2}-m_{\phi}^{2}\right)\tilde{\varsigma}\partial_{M}\theta_{\mathrm{m}}=0.
\]
Multiply it by $u^{M}$, and using the above EOMs, we have
\begin{align*}
\tilde{\eta} & =\tilde{T}\tilde{s}+\tilde{\mu}\tilde{\rho}+\left(\tilde{m}^{2}-m_{\phi}^{2}\right)\tilde{\varsigma}u^{M}\partial_{M}\theta_{\mathrm{m}}\\
 & =\begin{cases}
\tilde{T}\tilde{s}+\tilde{\mu}\tilde{\rho}+\left(\tilde{m}^{2}-m_{\phi}^{2}\right)\tilde{m}\tilde{\varsigma} & \qquad\left(\tilde{m}=u^{M}\partial_{M}\theta_{\mathrm{m}}\right)\\
\tilde{T}\tilde{s}+\tilde{\mu}\tilde{\rho} & \qquad\left(\tilde{m}=m_{\phi}\right)
\end{cases}.
\end{align*}
We define the thermal energy density of the incoherent fluid $\tilde{\varepsilon}$
via the thermodynamic relation 
\begin{equation}
\tilde{\varepsilon}+\tilde{p}=\tilde{T}\tilde{s}+\tilde{\mu}\tilde{\rho},
\end{equation}
i.e. $\tilde{\varepsilon}$ shall really be viewed as a short-hand
notation for $\tilde{T}\tilde{s}+\tilde{\mu}\tilde{\rho}-\tilde{p}$.
Using (\ref{eq:BulkFluidEOM_ConformalDeviation}), we have 
\begin{align}
\tilde{\eta} & =\tilde{\varepsilon}+\tilde{p}+\left(\tilde{m}^{2}-m_{\phi}^{2}\right)\tilde{m}\tilde{\varsigma}=\begin{cases}
\tilde{\varepsilon}+\tilde{p}+\frac{\partial\tilde{p}}{\partial\log\tilde{m}}+\frac{\chi}{\psi}\frac{\tilde{m}^{2}}{\kappa_{\mathrm{f}}^{2}} & \qquad\left(\tilde{m}=u^{M}\partial_{M}\theta_{\mathrm{m}}\right)\\
\tilde{\varepsilon}+\tilde{p} & \qquad\left(\tilde{m}=m_{\phi}\right)
\end{cases}\label{eq:BulkFluidEOM_EnthalpyDensity}\\
u_{M} & =\frac{1}{\tilde{\eta}}\left[-\tilde{s}\partial_{M}\theta_{\mathrm{s}}+\tilde{\rho}\left(\partial_{M}\theta-A_{M}\right)-\left(\tilde{m}^{2}-m_{\phi}^{2}\right)\tilde{\varsigma}\partial_{M}\theta_{\mathrm{m}}\right].\label{eq:BulkFluidEOM_Velocity}
\end{align}
In (\ref{eq:BulkFluidEOM_EnthalpyDensity}), $\tilde{\eta}$ is the
enthalpy density: it does not only contain the term $\tilde{\varepsilon}+\tilde{p}$
but also the additional term involving $\tilde{\varsigma}$ in the
quantum branch because $\tilde{m}$ is not a constant there. In (\ref{eq:BulkFluidEOM_Velocity}),
it is manifest now why $\theta$, $\theta_{\mathrm{s}}$ and $\theta_{\mathrm{m}}$
are collectively called the velocity-potentials: their gradients give
the distribution of the velocity field, in the same sense that the
gradient of a potential gives the field strength. From the Taub current
we defined earlier in (\ref{eq:TaubCurrent}) and $\xi_{M}=\mu_{\mathrm{h}}u_{M}$,
we also have 
\begin{align}
\big\langle\partial_{M}\vartheta\big\rangle & =q_{\phi}\left[\partial_{M}\theta-\frac{\tilde{s}}{\tilde{\rho}}\partial_{M}\theta_{\mathrm{s}}-\frac{\left(\tilde{m}^{2}-m_{\phi}^{2}\right)\tilde{\varsigma}}{\tilde{\rho}}\partial_{M}\theta_{\mathrm{m}}\right],\\
\mu_{\mathrm{h}} & =q_{\phi}\frac{\tilde{\eta}}{\tilde{\rho}}.
\end{align}
Here we can see the physical meaning of $\mu_{\mathrm{h}}$ is enthalpy
per charge. In fluid dynamics, it plays the analogous role of mass
in Newtonian dynamics. In the first equation for $\langle\partial_{M}\vartheta\rangle$,
if we consider the (super-)fluid dynamics for the condensate instead,
we will only have the $\partial_{M}\theta$ term appearing on the
right; the second term proportional to $\tilde{s}$ is a major difference
between a coherent (super-)fluid (which has vanishing entropy density)
and an incoherent thermal fluid. The third term with $\tilde{\varsigma}$
is a consequence of the mass renormalization effect.

Using (\ref{eq:FluidAction_OffShellVariation}), the variation of
$\chi$ and $\psi$ in the bulk yield the following equations for
them 
\begin{align}
\left[\nabla^{2}-\left(m_{\phi}^{2}-\tilde{m}^{2}\right)\right]\psi & =0,\\
\left[\nabla^{2}-\left(m_{\phi}^{2}-\tilde{m}^{2}\right)\right]\chi & =-4\kappa_{\mathrm{f}}^{2}\tilde{p}\psi.
\end{align}
The ratio of the two profile fields satisfies the following differential
equation
\begin{equation}
\left[\nabla^{2}+2\left(\nabla\log\psi\right)\cdot\nabla\right]\frac{\chi}{\psi}=-4\kappa_{\mathrm{f}}^{2}\tilde{p}.\label{eq:BulkFluidEOM_ProfileRatio_Covariant}
\end{equation}
It is more convenient to view the ratio of $\chi/\psi$ rather than
$\chi$ itself as an independent field. This is especially helpful
when taking some limit or discussing the dynamics in asymptotic regions.
Furthermore, if we view $\sqrt{\psi\chi}$ and $\psi/\chi$ rather
than $\psi$ and $\chi$ as independent fields, then their EOMs read
\begin{align}
\left[\nabla^{2}-m_{\phi}^{2}+\tilde{m}^{2}+\frac{1}{4}\left(\nabla\log\frac{\psi}{\chi}\right)^{2}+2\kappa_{\mathrm{f}}^{2}\tilde{p}\left(\frac{\psi}{\chi}\right)\right]\sqrt{\psi\chi} & =0,\\
\left[\nabla^{2}+2\left(\nabla\log\sqrt{\psi\chi}\right)\cdot\nabla\right]\log\frac{\psi}{\chi}-4\kappa_{\mathrm{f}}^{2}\tilde{p}\left(\frac{\psi}{\chi}\right) & =0.
\end{align}
Although these equations look more complicated than the previous equations
for $\psi$ and $\chi$, they will be more useful when discussing
bulk dynamics because in the following we will see the boundary conditions
are naturally written in terms of $\sqrt{\psi\chi}$ and $\psi/\chi$,
rather than $\psi$ and $\chi$.

\subsection{Boundary Conditions}

When deriving the above EOMs, we have generated a few boundary terms
by integration by parts. We now collect all the boundary terms in
the variation of the full off-shell action: 
\begin{align*}
 & \delta\left(S_{\mathrm{bulk}}^{\mathrm{fluid}}+S_{\mathrm{ct}}^{\mathrm{fluid}}\right)\\
= & \frac{1}{\kappa_{\phi}^{2}}\int_{\partial\mathcal{M}}d^{d}x\left\{ \frac{1}{2}\sqrt{-g}g^{zM}\partial_{M}\left(\psi+\chi\right)-\frac{\Delta_{\mathrm{ct}}(\lambda,\epsilon)}{R}\sqrt{-\gamma}\psi\right\} \delta\psi\\
 & +\frac{1}{2\kappa_{\phi}^{2}}\int_{\partial\mathcal{M}}d^{d}x\sqrt{-g}g^{zM}\left(\psi-\chi\right)\partial_{M}\delta\psi\\
 & -\int_{\partial\mathcal{M}}d^{d}x\sqrt{-g}\psi^{2}\tilde{s}u^{z}\delta\theta_{\mathrm{s}}+\int_{\partial\mathcal{M}}d^{d}x\sqrt{-g}\psi^{2}\tilde{\rho}u^{z}\delta\theta-\int_{\partial\mathcal{M}}d^{d}x\sqrt{-g}\psi^{2}\left(\tilde{m}^{2}-m_{\phi}^{2}\right)\tilde{\varsigma}u^{z}\delta\theta_{\mathrm{m}}.
\end{align*}
To make sure the whole equation vanishes, all three lines must vanish
separately. The vanishing of the last line implies 
\begin{equation}
u^{z}\Big|_{z=\epsilon}=0.
\end{equation}
This is nothing but a statement that the fluid can not flow through
the boundary, i.e. bulk flow streamlines near the boundary must
be tangential to the boundary. Vanishing of the second line yields
\begin{equation}
\left(\psi-\chi\right)\Big|_{z=\epsilon}=0.\label{eq:Fluid_BoundaryCondition_2_Original}
\end{equation}
Vanishing of the first line yields
\[
\left\{ \frac{1}{2}\sqrt{-g}g^{zM}\partial_{M}\left(\psi+\chi\right)-\frac{\Delta_{\mathrm{ct}}(\lambda,\epsilon)}{R}\sqrt{-\gamma}\psi\right\} \Big|_{z=\epsilon}=0,
\]
which, under the help of (\ref{eq:Fluid_BoundaryCondition_2_Original}),
can be written as
\begin{equation}
\left\{ \sqrt{-g}g^{zM}\partial_{M}\sqrt{\psi\chi}-\frac{\Delta_{\mathrm{ct}}(\lambda,\epsilon)}{R}\sqrt{-\gamma}\sqrt{\psi\chi}\right\} \Big|_{z=\epsilon}=0.\label{eq:Fluid_BoundaryCondition_1_Original}
\end{equation}
If we replace $\sqrt{\psi\chi}$ in this boundary condition by the
amplitude of the condensate field $|\phi_{0}|$, it is the same as
the boundary condition for the condensate in the absence of the source.

\subsection{Vacuum Polarization}

In quantum field theory in curved spacetime, an important quantity
for calculating the renormalized stress tensor is the vacuum polarization
$\langle\phi^{\dagger}(x)\phi(x)\rangle$ \cite{BirrellDavies:Book,Wald:Book}.
For us it can be obtained by taking  functional derivative of the
effective action with respect to the original mass square 
\[
\frac{1}{2\kappa_{\phi}^{2}}\langle\phi^{\dagger}(x)\phi(x)\rangle=-\frac{1}{\sqrt{-g}}\frac{\delta\Gamma_{\mathrm{fluid}}}{\delta m_{\phi}^{2}}.
\]
Using the full off-shell action (\ref{eq:FluidAction_OffShell_2}),
we have 
\[
\frac{1}{\sqrt{-g}}\frac{\delta\Gamma_{\mathrm{fluid}}}{\delta m_{\phi}^{2}}=\frac{\kappa_{\mathrm{f}}^{2}}{\kappa_{\phi}^{2}}\psi^{2}\tilde{\varsigma}\left(\tilde{m}-u^{M}\partial_{M}\theta_{\mathrm{m}}\right)-\frac{1}{2\kappa_{\phi}^{2}}\chi\psi.
\]
Putting it on-shell using (\ref{eq:BulkFluidEOM_ConformalDeviation}),
we have
\begin{equation}
\langle\phi^{\dagger}(x)\phi(x)\rangle=\begin{cases}
\quad\chi\psi & \qquad\left(\tilde{m}=u^{M}\partial_{M}\theta_{\mathrm{m}}\right)\\
-2\kappa_{\mathrm{f}}^{2}\psi^{2}\frac{\partial\tilde{p}}{\partial\tilde{m}^{2}} & \qquad\left(\tilde{m}=m_{\phi}\right)
\end{cases}.
\end{equation}
Thus we see in the quantum branch, the quantity $\sqrt{\psi\chi}$
is nothing but the amplitude of $\langle\phi^{\dagger}\phi\rangle$,
very much like $|\bar{\phi}|$ in the condensate case. In the condensate
case, the amplitude of the vacuum polarization $|\bar{\phi}|$ is
the holographic dual of the superconducting gap parameter of the field
theory. In the same sense, $\sqrt{\psi\chi}$ is the bulk dual of
the pseudogap parameter: the pseudogap exists when $\sqrt{\psi\chi}$
is non-trivial. It is more useful to view the product and ratio quantities
$\sqrt{\psi\chi}$ and $\psi/\chi$, rather than the individual fields
$\psi$ and $\chi$, as independent fields in the bulk analysis, because
the former two have more explicit physical meanings, as well as neater
boundary conditions (\ref{eq:Fluid_BoundaryCondition_1_Original})
and (\ref{eq:Fluid_BoundaryCondition_2_Original}).

\subsection{Conservation of Stress Tensor }

By taking functional derivative of the off-shell action (\ref{eq:FluidAction_OffShell_2})
with respect to $g_{MN}$ and $A_{M}$, and using the EOMs, the on-shell
stress tensor and charge current are 
\begin{align}
T_{\mathrm{fluid}}^{MN} & =\frac{\kappa_{\mathrm{f}}^{2}}{\kappa_{\phi}^{2}}\psi^{2}\left\{ \left(\tilde{\eta}-\tilde{p}\right)u^{M}u^{N}+\tilde{p}\left(g^{MN}+u^{M}u^{N}\right)\right\} \\
 & \quad+\frac{1}{2\kappa_{\phi}^{2}}\left\{ \left(\partial^{M}\chi\right)\left(\partial^{N}\psi\right)+\left(\partial^{N}\chi\right)\left(\partial^{M}\psi\right)-g^{MN}\left[\left(\partial^{P}\chi\right)\left(\partial_{P}\psi\right)+m_{\psi}^{2}\chi\psi\right]\right\} ,\nonumber \\
J_{\mathrm{fluid}}^{M} & =\frac{\kappa_{\mathrm{f}}^{2}}{\kappa_{\phi}^{2}}\psi^{2}\tilde{\rho}u^{M}.
\end{align}
Here we replace the subscript ``flct'' used in previous sections
by ``fluid'' to emphasize these are results from the fluid description,
but they are the same quantities. Here we see, due to the presence
of $\psi$ and $\chi$, the second line of $T_{\mathrm{fluid}}^{MN}$
introduces some anisotropic deviation from the standard isotropic
perfect fluid form of the first line. The pressure along the radial
direction and that along the transverse directions are different.
Such anisotropy has been seen in quantum field theory calculations
of the renormalized stress tensor, for example, that of the Hartle-Hawking
vacuum of Schwarzschild geometry \cite{Page:1982fm,Howard:1985yg}.
We also see that the total energy density due to the incoherent phase
is 
\begin{equation}
\tilde{\varepsilon}_{\mathrm{phase}}=\tilde{\eta}-\tilde{p}=\begin{cases}
\tilde{\varepsilon}+\frac{\partial\tilde{p}}{\partial\log\tilde{m}}+\frac{\chi}{\psi}\frac{\tilde{m}^{2}}{\kappa_{\mathrm{f}}^{2}} & \qquad\left(\tilde{m}=u^{M}\partial_{M}\theta_{\mathrm{m}}\right)\\
\tilde{\varepsilon} & \qquad\left(\tilde{m}=m_{\phi}\right)
\end{cases},
\end{equation}
and $\tilde{\eta}$ has the meaning of enthalpy density $\tilde{\eta}=\tilde{\varepsilon}_{\mathrm{phase}}+\tilde{p}$.
Notice that for the incoherent phase part, the enthalpy density is
\emph{still} the sum of the energy density $\tilde{\varepsilon}_{\mathrm{phase}}$
and the pressure $\tilde{p}$ of the incoherent fluid, but the former
is different from the thermal energy density $\tilde{\varepsilon}$
just computed from the EOS in the quantum branch: there is an additional
contribution to the total energy density in the form of $\left(\tilde{m}^{2}-m_{\phi}^{2}\right)\tilde{m}\tilde{\varsigma}$
given by (\ref{eq:BulkFluidEOM_ConformalDeviation}), due to the fact
that $\tilde{m}$ is not a constant but a local field. This is a consequence
of the mass renormalization effect; or equivalently, it can be viewed
as the part of Casimir energy density due to the quantum potential
$U_{\mathrm{Q}}$ discussed earlier.

We now show that the EOMs lead to the conservation of stress tensor,
under a certain condition, so as to agree with what is expected from
the Bianchi identity for the Einstein equation. To complete the circle,
we have to add the Maxwell sector.%
\footnote{For simplicity, we ignore the condensate part of the scalar field.
Adding it does not change any conclusion we will reach.%
} The action is 
\[
S_{\mathrm{bulk}}^{\mathrm{Maxwell}}\left[A_{M}\right]=-\frac{1}{4e_{A}^{2}}\int_{\mathcal{M}}d^{d+1}x\sqrt{-g}F_{MN}F^{MN},
\]
where $F_{MN}=\partial_{M}A_{N}-\partial_{N}A_{M}$ and $e_{A}$ is
the coupling constant of the Maxwell sector. The Maxwell equation
reads 
\[
\nabla^{M}F_{MN}=-e_{A}^{2}J_{N}^{\mathrm{fluid}}.
\]
The Maxwell field's contribution to the stress tensor is 
\[
T_{MN}^{\mathrm{Maxwell}}=\frac{1}{e_{A}^{2}}\left(F_{MP}F_{N}^{\phantom{N}P}-\frac{1}{4}g_{MN}F^{2}\right).
\]
Using $\left[\nabla_{P},\nabla_{N}\right]A_{Q}=-R_{\phantom{M}QPN}^{M}A_{M}$,
the cyclic identities $R_{\phantom{M}NPQ}^{M}+R_{\phantom{M}QNP}^{M}+R_{\phantom{M}PQN}^{M}=0$
and $F^{PQ}\left(R_{\phantom{M}NPQ}^{M}+2R_{\phantom{M}PQN}^{M}\right)=0$
and the Maxwell equation, we can show
\begin{equation}
\nabla^{M}T_{MN}^{\mathrm{Maxwell}}=J_{\mathrm{fluid}}^{M}F_{MN}.
\end{equation}
From the fluid (scalar fluctuation) part, using $\psi$ and $\chi$'s
EOMs, we have 
\begin{align}
\nabla^{M}T_{MN}^{\mathrm{fluid}} & =\frac{\kappa_{\mathrm{f}}^{2}}{\kappa_{\phi}^{2}}\psi^{2}\Big\{\nabla_{N}\tilde{p}+\left(u^{M}\nabla_{M}\right)\left(\tilde{\eta}u_{N}\right)+\left(\tilde{\eta}u_{N}\right)\left[\left(\nabla^{M}u_{M}\right)+\left(u^{M}\nabla_{M}\log\psi^{2}\right)\right]\nonumber \\
 & \qquad+\frac{\chi}{\psi}\frac{\tilde{m}^{2}}{\kappa_{\mathrm{f}}^{2}}\left(\nabla_{N}\log\tilde{m}\right)\Big\}.
\end{align}
Using the last two equations and $T_{MN}^{\mathrm{matter}}=T_{MN}^{\mathrm{Maxwell}}+T_{MN}^{\mathrm{fluid}}$,
we have
\begin{align*}
\nabla^{M}T_{MN}^{\mathrm{matter}} & =\frac{\kappa_{\mathrm{f}}^{2}}{\kappa_{\phi}^{2}}\psi^{2}\Big\{\nabla_{N}\tilde{p}+\left(u^{M}\nabla_{M}\right)\left(\tilde{\eta}u_{N}\right)+\frac{\chi}{\psi}\frac{\tilde{m}^{2}}{\kappa_{\mathrm{f}}^{2}}\left(\nabla_{N}\log\tilde{m}\right)\\
 & \qquad+\left(\tilde{\eta}u_{N}\right)\left[\left(\nabla^{M}u_{M}\right)+\left(u^{M}\nabla_{M}\log\psi^{2}\right)\right]+\frac{\kappa_{\phi}^{2}}{\kappa_{\mathrm{f}}^{2}\psi^{2}}F_{MN}J_{\mathrm{fluid}}^{M}\Big\}.
\end{align*}
The conservation of stress tensor requires the right hand side to
vanish, which yields
\begin{equation}
\nabla_{N}\tilde{p}+\left[\frac{d}{d\tau}+\left(\nabla_{M}u^{M}\right)+\left(\frac{d}{d\tau}\log\psi^{2}\right)\right]\tilde{\eta}_{N}+\frac{\chi}{\psi}\left(\nabla_{N}\frac{\tilde{m}^{2}}{2\kappa_{\mathrm{f}}^{2}}\right)+F_{MN}\tilde{J}_{\mathrm{fluid}}^{M}=0,\label{eq:BulkFluidEOM_TOV_Covariant}
\end{equation}
where we have defined the fluid's enthalpy current $\tilde{\eta}^{M}\equiv\tilde{\eta}u^{M}$
and the time derivative in local inertial frame $d/d\tau\equiv u^{M}\nabla_{M}$.
This equation is the covariant form of a generalized version of the
Tolman-Oppenheimer-Volkoff (TOV) equation. 

\bigskip{}


\section{Hydrostatic Equilibrium Configuration}

In static AdS black hole background, the hydrostatic fluid velocity
takes the following form
\begin{equation}
u^{t}(z)=\frac{1}{\sqrt{-g_{tt}(z)}},\qquad u^{I}=0,
\end{equation}
and all quantities except the velocity-potentials are functions of
$z$ only. We assume $A_{I}=0$ and the metric is diagonal. First
of all, in hydrostatic configuration, the bulk EOMs $\nabla_{M}\left(\psi^{2}\tilde{s}u^{M}\right)=0$,
$\nabla_{M}\left(\psi^{2}\tilde{\rho}u^{M}\right)=0$ and $\nabla_{M}\left[\psi^{2}\left(\tilde{m}^{2}-m_{\phi}^{2}\right)\tilde{\varsigma}u^{M}\right]=0$
are all trivially satisfied. Thus the entropy and charge currents
are identically conserved.

\subsection{Velocity-Potentials and a Consistency Constraint}

From (\ref{eq:BulkFluidEOM_Thermasy}) and (\ref{eq:BulkFluidEOM_Clebsch}),
we can write 
\begin{align*}
\theta_{\mathrm{s}}(x) & =\left[\sqrt{-g_{tt}(z)}\,\tilde{T}(z)\right]t+\theta_{\mathrm{s}}^{(0)}(z,\vec{x}),\\
\theta(x) & =\left[-\sqrt{-g_{tt}(z)}\,\tilde{\mu}(z)+A_{t}(z)\right]t+\theta^{(0)}(z,\vec{x}),\\
\theta_{\mathrm{m}}(x) & =\left[\sqrt{-g_{tt}(z)}\,\tilde{m}(z)\right]t+\theta_{\mathrm{m}}^{(0)}(z,\vec{x}).
\end{align*}
Here we will assume we are in the quantum branch where $\tilde{m}=u^{M}\partial_{M}\theta_{\mathrm{m}}$.
If we are in the classical branch where $\tilde{m}=m_{\phi}$, the
last equation for $\theta_{\mathrm{m}}$ will become irrelevant. Then
(\ref{eq:BulkFluidEOM_Velocity}) gives
\begin{align*}
u_{i} & =\frac{1}{\tilde{\eta}}\left\{ -\tilde{s}\partial_{i}\theta_{\mathrm{s}}^{(0)}(z,\vec{x})+\tilde{\rho}\left[\partial_{i}\theta^{(0)}(z,\vec{x})-A_{i}(z)\right]-\left[\tilde{m}(z)^{2}-m_{\phi}^{2}\right]\tilde{\varsigma}\partial_{i}\theta_{\mathrm{m}}^{(0)}(z,\vec{x})\right\} ,\\
u_{z} & =\frac{t}{\tilde{\eta}}\Big\{-\tilde{s}\partial_{z}\left[\sqrt{-g_{tt}}\,\tilde{T}(z)\right]+\tilde{\rho}\partial_{z}\left[-\sqrt{-g_{tt}}\,\tilde{\mu}(z)+A_{t}(z)\right]\\
 & \qquad-\left[\tilde{m}(z)^{2}-m_{\phi}^{2}\right]\tilde{\varsigma}(z)\partial_{z}\left[\sqrt{-g_{tt}}\,\tilde{m}(z)\right]\Big\}\\
 & \quad+\frac{1}{\tilde{\eta}}\left\{ -\tilde{s}\partial_{z}\theta_{\mathrm{s}}^{(0)}(z,\vec{x})+\tilde{\rho}\left[\partial_{z}\theta^{(0)}(z,\vec{x})-A_{z}(z)\right]-\left[\tilde{m}(z)^{2}-m_{\phi}^{2}\right]\tilde{\varsigma}\partial_{z}\theta_{\mathrm{m}}^{(0)}(z,\vec{x})\right\} .
\end{align*}
To ensure $u_{I}=0$ so as to be self-consistent, the three $\left\{ \ldots\right\} $
in the above equations must vanish separately: 
\begin{align*}
\tilde{s}(z)\partial_{z}\left[\sqrt{-g_{tt}}\,\tilde{T}(z)\right]+\tilde{\rho}(z)\partial_{z}\left[\sqrt{-g_{tt}}\,\tilde{\mu}(z)-A_{t}(z)\right]+\left[\tilde{m}(z)^{2}-m_{\phi}^{2}\right]\tilde{\varsigma}(z)\partial_{z}\left[\sqrt{-g_{tt}}\,\tilde{m}(z)\right] & =0,\\
-\tilde{s}(z)\partial_{z}\theta_{\mathrm{s}}^{(0)}(z)+\tilde{\rho}(z)\left[\partial_{z}\theta^{(0)}(z)-A_{z}(z)\right]-\left[\tilde{m}(z)^{2}-m_{\phi}^{2}\right]\tilde{\varsigma}\partial_{z}\theta_{\mathrm{m}}^{(0)}(z,\vec{x}) & =0,\\
-\tilde{s}(z)\partial_{i}\theta_{\mathrm{s}}^{(0)}(z,\vec{x})+\tilde{\rho}(z)\left[\partial_{i}\theta^{(0)}(z,\vec{x})-A_{i}(z)\right]-\left[\tilde{m}(z)^{2}-m_{\phi}^{2}\right]\tilde{\varsigma}\partial_{i}\theta_{\mathrm{m}}^{(0)}(z,\vec{x}) & =0.
\end{align*}
The last two equations are not really relevant, but the first equation
is highly non-trivial. Equivalently, for the first equation, we can
say that we have a non-trivial bulk EOM from $u_{z}=0$:
\begin{equation}
\tilde{s}\partial_{z}\left[\sqrt{-g_{tt}}\,\tilde{T}\right]+\tilde{\rho}\partial_{z}\left[\sqrt{-g_{tt}}\,\tilde{\mu}-A_{t}\right]+\left(\tilde{m}^{2}-m_{\phi}^{2}\right)\tilde{\varsigma}\partial_{z}\left[\sqrt{-g_{tt}}\,\tilde{m}\right]=0.\label{eq:BulkEOM_Fluid_uz}
\end{equation}
Using the EOMs $\tilde{s}=\partial\tilde{p}/\partial\tilde{T}$, $\tilde{\rho}=\partial\tilde{p}/\partial\tilde{\mu}$
and (\ref{eq:BulkFluidEOM_EnthalpyDensity}), the above equation can
be written as
\begin{equation}
\tilde{\eta}\left(-\frac{\partial\log\sqrt{-g_{tt}}}{\partial z}\right)=\sum_{\tilde{X}=\tilde{T},\tilde{\mu},\tilde{m}}\frac{\partial\tilde{p}}{\partial\tilde{X}}\frac{\partial\tilde{X}}{\partial z}-\frac{\tilde{\rho}}{\sqrt{-g_{tt}}}\frac{\partial A_{t}}{\partial z}+\frac{\chi}{\psi}\frac{\partial}{\partial z}\frac{\tilde{m}^{2}}{2\kappa_{\mathrm{f}}^{2}},\label{eq:BulkEOM_Fluid_uz=00003D0}
\end{equation}
where the enthalpy density $\tilde{\eta}$ is given by (\ref{eq:BulkFluidEOM_EnthalpyDensity}). 

If we are in the $\tilde{m}=m_{\phi}$ classical branch, we will just
have the equation
\[
\tilde{s}\partial_{z}\left[\sqrt{-g_{tt}}\,\tilde{T}\right]+\tilde{\rho}\partial_{z}\left[\sqrt{-g_{tt}}\,\tilde{\mu}-A_{t}\right]=0
\]
instead of (\ref{eq:BulkEOM_Fluid_uz}), then
\[
\tilde{\eta}\left(-\frac{\partial\log\sqrt{-g_{tt}}}{\partial z}\right)=\sum_{\tilde{X}=\tilde{T},\tilde{\mu}}\frac{\partial\tilde{p}}{\partial\tilde{X}}\frac{\partial\tilde{X}}{\partial z}-\frac{\tilde{\rho}}{\sqrt{-g_{tt}}}\frac{\partial A_{t}}{\partial z}
\]
instead of (\ref{eq:BulkEOM_Fluid_uz=00003D0}). Since this is just
a special case of (\ref{eq:BulkEOM_Fluid_uz=00003D0}), we will not
mention this case separately in the following.

We now show that the above equation (\ref{eq:BulkEOM_Fluid_uz=00003D0})
is exactly the same as the non-vanishing component of the covariant
TOV equation (\ref{eq:BulkFluidEOM_TOV_Covariant}) derived from stress
tensor conservation, under a certain consistency condition. Notice
for (\ref{eq:BulkFluidEOM_TOV_Covariant}), in our current case, we
have $d\tilde{\eta}_{N}/d\tau=\Gamma_{tN}^{t}\tilde{\eta}$, then
the $z$-component of (\ref{eq:BulkFluidEOM_TOV_Covariant}) reads
\[
\frac{\partial\tilde{p}}{\partial z}+\tilde{\eta}\Gamma_{tz}^{t}+\frac{\chi}{\psi}\frac{\partial}{\partial z}\frac{\tilde{m}^{2}}{2\kappa_{\mathrm{f}}^{2}}-\tilde{\rho}u^{t}\frac{\partial A_{t}}{\partial z}=0.
\]
Using $\Gamma_{tz}^{t}=\partial_{z}\log\sqrt{-g_{tt}}$ it becomes
\begin{equation}
\tilde{\eta}\left(-\frac{\partial\log\sqrt{-g_{tt}}}{\partial z}\right)=\frac{\partial\tilde{p}}{\partial z}-\frac{\tilde{\rho}}{\sqrt{-g_{tt}}}\frac{\partial A_{t}}{\partial z}+\frac{\chi}{\psi}\frac{\partial}{\partial z}\frac{\tilde{m}^{2}}{2\kappa_{\mathrm{f}}^{2}}.\label{eq:TOVEquation_Conservation}
\end{equation}
Now if we compare this equation with (\ref{eq:BulkEOM_Fluid_uz=00003D0}),
we find they are almost the same except for the first terms on the
right hand side. Thus for consistency, we shall require
\[
\frac{\partial\tilde{p}}{\partial z}=\sum_{\tilde{X}=\tilde{T},\tilde{\mu},\tilde{m}}\frac{\partial\tilde{p}}{\partial\tilde{X}}\frac{\partial\tilde{X}}{\partial z},
\]
which means the $z$-dependence in the EOS takes the following form
\begin{equation}
\tilde{p}(z)=\tilde{p}\left[\tilde{T}(z),\tilde{\mu}(z),\tilde{m}(z),h_{\alpha}\right].
\end{equation}
This is a statement that the EOS may depend on additional parameters
$h_{\alpha}$ (dimensionful or dimensionless), but these parameters
must be constant, not local functions, and the only locally varying
parameters are $\tilde{T}$, $\tilde{\mu}$ and $\tilde{m}$. In fact,
this is the consequence of the principle of the velocity-potential
representation: any locally varying field in the EOS must have its
own corresponding velocity-potential (the $\theta$'s), otherwise
the formalism is not self-consistent! Had we not introduced the potential
$\theta_{\mathrm{m}}$ for $\tilde{m}$, we would not reach a consistent
result either. Here we see under the above consistency condition of
the EOS, we can derive the TOV equation from two seemingly different
approaches: one is the stress tensor conservation shown in the previous
section, the other the $u_{z}=0$ condition due to the velocity-potential
formalism shown in this section.

\subsection{Bulk Equations of Motion and Stress Tensor}

Here we only consider the quantum  branch $\tilde{m}=u^{M}\partial_{M}\theta_{\mathrm{m}}$.
We now collect all non-trivial bulk EOMs for hydrostatic configurations
in the following. EOMs for $\psi$ and $\chi$ are
\begin{align}
\left[\nabla^{2}-m_{\phi}^{2}+\tilde{m}(z)^{2}\right]\psi(z) & =0,\label{eq:BulkEOM_Fluid_RadialProfile_psi}\\
\left[\nabla^{2}-m_{\phi}^{2}+\tilde{m}(z)^{2}\right]\chi(z) & =-4\kappa_{\mathrm{f}}^{2}\tilde{p}(z)\psi(z).\label{eq:BulkEOM_Fluid_RadialProfile_chi}
\end{align}
The TOV equation for static fluid with $u_{z}=0$ reads 
\begin{equation}
\tilde{\eta}(z)\left[-\frac{\partial\log\sqrt{-g_{tt}(z)}}{\partial z}\right]=\frac{\partial\tilde{p}(z)}{\partial z}-\frac{\tilde{\rho}(z)}{\sqrt{-g_{tt}(z)}}\frac{\partial A_{t}(z)}{\partial z}+\frac{\chi(z)}{\psi(z)}\frac{\partial}{\partial z}\frac{\tilde{m}(z)^{2}}{2\kappa_{\mathrm{f}}^{2}},\label{eq:BulkEOM_Fluid_TOV}
\end{equation}
where 
\begin{equation}
\tilde{\eta}=\tilde{\varepsilon}+\tilde{p}+\frac{\partial\tilde{p}}{\partial\log\tilde{m}}+\frac{\chi}{\psi}\frac{\tilde{m}^{2}}{\kappa_{\mathrm{f}}^{2}}.
\end{equation}
In the above equations the functional forms of all $\tilde{X}(z)=\tilde{X}\left[\tilde{T}(z),\tilde{\mu}(z),\tilde{m}(z),h_{\alpha}\right]$
where $\tilde{X}=\tilde{\varepsilon},\tilde{p},\tilde{\rho}$ are
all given by the EOS and $h_{\alpha}$ are some additional constant
parameters in the EOS. Here the TOV equation (\ref{eq:BulkEOM_Fluid_TOV})
plays the role of Newton's second law in the fluid dynamics: the enthalpy
density $\tilde{\eta}$ is the analog of mass; $-\partial_{z}\log\sqrt{-g_{tt}(z)}$
is the gravitational acceleration (positive when pointing inward);
$\partial_{z}\tilde{p}$ is the buoyant force, the thermal force that
maintains the balance of a star in astrophysics; the term involving
$\tilde{\rho}$ is the electric force (this term is positive when
the force is repulsive and points outward); the last term involving
derivative of $\tilde{m}^{2}$ is the quantum force due to mass renormalization
effect that we mentioned earlier: what is behind the $z$-derivative
is precisely the quantum potential $U_{\mathrm{Q}}$.

The fluid will contribute to bulk Einstein and Maxwell equations through
its stress tensor and charge current. The non-vanishing component
of the fluid's charge current is
\begin{equation}
J_{\mathrm{fluid}}^{t}=\frac{\tilde{\rho}}{\sqrt{-g_{tt}}}\frac{\kappa_{\mathrm{f}}^{2}}{\kappa_{\phi}^{2}}\psi^{2}.
\end{equation}
The non-vanishing components of the fluid's stress tensor are:%
\footnote{Notice 
\[
\left(\frac{\partial\chi}{\partial z}\right)\left(\frac{\partial\psi}{\partial z}\right)=\left(\frac{\partial}{\partial z}\sqrt{\psi\chi}\right)^{2}-\frac{1}{4}\left(\psi\chi\right)\left(\frac{\partial}{\partial z}\log\frac{\psi}{\chi}\right)^{2}.
\]
} 
\begin{align}
-\left(T_{t}^{t}\right)_{\mathrm{fluid}} & =\frac{\kappa_{\mathrm{f}}^{2}}{\kappa_{\phi}^{2}}\psi^{2}\left(\tilde{\varepsilon}+\frac{\partial\tilde{p}}{\partial\log\tilde{m}}+\frac{\chi}{\psi}\frac{\tilde{m}^{2}}{\kappa_{\mathrm{f}}^{2}}\right)+\frac{1}{2\kappa_{\phi}^{2}}\left[g^{zz}\left(\frac{\partial\chi}{\partial z}\right)\left(\frac{\partial\psi}{\partial z}\right)+\left(m_{\phi}^{2}-\tilde{m}^{2}\right)\chi\psi\right],\\
\left(T_{i}^{i}\right)_{\mathrm{fluid}} & =\frac{\kappa_{\mathrm{f}}^{2}}{\kappa_{\phi}^{2}}\psi^{2}\tilde{p}-\frac{1}{2\kappa_{\phi}^{2}}\left[g^{zz}\left(\frac{\partial\chi}{\partial z}\right)\left(\frac{\partial\psi}{\partial z}\right)+\left(m_{\phi}^{2}-\tilde{m}^{2}\right)\chi\psi\right],\\
\left(T_{z}^{z}\right)_{\mathrm{fluid}} & =\frac{\kappa_{\mathrm{f}}^{2}}{\kappa_{\phi}^{2}}\psi^{2}\tilde{p}+\frac{1}{2\kappa_{\phi}^{2}}\left[g^{zz}\left(\frac{\partial\chi}{\partial z}\right)\left(\frac{\partial\psi}{\partial z}\right)-\left(m_{\phi}^{2}-\tilde{m}^{2}\right)\chi\psi\right],
\end{align}
where $i$ index is not summed in the above expression. We have seen
that due to isotropy in the transverse spatial directions, the stress
tensor has only three independent diagonal components: the temporal,
the radial and the transverse ones that are listed explicitly in the
above equations. A slightly different but also useful parametrization
of the stress tensor is given by the following three linear combinations
of the above components. The first one is the trace of the stress
tensor:
\begin{align}
\kappa_{\phi}^{2}g_{MN}T_{\mathrm{fluid}}^{MN} & =\kappa_{\mathrm{f}}^{2}\psi^{2}\left(-\tilde{\varepsilon}+d\cdot\tilde{p}-\frac{\partial\tilde{p}}{\partial\log\tilde{m}}\right)\nonumber \\
 & \quad-m_{\phi}^{2}\chi\psi-\frac{d-1}{2}\left[g^{zz}\left(\frac{\partial\chi}{\partial z}\right)\left(\frac{\partial\psi}{\partial z}\right)+\left(m_{\phi}^{2}-\tilde{m}^{2}\right)\chi\psi\right].
\end{align}
The second one can be viewed as a measurement of anisotropy between
temporal direction and spatial directions, and is related to the enthalpy
density (notice there is no sum of $i$ in the following) 
\begin{equation}
\kappa_{\phi}^{2}\left(T_{t}^{t}-T_{i}^{i}\right)_{\mathrm{fluid}}=-\kappa_{\mathrm{f}}^{2}\psi^{2}\tilde{\eta}.
\end{equation}
The third one is the measurement of anisotropy between radial direction
and transverse spatial directions (no sum of $i$)
\begin{equation}
\kappa_{\phi}^{2}\left(T_{z}^{z}-T_{i}^{i}\right)_{\mathrm{fluid}}=g^{zz}\left(\frac{\partial\chi}{\partial z}\right)\left(\frac{\partial\psi}{\partial z}\right),
\end{equation}
and this is a pure quantum effect due to vacuum polarization.

\subsection{Degrees of Freedom and Boundary Conditions}

Now we have a set of EOMs, including (\ref{eq:BulkEOM_Fluid_RadialProfile_psi}),
(\ref{eq:BulkEOM_Fluid_RadialProfile_chi}) and (\ref{eq:BulkEOM_Fluid_TOV}).
Let us count how many degrees of freedom we have and what corresponding
boundary conditions we have to impose.
\begin{itemize}
\item The independent components of Einstein equation, which are not listed
explicitly here, completely determine all the metric components when
appropriate boundary/horizon conditions are imposed. So does the Maxwell
equation to the gauge field. Thus in the following, for the purpose
of counting degrees of freedom alone, we will think of the metric and
gauge field as already fixed.
\item The boundary condition $u^{z}\big|_{z=\epsilon}=0$ is already satisfied
by the hydrostatic ansatz.
\item The EOMs for both $\psi(z)$ and $\chi(z)$, (\ref{eq:BulkEOM_Fluid_RadialProfile_psi})
and (\ref{eq:BulkEOM_Fluid_RadialProfile_chi}), are second order
differential equations, which determine them up to \emph{two} integration
constants \emph{for each}. We shall impose the following four boundary
conditions to fix them:
\begin{gather}
\left[z\frac{\partial}{\partial z}-\Delta_{\mathrm{ct}}(\lambda,\epsilon)\right]\sqrt{\psi(z)\chi(z)}\Big|_{z=\epsilon}=0,\label{eq:Fluid_BoundaryCondition_1}\\
\sqrt{\psi(z)\chi(z)}\Big|_{z\rightarrow z_{\mathrm{h}}}\textrm{ is regular},
\end{gather}
and
\begin{gather}
\frac{\chi(z)}{\psi(z)}\Big|_{z=\epsilon}=1,\label{eq:Fluid_BoundaryCondition_2}\\
\frac{\chi(z)}{\psi(z)}\Big|_{z\rightarrow z_{\mathrm{h}}}\textrm{ is regular}.
\end{gather}
Here the two conditions at the boundary are from (\ref{eq:Fluid_BoundaryCondition_1_Original})
and (\ref{eq:Fluid_BoundaryCondition_2_Original}). It is interesting
to notice that the condition (\ref{eq:Fluid_BoundaryCondition_1})
takes the same form as (\ref{eq:BoundaryCondition_Deformed_general})
when $J_{b}=0$ in the latter.
\item For the fluid part, we are left with a single TOV equation (\ref{eq:BulkEOM_Fluid_TOV}).
Since it is first order in derivatives, we need to impose one boundary
condition to fix \emph{one} integration constant. We can view the
TOV equation as an equation for $\tilde{T}(z)$, which determines
it in terms the other two unknown functions $\tilde{\mu}(z)$ and
$\tilde{m}(z)$. Since we want the stress tensor to be regular at
the horizon to avoid infinite backreactions, this requires all thermal
functions, particularly the pressure $\tilde{p}$, charge density
$\tilde{\rho}$ and $\tilde{m}$ to be regular at the horizon. By
inspecting the near horizon limit of the TOV equation (\ref{eq:BulkEOM_Fluid_TOV}),
we find the regularity can only be achieved if its left hand is regular,
which means the enthalpy density $\tilde{\eta}$ must vanish no slower
than $O\left(z-z_{\mathrm{h}}\right)$:
\begin{equation}
\tilde{\eta}(z)\Big|_{z\rightarrow z_{\mathrm{h}}}\sim O\left(z-z_{\mathrm{h}}\right).
\end{equation}
This can be viewed as a horizon condition that fixes the integration
constant of the TOV equation. In some cases, this can be viewed as
a statement of setting the temperature of the fluid in certain frame
to the Hawking temperature (for example, in \cite{Loranz:1995gc}).
But we will not make this statement here, because this may not look
transparent and illuminating in our current fluid formalism. Rather,
we will just view this as a statement of regularity at the horizon.
\end{itemize}
Obviously, from the last entry of the above counting, we see that
the fluid part of the dynamics is not completely deterministic so
far, because we have used up all the non-trivial equations but still
have two undetermined functions $\tilde{\mu}(z)$ and $\tilde{m}(z)$.
To completely determine the dynamics, we need to supply two more bulk
equations. In fact, it is a well known fact that if the EOS is more-than-one-dimensional
(i.e. depends on more than one local function: three for our case),
the fluid dynamics is not fully deterministic and additional EOMs
have to be supplied from elsewhere, by some models or by going to
more microscopic levels such as the kinetic theory or even quantum
field theory. We will discuss one of the two additional equations
in the next section by considering charge dynamics, which determines
$\tilde{\mu}(z)$. The discussion of the second equation will be left
to the follow-up paper \cite{Wu:2016_Note3}; it involves the conformal
anomaly and determines $\tilde{m}(z)$.

\bigskip{}


\section{Fluid Dynamics of Charges}

So far, our fluid dynamics applies to both charged and neutral fluids.
In this section, we discuss some aspects of the dynamics that is only
relevant to charged fluids, and show where the pairing fluctuation
pseudogap in BCS-BEC crossover can arise in this bulk fluid picture.

\subsection{Chemical Potential in Thermal Equilibrium}

In this subsection, we supply one more bulk EOM to determine the ratio
$\tilde{\mu}/\tilde{T}$ for the charged fluid.

First, we want to impose the following physical requirement: the
electric force experienced by the bulk fluid always points outward
toward the boundary, i.e. the second term on the right hand side
of the TOV equation (\ref{eq:BulkEOM_Fluid_TOV}) is positive everywhere
outside the horizon. This implies
\begin{equation}
\tilde{\rho}A_{t}\geqslant0.
\end{equation}
This is simply a statement that the charge of the fluid outside the
horizon is everywhere the same as the the charge of the black hole.
This is a very physical assumption since any local charge density
of the opposite sign will not be stable and will be neutralized by
the opposite sign charge density around it, and a global charge density
of opposite sign will be more likely to be eaten up by the black hole
and hence it is less stable than the charge density of the same sign.
In terms of the dual field theory language, this is simply the fact
that an incoherent Cooper pair carries a charge of the same sign as
that of the elementary fermion (because the former is made of a pair
of the latter). Notice the sign of $\tilde{\rho}$ is proportional
to the sign of $\tilde{\mu}$, while independent of the sign of $q_{\phi}$.%
\footnote{For example, for free charged particles in flat spacetime, for small
chemical potential, we have $\tilde{\rho}\sim q_{\phi}^{2}\tilde{\mu}$.
In general, $q_{\phi}^{-1}\tilde{\rho}$ is proportional to an odd
function of $q_{\phi}\tilde{\mu}$.%
} We thus have
\begin{equation}
\tilde{\mu}(z)A_{t}(z)\geqslant0\qquad\qquad z\in\left[\epsilon,z_{\mathrm{h}}\right].
\end{equation}

The main idea here is the approximation widely used in many different
contexts of physics: in equilibrium configurations, the local chemical
potential is determined by the local gauge field associated with the
same gauge symmetry, i.e. $\tilde{\mu}$ will be determined by $A_{M}$.
In flat spacetime, this can be simply expressed as $\tilde{\mu}=A_{t}$.
The corresponding curved spacetime version in our notation is sometimes
written as
\[
\tilde{\mu}(z)=u^{M}(z)A_{M}(z)=\frac{1}{\sqrt{-g_{tt}(z)}}A_{t}(z)
\]
in the literature. It has been used, for example, for holographic
electron stars in \cite{Hartnoll:2010gu}. However, there is an obvious
problem with it: it is not gauge invariant! An equivalent way of thinking
of it is that the above relation can be obtained by setting the Clebsch
potential $\theta$ to be time-independent in (\ref{eq:BulkFluidEOM_Clebsch}),
but it is hard to see why this can be generally true on a physical
ground. To cure this problem, we take a different perspective. If
the fluid is slightly out of equilibrium, first order hydrodynamics
\cite{Jensen:2011xb} tells us that the charge current will have an
additional term
\[
\tilde{J}_{M}^{\mathrm{diss}}=\sigma\left(F_{MN}u^{N}-\tilde{T}\nabla_{M}\frac{\tilde{\mu}}{\tilde{T}}\right),
\]
where $\sigma$ is the conductivity. This term is dissipative in nature
and contributes to entropy production, thus for equilibrium configurations,
it shall vanish identically. This yields a gauge invariant condition
\begin{equation}
\nabla_{M}\frac{\tilde{\mu}}{\tilde{T}}=\frac{1}{\tilde{T}}F_{MN}u^{N}.
\end{equation}
In fact, this equation can be derived using the Boltzmann-Vlasov equation
in curved spacetime \cite{Debbasch:2009(I),Debbasch:2009(II)} for
equilibrium configurations \cite{Romatschke:2012up}. It is also given
in eq. (11) of \cite{Jensen:2012jh} if the proper acceleration $a_{\mu}$
is canceled in the first two equations there. Its only non-trivial
component is the $z$-component
\begin{equation}
\frac{\partial}{\partial z}\frac{\tilde{\mu}(z)}{\tilde{T}(z)}=\frac{1}{\sqrt{-g_{tt}(z)}\tilde{T}(z)}\frac{\partial A_{t}(z)}{\partial z}.
\end{equation}
It can be directly integrated to give the solution
\begin{equation}
\frac{\tilde{\mu}(z)}{\tilde{T}(z)}=\frac{\tilde{\mu}(z)}{\tilde{T}(z)}\Big|_{z=z_{0}}+\int_{z_{0}}^{z}d\xi\frac{1}{\sqrt{-g_{tt}(\xi)}\tilde{T}(\xi)}\frac{\partial A_{t}(\xi)}{\partial\xi},\label{eq:ChemicalPotential_Solution}
\end{equation}
where $z_{0}$ is a constant that can be chosen to be either $z_{\mathrm{h}}$
or $\epsilon$ depending on where we want to impose the boundary condition.
In flat spacetime when $g_{tt}=-1$ and $\tilde{T}$ is constant,
this solution reduces to the relation $\tilde{\mu}=A_{t}$. Now the
TOV equation (\ref{eq:BulkEOM_Fluid_TOV}) can be written as
\begin{equation}
\tilde{\eta}(z)\left[-\frac{\partial\log\sqrt{-g_{tt}(z)}}{\partial z}\right]=\frac{\partial\tilde{p}(z)}{\partial z}+\tilde{\mu}(z)\tilde{\rho}(z)\left[-\frac{\partial}{\partial z}\log\frac{\tilde{\mu}(z)}{\tilde{T}(z)}\right]+\frac{\chi(z)}{\psi(z)}\frac{\partial}{\partial z}\frac{\tilde{m}(z)^{2}}{2\kappa_{\mathrm{f}}^{2}}.
\end{equation}

\subsection{Bulk Plasma Oscillation and Pseudogap in AC Conductivity}

Now we look at a set of small linear perturbations in the bulk. We
look at the vectorial sector which includes small variations of $u_{x}(t,z)$,
$A_{x}(t,z)$, $g_{tx}(t,z)$ and $g_{zx}(t,z)$, where $x$ is one
of the transverse spatial directions. These set of fields decouple
from perturbations of other bulk fields at linear level due to the
$SO(d-1)$ rotational symmetry in the transverse spatial directions.
We want to show how the pseudogap in the longitudinal AC conductivity
of the dual field theory can be related to the bulk plasma oscillation
of the incoherent fluid. We will work in the gauge $g_{zx}=0$.

The Einstein and Maxwell equations take the covariant forms $G_{MN}+\Lambda g_{MN}=\kappa_{g}^{2}T_{MN}$
and $\nabla^{M}F_{MN}=-e_{A}^{2}J_{N}$. Varying (\ref{eq:BulkFluidEOM_TOV_Covariant})
yields
\[
\tilde{\eta}\frac{\partial u_{x}}{\partial t}+\frac{\kappa_{\mathrm{f}}^{2}}{\kappa_{\phi}^{2}}\psi^{2}\tilde{\rho}\frac{\partial A_{x}}{\partial t}=0.
\]
The $zx$-component of the Einstein equations is
\[
\frac{\partial^{2}g_{tx}}{\partial z\partial t}-\frac{\partial\log g_{\perp}}{\partial z}\frac{\partial g_{tx}}{\partial t}=-\frac{2\kappa_{g}^{2}}{e_{A}^{2}}\frac{\partial A_{t}}{\partial z}\frac{\partial A_{x}}{\partial t}.
\]
The $x$-component of the Maxwell equation is
\begin{gather*}
\frac{1}{g^{\perp}}\frac{\partial}{\partial z}\left(g^{zz}g^{\perp}\frac{\partial A_{x}}{\partial z}\right)+g^{zz}\frac{\partial\log\sqrt{-g}}{\partial z}\frac{\partial A_{x}}{\partial z}+g^{tt}\frac{\partial^{2}A_{x}}{\partial t^{2}}+e_{A}^{2}\frac{\kappa_{\mathrm{f}}^{2}}{\kappa_{\phi}^{2}}\psi^{2}\tilde{\rho}u_{x}\\
+g^{tt}g^{zz}\frac{\partial A_{t}}{\partial z}\left(\frac{\partial\log g_{\perp}}{\partial z}g_{tx}-\frac{\partial g_{tx}}{\partial z}\right)=0,
\end{gather*}
where $g^{\perp}(z)=1/g_{\perp}(z)$. Using the previous two equations
to cancel $u_{x}$ and $g_{tx}$ in the above equation and Fourier
transforming $t$ to frequency $\omega$, we have
\begin{equation}
\left[\frac{\partial^{2}}{\partial z^{2}}+\frac{\partial\log\left(g^{zz}g^{\perp}\sqrt{-g}\right)}{\partial z}\frac{\partial}{\partial z}+\frac{-g^{tt}}{g^{zz}}\omega^{2}+g^{tt}\frac{2\kappa_{g}^{2}}{e_{A}^{2}}\left(\frac{\partial A_{t}}{\partial z}\right)^{2}-e_{A}^{2}\frac{\kappa_{\mathrm{f}}^{4}}{\kappa_{\phi}^{4}}\frac{\psi^{4}\tilde{\rho}^{2}}{\tilde{\eta}g^{zz}}\right]\frac{\partial A_{x}}{\partial t}=0.\label{eq:PerturbativeEOM_Ax}
\end{equation}
We now define a new field variable and a new radial coordinate (the
tortoise coordinate) 
\begin{equation}
\mathcal{A}\equiv\left(g_{\perp}\right)^{\frac{d-3}{4}}\frac{\partial A_{x}}{\partial t},\qquad r=\int_{0}^{z}\sqrt{\frac{g_{zz}(z^{\prime})}{-g_{tt}(z^{\prime})}}dz^{\prime}.
\end{equation}
For the new radial coordinate, we have
\begin{equation}
r\Big|_{z\rightarrow\epsilon}=z,\qquad r\Big|_{z\rightarrow z_{\mathrm{h}}}=-\frac{1}{4\pi T_{\mathrm{H}}}\log\left(z_{\mathrm{h}}-z\right),
\end{equation}
where $T_{\mathrm{H}}$ is the Hawking temperature. Then (\ref{eq:PerturbativeEOM_Ax})
can be written as a one-dimensional Schrödinger equation  
\begin{equation}
\left[-\frac{\partial^{2}}{\partial r^{2}}+\mathcal{V}(r)\right]\mathcal{A}=\omega^{2}\mathcal{A},\label{eq:PerturbativeEOM_Ax_Schroedinger}
\end{equation}
where the potentials are
\begin{align}
\mathcal{V} & =\mathcal{V}_{\mathrm{fluid}}+\mathcal{V}_{\mathrm{gauge}}+\mathcal{V}_{\mathrm{grav}},\\
\mathcal{V}_{\mathrm{fluid}} & =\left(-g_{tt}\right)\frac{e_{A}^{2}\tilde{\rho}^{2}}{\tilde{\eta}}\left(\frac{\kappa_{\mathrm{f}}}{\kappa_{\phi}}\psi\right)^{4},\\
\mathcal{V}_{\mathrm{gauge}} & =g^{zz}\frac{2\kappa_{g}^{2}}{e_{A}^{2}}\left(\frac{\partial A_{t}}{\partial z}\right)^{2},\\
\mathcal{V}_{\mathrm{grav}} & =\frac{d-3}{4g_{\perp}^{\frac{d-3}{4}}}\sqrt{\frac{-g_{tt}}{g_{zz}}}\frac{\partial}{\partial z}\left(g_{\perp}^{\frac{d-3}{4}}\sqrt{\frac{-g_{tt}}{g_{zz}}}\frac{\partial\log g_{\perp}}{\partial z}\right)\nonumber \\
 & =\frac{d-3}{4}g_{\perp}^{-\frac{d-3}{4}}\frac{\partial}{\partial r}\left(g_{\perp}^{\frac{d-3}{4}}\frac{\partial\log g_{\perp}}{\partial r}\right).
\end{align}
Near the boundary, it is reasonable to assume that $\psi^{2}\tilde{\rho}$
falls off fast enough, i.e. the fluid does not have a high charge
density near the boundary. Then $\mathcal{V}_{\mathrm{fluid}}$ does
not change the near-boundary behavior and is subleading to $\omega^{2}$.%
\footnote{A more detailed discussion on the near-boundary behavior will be presented
in \cite{Wu:2016_Note3}. Our assumption of fast enough fall-off near
the boundary agrees with that in \cite{Faulkner:2010gj}.%
} Thus the first two terms in (\ref{eq:PerturbativeEOM_Ax}) dominate.
The two independent solutions near the boundary are 
\begin{equation}
\frac{\partial A_{x}}{\partial t}(\omega,z)\Big|_{z\rightarrow\epsilon}=\alpha_{0}(\omega)+\alpha_{1}(\omega)\, z^{d-2}.
\end{equation}
Equivalently, we have
\[
\mathcal{V}_{\mathrm{grav}}\Big|_{z\rightarrow\epsilon}=\frac{\left(d-1\right)\left(d-3\right)}{4z^{2}}=\frac{\left(d-1\right)\left(d-3\right)}{4r^{2}},\qquad\mathcal{V}_{\mathrm{gauge}}\Big|_{z\rightarrow\epsilon}\sim O\left(z^{2}\right).
\]
(\ref{eq:PerturbativeEOM_Ax_Schroedinger}) near the boundary reads
\[
\left[\frac{\partial^{2}}{\partial r^{2}}-\frac{\left(d-1\right)\left(d-3\right)}{4r^{2}}+\omega^{2}\right]\mathcal{A}=0,
\]
which has solutions
\begin{align}
\mathcal{A}\Big|_{r\rightarrow\epsilon} & \sim\sqrt{r}\left\{ H_{\frac{d}{2}-1}^{(1)}\left(\omega r\right)-\mathscr{R}(\omega)\, H_{\frac{d}{2}-1}^{(2)}\left(\omega r\right)\right\} ,\nonumber \\
\mathscr{R}(\omega) & =\frac{i\pi\left(\frac{\omega}{2}\right)^{d-2}\alpha_{0}(\omega)-\Gamma\left(\frac{d}{2}\right)\Gamma\left(\frac{d}{2}-1\right)\alpha_{1}(\omega)}{i\pi\left(\frac{\omega}{2}\right)^{d-2}\alpha_{0}(\omega)+\Gamma\left(\frac{d}{2}\right)\Gamma\left(\frac{d}{2}-1\right)\alpha_{1}(\omega)}.
\end{align}
where $\sqrt{r}H_{\frac{d}{2}-1}^{(1)}\left(\omega r\right)$ is the
mode falling toward the interior of the bulk and $\sqrt{r}H_{\frac{d}{2}-1}^{(2)}\left(\omega r\right)$
the one coming out toward the boundary.

Near the horizon, $\psi$ and $\tilde{\rho}$ are regular. We have
\[
\mathcal{V}_{\mathrm{fluid}}\Big|_{z\rightarrow z_{\mathrm{h}}}\sim O\left(z_{\mathrm{h}}-z\right)\sim\mathcal{V}_{\mathrm{gauge}}\Big|_{z\rightarrow z_{\mathrm{h}}},\qquad\mathcal{V}_{\mathrm{grav}}\Big|_{z\rightarrow z_{\mathrm{h}}}\sim O\left[\left(d-3\right)\left(z_{\mathrm{h}}-z\right)\right],
\]
which are all vanishing near the horizon and subleading to $\omega^{2}$.
Then (\ref{eq:PerturbativeEOM_Ax_Schroedinger}) near the horizon
reads
\[
\left(\frac{\partial^{2}}{\partial r^{2}}+\omega^{2}\right)\mathcal{A}=0,
\]
which has solutions
\begin{equation}
\mathcal{A}\Big|_{r\rightarrow\infty}=\alpha_{-}(\omega)e^{i\omega r}+\alpha_{+}(\omega)e^{-i\omega r}.
\end{equation}
$\alpha_{-}$ is the mode falling into the horizon and $\alpha_{+}$
the one coming out of horizon. We shall eliminate the outgoing mode
by setting 
\begin{equation}
\alpha_{+}(\omega)=0.
\end{equation}
It is well known, particularly for $d=3$ \cite{Horowitz:2009ij},
that the problem of calculating the AC conductivity $\sigma(\omega)$
of the dual field theory can be mapped to a scattering problem of
the one-dimensional Schrödinger equation (\ref{eq:PerturbativeEOM_Ax_Schroedinger})
with potential $\mathcal{V}(r)$. In general dimensions, the AC conductivity
is related to the complex reflection amplitude $\mathscr{R}$ as 
\begin{equation}
\sigma(\omega)\sim-\frac{i}{\omega}\frac{\alpha_{1}(\omega)}{\alpha_{0}(\omega)}\sim\left(\frac{\omega}{2}\right)^{d-3}\frac{1-\mathscr{R}(\omega)}{1+\mathscr{R}(\omega)},
\end{equation}
up to numeric factors and factors of the AdS radius $R$. Particularly,
the real part
\begin{equation}
\mathfrak{Re}\sigma(\omega)\sim\left(\frac{\omega}{2}\right)^{d-3}\frac{\big|\mathscr{T}(\omega)\big|^{2}}{\big|1+\mathscr{R}(\omega)\big|^{2}}
\end{equation}
is proportional to the real transmission coefficient $\big|\mathscr{T}(\omega)\big|^{2}=1-\big|\mathscr{R}(\omega)\big|^{2}=\big|\alpha_{-}(\omega)\big|^{2}$.

The potential $\mathcal{V}_{\mathrm{gauge}}$ is produced by the backreaction
to the metric, which is negligible in the probe limit $e_{A}\rightarrow\infty$.
$\mathcal{V}_{\mathrm{grav}}$ is identically zero when $d=3$. In
the following, we will focus on the simpler case of $d=3$,%
\footnote{For $d=4$, there is an additional $\omega$ factor appearing in the
above expressions for $\sigma(\omega)$, and the potential $\mathcal{V}_{\mathrm{grav}}$
is divergent near the boundary, both facts making the analysis of
the qualitative behavior of AC conductivity using the scattering analogy
less intuitive.%
} in the regime where $\mathcal{V}_{\mathrm{gauge}}$ does not play
an important role either. For the more generic case, the qualitative picture
may still be true, but the argument in the following based on the scattering
analog will be less transparent. Now, as $\mathcal{V}_{\mathrm{gauge}}$
and $\mathcal{V}_{\mathrm{grav}}$ are both negligible, it is easy
to recognize that it is the potential $\mathcal{V}_{\mathrm{fluid}}$
that damps the transmission amplitude in the Schrödinger scattering
problem and thus produces the pseudogap observed in the AC conductivity.
This is not surprising because pseudogap is dual to the incoherent
charged fluid in the bulk and $\mathcal{V}_{\mathrm{fluid}}$ is proportional
to its charge density $\tilde{\rho}$. Suppose $\mathcal{V}_{\mathrm{fluid}}$
reaches its maximum at $z=z_{*}$ ($r=r_{*}$), then we can define
a corresponding frequency $\omega_{*}$ as 
\begin{equation}
\omega_{*}^{2}=\mathcal{V}_{\mathrm{fluid}}\left(z_{*}\right)=-g_{tt}(z)\frac{e_{A}^{2}\tilde{\rho}^{2}(z)}{\tilde{\eta}(z)}\left[\frac{\kappa_{\mathrm{f}}}{\kappa_{\phi}}\psi(z)\right]^{4}\Big|_{z=z_{*}}.
\end{equation}
This $\omega_{*}$ is a rough estimation of the pairing fluctuation
pseudogap. (\ref{eq:PerturbativeEOM_Ax_Schroedinger}) defined a one-dimensional
problem of a photon scattering off a charged plasma with potential
$\mathcal{V}$. If $\omega>\omega_{*}$, the photon can go through
the plasma with little damping, and the conductivity is large. If
$\omega<\omega_{*}$, inside the potential $\mathcal{V}$, the photon's
local frequency (in the WKB approximation) is imaginary and its probability
density decays. The smaller $\omega$ is, the stronger the decay is,
and the smaller the conductivity is. Thus across the region where
$\omega\sim\omega_{*}$, the behavior of the AC conductivity $\sigma(\omega)$
will change qualitatively. This shows $\omega_{*}$ can be an estimation
of the pseudogap.

A physical picture emerging from this bulk analysis is the following.
As the incoherent charged fluid is free to move and backreacts to
the electric field, it can be viewed as a plasma, much like the electron
gas in metals. Measuring AC conductivity in the dual field theory
is like shining a beam of light through this bulk plasma from the
boundary and seeing how much it can get through to the other side
(the horizon). We know that the plasma will oscillate under the driving
of this external AC electric field, and it also has an intrinsic frequency,
the plasma frequency $\omega_{\mathrm{plasma}}$, which is proportional
to the charge to mass ratio ($\tilde{\rho}^{2}/\mathrm{\tilde{\eta}}$
in our case) of the plasma. Let us think of metals as a familiar example.
When the light has higher frequency than $\omega_{\mathrm{plasma}}$,
metals are essentially transparent to light. On the contrary, if the
frequency is lower than $\omega_{\mathrm{plasma}}$, the light is
quickly damped inside metals and reflected: this is why metals appear
shiny under visible lights. What happens in the bulk is very similar.
Our $\omega_{*}$ is essentially the plasma frequency of the incoherent
fluid, and the pseudogap in AC conductivity is related to damping
due to plasma oscillation in the holographic bulk. This shows how
the incoherent charged fluid in the bulk is capable of producing a
pseudogap in the AC conductivity of the dual field theory. Of course,
a more accurate analysis on the pseudogap and how soft or hard it
is has to rely on numeric calculations.

\bigskip{}


\section{Summary and Remarks}

In this paper we constructed a holographic model for the pseudogap phase
in high temperature superfluidity. This phenomenon is predicted by
the BCS-BEC crossover scenario and subsequently observed in recent
cold atom experiments. In this phase there exists a gap in the system
but the $U(1)$ symmetry is not broken nor is superfluidity developed.
Unlike the pseudogap phase in cuprate superconductivity, which has
been largely attributed to competing orders and can be modeled in
holography by introducing more bulk fields which develop their own
condensates, the pseudogap we study here originates from incoherent
Cooper pairing and defies the introduction of any other order and
additional bulk field. Using the Abelian Higgs model of holographic
superconductors as an example, we propose that the pseudogap is realized
in the bulk as the incoherent fluctuations of the charged scalar.
The fluctuations deplete the condensate, form a non-trivial bulk profile
like a normal fluid while still preserving the $U(1)$ symmetry in the
field theory via phase decoherence effect. We develop an upgraded
version of perfect fluid dynamics to serve as an effective theory
for these bulk fluctuations. It includes a pair of real radial profile
functions $\psi$ and $\chi$ which serve a triple role: (1) they
inherit the boundary conditions from the scalar; (2) they encode the
renormalization effect due to curved spacetime and shifts the negative
mass square of the scalar field to a non-negative value; and (3) their
combination $\sqrt{\psi\chi}$ is related to the real pseudogap parameter,
much like the profile of the bulk condensate is dual to the superconducting
order parameter. The pseudogap energy is related to the plasma oscillation
of this bulk fluid. 

We suggest that the scalar double-trace deformation in AdS/CFT can be used as
the holographic counterpart of the phenomenological 4-fermi interaction
used in condensed matter field theories which serves as the external
knob in the theory of BCS-BEC crossover to control the strength of
Cooper pairing. Both of them shall be viewed as low energy effective
operators slightly irrelevant in the IR, which are generated by the RG
flow from the UV. The deformed and undeformed effective actions are related
by a general relation given by (\ref{eq:GeneratingFunctional_HSTransformed})
via a path integral over the Hubbard-Stratonovich auxiliary field.
The holographic duality can be viewed as a second Hubbard-Stratonovich
transformation allowing one to integrate out the previous auxiliary field
completely. It is the presence of the double-trace deformation that
elevates the effect of fluctuations and enhances the existence of pseudogap. 

In this paper, we have built up a holographic framework and written
down the bulk dynamics. To actually solve the model, an EOS for the
bulk fluid has to be supplied. The form of EOS cannot take classical
WKB form as previously employed in holographic electron stars. The
typical wavelength of the fluid is comparable to the geometric scales
and the EOS receives non-negligible quantum corrections due to the renormalization
effect in curved spacetime. This is crucial for getting regular hydrostatic
solutions for bosonic matter in the presence of a black hole. Moreover,
since our fluid EOS is three-dimensional, while so far we have only
written down two EOMs for them (the TOV equation and the chemical
potential equation), an additional EOM has to be supplied. These issues
will be discussed in a follow-up paper \cite{Wu:2016_Note3}.

In the current model, the bulk scalar is linear and we have been focusing
on the fluctuations of this scalar alone. It is interesting to see
how our fluid dynamics will be modified by non-linear effects such
as a self-interaction of the scalar. Such non-linear terms can be
either part of the full bulk action or an effective description of
terms generated via loop effects. Phenomenologically they are related
to the residual interactions of Cooper pairs. It may be useful
to study the fluctuations of other bulk fields, such as the $U(1)$
gauge field, in a similar fashion. This might be either interesting
on its own for exploring new phases, or as a necessary part of our
current model for consistency and completeness. We will leave the
discussions on these topics for the future. We hope the model and
methods presented in this paper can serve as a holographic paradigm
for studying phases involving fluctuations in strongly coupled quantum
field theories.

\bigskip{}


\addcontentsline{toc}{section}{Acknowledgments}

\acknowledgments

We thank Daniel Dessau, Daniel Grumiller, Takaaki Ishii, Elias Kiritsis,
Sergej Moroz, Paul Romatschke, Jakob Salzer and Xiao Yin for useful
discussions. C.W. thanks Kathryn Levin for her lectures given at the
University of Chicago in 2014 and 2015 which inspired this project,
and acknowledges an early research collaboration with Leopoldo Pando-Zayas
and Diana Vaman on related issues of double-trace deformation. C.W.
also thanks the hospitality of the organizers and the feedback of
the participants of the Shanghai Workshop on Gauge/Gravity Duality
and its Applications, held at Shanghai University in August 2016,
where a preliminary talk of this paper was given. This work was supported by the Department of Energy under Grant No.~DE-FG02-91-ER-40672.  O.H.\ was also supported by a Dissertation Completion Fellowship from the Graduate School at the University of Colorado Boulder. 

\bigskip{}

\appendix


\section{Velocity-Potential Representation of Classical Fluid Dynamics }

We review the velocity-potential formalism of fluid dynamics for a
relativistic time-like perfect fluid (TPF). This approach was first
introduced by Schutz \cite{Schutz:1970my}. A useful review on this
topic is given in \cite{Brown:1992kc}.

\subsection{Covariant Off-Shell Fluid Action}

For a perfect fluid in flat spacetime, the Lagrangian density is just
the pressure $p$ in co-moving frame. This is the characteristic function
of the grand canonical ensemble, which is a function of temperature,
chemical potential and particle mass: $p=p\left[T,\mu,m\right]$.
The on-shell perfect fluid action is
\begin{equation}
S_{\mathrm{on-shell}}^{\mathrm{TPF}}\left[T,\mu,m\right]=\int d^{d+1}x\sqrt{-g}\, p\left[T,\mu,m\right].
\end{equation}
The specific form of $p\left[T,\mu,m\right]$ is given by the EOS.
In \cite{Schutz:1970my}, the independent variables are entropy $S$
and chemical potential $\mu$, while in Section 4 of \cite{Hartnoll:2010gu}
they first assume the independent variables are $\rho$ and $s$,
and then change to $\mu$ and $s$. These are all equivalent by Legendre
transformations and the thermodynamic relation $\varepsilon+p=Ts+\mu\rho$,
where $\varepsilon$, $s$ and $\rho$ are the energy density, entropy
density and charge density. A general discussion on this can be found
in Section 6 of \cite{Brown:1992kc}. Since the Lagrangian density
is just $p$ and in the thermodynamic relation only $T$ and $\mu$
are Lagrange multipliers, while all other functions are expectation
values of operators in quantum field theory, using $p=p\left[T,\mu,m\right]$
is a convenient choice.

All following discussions will be general, applying both to bosons
and fermions. Their only difference from the fluid point of view is
the form of their EOSs, which we assume to be general just as $p=p\left[T,\mu,m\right]$
in this section.

The above action still does not give the right EOMs if $T$ and $\mu$
are treated as field variables, because the EOMs from varying $T$
and $\mu$ are just the vanishing of entropy density and charge density,
which are wrong. Furthermore, we want to express it in arbitrary frame
(characterized by velocity $u^{M}$) in a covariant way. Thus we do
the following transformation and re-express $T$ and $\mu$ in a covariant
way in terms of the velocity $u^{M}$ and the velocity-potentials
$\theta_{\mathrm{s}}$ and $\theta$ \cite{Schutz:1970my}: 
\begin{align}
T & =u^{M}\partial_{M}\theta_{\mathrm{s}},\label{eq:VelocityPotential_Thermasy}\\
\mu & =u^{M}\left(-\partial_{M}\theta+A_{M}\right),\label{eq:VelocityPotential_Clebsch}
\end{align}
where $\theta_{\mathrm{s}}$ is called ``thermasy'' and $\theta$
the Clebsch potential. At this moment, these can be viewed as just
field redefinitions from $T$ and $\mu$ to $\theta_{\mathrm{s}}$
and $\theta$. Notice $\theta$ is a Stückelberg field, which transforms
under $U(1)$ gauge transformation since it couples to $A_{M}$. When
there are vortices in the fluid, more velocity-potentials will be
needed \cite{Brown:1992kc}. But this is not quite relevant for us,
thus throughout this note we will assume our fluid is \emph{curl-less}.
We will enforce the above two relations in the action by two Lagrange
multipliers, which turn out to be just the entropy density $s$ and
charge density $\rho$. The velocity field $u^{M}$ has the standard
time-like normalization 
\begin{equation}
g_{MN}u^{M}u^{N}=-1.\label{eq:Velocity_Normalization_TimeLike}
\end{equation}
This will be enforced in the action by a Lagrange multiplier $\eta$,
which turns out to be the enthalpy density $\varepsilon+p$. Here
we view $u^{M}$ with \emph{upper} index and $A_{M}$ with \emph{lower}
indices as elementary fields. 

The off-shell perfect fluid action is
\begin{align}
 & S_{\mathrm{off-shell}}^{\mathrm{TPF}}\left[\theta_{\mathrm{s}},\theta,u^{M},T,\mu,s,\rho,\eta;g_{MN},A_{M}\right]\nonumber \\
= & \int d^{d+1}x\sqrt{-g}\Big\{ p\left[T,\mu,m\right]+\frac{1}{2}\eta\left(u^{M}u_{M}+1\right)\label{eq:FluidAction_OffShell_TimeLike}\\
 & \qquad-s\left(T-u^{M}\partial_{M}\theta_{\mathrm{s}}\right)-\rho\left[\mu+u^{M}\left(\partial_{M}\theta-A_{M}\right)\right]\Big\}.\nonumber 
\end{align}

\subsection{Physical Meaning of Velocity-Potentials}

To understand the physical meaning of $\theta_{\mathrm{s}}$ and $\theta$,
let $\ell$ denote the world-line of a small element of the fluid
with affine parameter $\tau$, i.e. 
\[
\frac{d\varphi}{d\tau}=u^{M}\partial_{M}\varphi,\qquad u^{M}=\frac{dx^{M}}{d\tau},
\]
for any scalar $\varphi$. Then (\ref{eq:VelocityPotential_Thermasy})
and (\ref{eq:VelocityPotential_Clebsch}) can be written in an integral
form as 
\begin{align}
\theta_{\mathrm{s}}(\ell) & =\int_{\ell}Td\tau,\\
\theta(\ell) & =-\int_{\ell}\mu d\tau+\int_{\ell}A_{M}dx^{M},
\end{align}
These expressions also suggest that when non-vanishing, $\theta_{\mathrm{s}}$
and $\theta$ will probably have some linear dependence on time. In
the sixth paragraph of the Introduction section of \cite{Brown:1992kc},
the author gives an analogous interpretation of these velocity-potentials
as the Lagrangian coordinates of the ``fluid space'', just as the
position coordinates in Lagrangian mechanics, and each one has a gauge
freedom (since only their derivatives appear in the above transformations
of $T$ and $\mu$) that is related to a global symmetry transformation
due to certain physical properties of the fluid. Each set of values
of the velocity-potentials can be viewed as a position vector in the
``fluid space'' that labels a sub-manifold isomorphic to a hypersurface
perpendicular to the fluid's streamlines.

\subsection{Equations of Motion and On-Shell Velocity}

We now look at the EOMs derived from the above action. 

The EOMs obtained by varying $s$, $\rho$ and $\eta$ are the velocity-potential
definitions (\ref{eq:VelocityPotential_Thermasy}), (\ref{eq:VelocityPotential_Clebsch})
and the velocity normalization (\ref{eq:Velocity_Normalization_TimeLike}).
Variations with respect to $T$ and $\mu$ simply give the thermodynamic
relations in grand canonical ensemble 
\begin{equation}
s=\frac{\partial p}{\partial T},\qquad\rho=\frac{\partial p}{\partial\mu}.\label{eq:ThermoRelation_Entropy_and_Density}
\end{equation}
The EOM for $u^{M}$ is 
\begin{equation}
u_{M}=\frac{1}{\eta}\left[-s\partial_{M}\theta_{\mathrm{s}}+\rho\left(\partial_{M}\theta-A_{M}\right)\right].\label{eq:Velocity_OnShell_Eq}
\end{equation}
Multiply it by $u^{M}$ and use (\ref{eq:Velocity_Normalization_TimeLike}),
(\ref{eq:VelocityPotential_Thermasy}) and (\ref{eq:VelocityPotential_Clebsch}),
we obtain the standard thermodynamic relation 
\begin{equation}
\eta=Ts+\mu\rho=\varepsilon+p,\label{eq:LagrangeMultiplier_EnthalpyDensity}
\end{equation}
thus the on-shell value of the Lagrange multiplier $\eta$ is just
the enthalpy density. Then the on-shell value for the velocity field
is 
\begin{equation}
u_{M}=\frac{1}{\varepsilon+p}\left[-s\left(\partial_{M}\theta_{\mathrm{s}}\right)+\rho\left(\partial_{M}\theta-A_{M}\right)\right].\label{eq:Velocity_OnShell_Eq_1}
\end{equation}
This equation tells us that the two velocity-potentials $\theta_{\mathrm{s}}$
and $\theta$ are responsible for the configurations of the velocity
field, hence they get their names. Using this equation, one can further
develop a Hamiltonian description for the fluid from the action (\ref{eq:FluidAction_OffShell_TimeLike}),
as discussed in Section 3 of \cite{Brown:1992kc}. Notice that when
imposing initial or boundary conditions, one has to make sure the
the resulting $u^{M}$ given by the above equation is physical. The
two velocity-potentials appearing here describe velocity field $u^{M}$
without vorticity. In the presence of vorticity, a third velocity-potential
will be needed, which is the stream-line integral of the helicity
(or angular momentum); see Section 4 in \cite{Brown:1992kc}. 

From the above equation, we can also see through the Taub current
that we defined earlier in (\ref{eq:TaubCurrent}) and $\xi_{M}=\mu_{\mathrm{h}}u_{M}$
that 
\begin{align}
\frac{1}{q_{\phi}}\big\langle\partial_{M}\vartheta\big\rangle & =\partial_{M}\theta-\frac{s}{\rho}\partial_{M}\theta_{\mathrm{s}},\\
\mu_{\mathrm{h}} & =q_{\phi}\frac{\eta}{\rho}=q_{\phi}\frac{\varepsilon+p}{\rho},
\end{align}
where $s/\rho$ is entropy per charge. In the first equation above,
we see $q_{\phi}^{-1}\big\langle\partial_{M}\vartheta\big\rangle$
differs from the purely gauge part $\partial_{M}\theta$ only when
$s/\rho$ (i.e. entropy per charge) is non vanishing. For condensate,
which is a pure state, it always has vanishing entropy density: $s=0$.
For fluctuations/excitations, due to decoherence at finite temperature,
it always has $s\neq0$, thus $-(s/\rho)\partial_{M}\theta_{\mathrm{s}}$
contributes to $\big\langle\partial_{M}\vartheta\big\rangle$. This
term is the difference between condensate and excitations in a fluid
description.

The EOMs obtained by varying $\theta$ and $\theta_{\mathrm{s}}$
are first order differential equations: 
\begin{align}
\nabla_{M}\left(su^{M}\right) & =0,\\
\nabla_{M}\left(\rho u^{M}\right) & =0.
\end{align}
These are just the conservation of entropy current and charge current.
The entropy current is conserved here because we are dealing with
perfect fluid and in this case no dissipation is allowed in relativistic
fluid dynamics.%
\footnote{See footnote 8 in \cite{Schutz:1970my}.%
} Since we have viewed $\theta$ and $\theta_{\mathrm{s}}$ as Lagrangian
coordinates and they do not appear directly in the action, but only
through their derivatives in (\ref{eq:VelocityPotential_Thermasy})
and (\ref{eq:VelocityPotential_Clebsch}), the above conservation
equations can be viewed as a consequence of the translational invariance
of the fluid action (\ref{eq:FluidAction_OffShell_TimeLike}) in the
space of these Lagrangian coordinates. For a detailed discussion,
see Section 2.2 in \cite{Brown:1992kc}.

\subsection{Stress Tensor and Charge Current}

Taking functional derivatives of (\ref{eq:FluidAction_OffShell_TimeLike})
with respect to $g_{MN}$ and $A_{M}$, we obtain the off-shell expressions
for stress tensor and current. 
\[
T^{MN}=\eta u^{M}u^{N}+g^{MN}\left[p+\frac{1}{2}\eta\left(u^{2}+1\right)\right],\qquad J^{M}=\rho u^{M}.
\]
By inserting (\ref{eq:Velocity_Normalization_TimeLike}), the on-shell
stress tensor and current are of the standard perfect fluid form 
\begin{equation}
T^{MN}=\left(\varepsilon+p\right)u^{M}u^{N}+pg^{MN},\qquad J^{M}=\rho u^{M}.
\end{equation}

\bigskip{}


\section{Classical Branch and Conventional Fluid Dynamics }

Here we show that our new fluid dynamics reduces to the conventional
formalism in the classical branch 
\begin{equation}
\tilde{m}=m_{\phi}.
\end{equation}
Clearly, such a formalism only exists when $m_{\phi}\geqslant0$ as
we have required $\tilde{m}\geqslant0$, thus we will make this assumption
in this subsection. In this branch, the quantum potential $U_{\mathrm{Q}}=0$.
Now we have 
\begin{equation}
\tilde{\eta}=\tilde{\varepsilon}+\tilde{p},
\end{equation}
and (\ref{eq:BulkEOM_Fluid_RadialProfile_psi}) becomes simply 
\[
\nabla^{2}\psi=0,
\]
which has a trivial solution 
\[
\frac{\kappa_{\mathrm{f}}}{\kappa_{\phi}}\psi=1.
\]
Such a solution will not satisfy the boundary condition (\ref{eq:Fluid_BoundaryCondition_1})
unless 
\begin{equation}
\Delta_{\mathrm{ct}}(\lambda,\epsilon)=0,
\end{equation}
but for general relativity considered in most textbooks, this is a
reasonable boundary condition. In AdS/CFT correspondence, we have
a time-like boundary in the asymptotically AdS regime, thus have the
boundary condition (\ref{eq:Fluid_BoundaryCondition_1}). In textbook
general relativity and astrophysics, we typically consider asymptotically
flat spacetime, where the asymptotic boundary is light-like and does
not need a boundary condition with a non-trivial $\Delta_{\mathrm{ct}}$.

Now (\ref{eq:BulkEOM_Fluid_RadialProfile_chi}) reduces to
\[
\nabla^{2}\chi=-4\kappa_{\mathrm{f}}^{2}\tilde{p},
\]
which in principle should have a non-trivial solution for $\chi$.
But whatever solution $\chi$ has it does not really matter, because
$\chi$ drops off in all equations. Now in this branch the on-shell
stress tensor and charge current simply reduce to the textbook form
\begin{equation}
T_{\mathrm{fluid}}^{MN}=\left(\tilde{\varepsilon}+\tilde{p}\right)u^{M}u^{N}+\tilde{p}g^{MN},\qquad J_{\mathrm{fluid}}^{MN}=\tilde{\rho}u^{M}.
\end{equation}
The TOV equation (\ref{eq:BulkEOM_Fluid_TOV}) also reduces to the
conventional form
\begin{equation}
\left(\tilde{\varepsilon}+\tilde{p}\right)\frac{\partial\log\sqrt{-g_{tt}}}{\partial z}+\frac{\partial\tilde{p}}{\partial z}-\frac{\tilde{\rho}}{\sqrt{-g_{tt}}}\frac{\partial A_{t}}{\partial z}=0,
\end{equation}
which is the only non-trivial equation ($z$-component) from the stress
tensor conservation 
\begin{equation}
\nabla_{M}T_{\mathrm{fluid}}^{MN}+F_{MN}J_{\mathrm{fluid}}^{M}=0
\end{equation}
using $u^{t}=1/\sqrt{-g_{tt}}$.

The above analysis is pure mathematics, it does not tell us when the
$\tilde{m}=m_{\phi}$ classical branch is dynamically preferred. Physically
we can expect this will happen, or the two branches will become almost
degenerate, when the typical wavelength (energy) of the quantum modes
of the scalar field is much smaller (higher) than the typical length
(energy) scale set by the curvature of the geometry, i.e. in the WKB
limit. This is the regime when for the scattering between the scalar
field and gravitons, the energy of the scalar is much larger than
the momentum transfer in the scattering. The scalar is hard and its
mass renormalization due to the loop corrections of graviton and gauge
field is negligible. This mass renormalization due to scattering with
gravitons (and possibly the gauge field as well) is the main underlying
microscopic origin of the locally varying $\tilde{m}$ and non-trivial
$\psi$ in the quantum branch $\tilde{m}=u^{M}\partial_{M}\theta_{\mathrm{m}}$.
Thus when the typical energy scale of the fluid (i.e. scalar fluctuations)
is much higher than that of the geometry, the potential becomes vary
shallow and the geometry becomes almost flat to the fluid, thus $\psi$
becomes trivial and $\tilde{m}\simeq m_{\phi}$. Now quantum effects
such as the vacuum polarization are negligible and the fluid is almost
classical. In this case the EOS also depends only on $\tilde{T}$,
$\tilde{\mu}$ and $\tilde{m}$ as can be calculated using standard
thermal ensembles in flat spacetime, and does not depend on additional
dimensionful parameters related to the geometry, since these parameters
only enter the EOS through quantum corrections.

\bigskip{}


\end{document}